\documentclass[11pt,hyper,letterpaper]{article}
\pdfoutput=1
\usepackage{jheppub}
\usepackage{amsmath,amssymb,amsfonts,amscd,bm}
\usepackage{graphicx, wrapfig}
\usepackage{multirow}
\usepackage{verbatim}
\usepackage{appendix}
\usepackage{fancybox}

\graphicspath{{./figures/}}

\usepackage{url}
\usepackage{float}

\newcommand{\bea}{\begin{eqnarray}}
\newcommand{\eea}{\end{eqnarray}}
\newcommand{\be}{\begin{equation}}
\newcommand{\ee}{\end{equation}}

\newcommand{\WT}[1]{{\overset{\raisebox{-.1cm}{{\scriptsize $\sim$}}}{#1}}}

\newcommand{\ol}{\overline}

\newcommand{\Z}{{\mathbb Z}}
\newcommand{\R}{{\mathbb R}}
\newcommand{\C}{{\mathbb C}}

\newcommand{\Q}{{\mathbb Q}}

\newcommand{\Li}{{\rm Li}}

\def\Tr{{\rm Tr \,}}

\def\k{\kappa}

\def\frak{\mathfrak}

\def\tilde{\widetilde}
\def\hat{\widehat}

\def\CA{{\mathcal A}}

\def\CH{{\mathcal H}}
\def\CI{{\mathcal I}}

\def\CL{{\mathcal L}}
\def\CM{{\mathcal M}}
\def\CN{{\mathcal N}}
\def\CO{{\mathcal O}}
\def\CP{{\mathcal P}}

\def\CR{{\mathcal R}}
\def\CS{{\mathcal S}}

\def\CW{{\mathcal W}}

\def\CZ{{\mathcal Z}}

\def\wt{\widetilde}

\newcommand{\cp}{{\mathbb{C}}{\mathbf{P}}}

\renewcommand{\hat}{\widehat}

\title{3d-3d Correspondence Revisited}

\author[1]{Hee-Joong Chung}
\author[2]{Tudor Dimofte}
\author[1,3]{Sergei Gukov}
\author[1,4]{Piotr Su\l kowski}

\affiliation[1]{California Institute of Technology, Pasadena, CA 91125, USA}
\affiliation[2]{Institute for Advanced Study, Einstein Dr., Princeton, NJ 08540, USA}
\affiliation[3]{Max-Planck-Institut f\"ur Mathematik, Vivatsgasse 7, D-53111 Bonn, Germany}
\affiliation[4]{Faculty of Physics, University of Warsaw, ul. Ho{\.z}a 69, 00-681 Warsaw, Poland}

\abstract{In fivebrane compactifications on 3-manifolds, we point out the importance of {\it all} flat connections in the proper definition of the effective 3d $\CN=2$ theory. The Lagrangians of some theories with the desired properties can be constructed with the help of homological knot invariants that categorify colored Jones polynomials.
Higgsing the full 3d theories constructed this way recovers theories found previously by Dimofte-Gaiotto-Gukov.
We also consider the cutting and gluing of 3-manifolds along smooth boundaries and the role played by all flat connections in this operation.
\\
\\
\\
\\
\\
\\
\\
{\tt CALT 68-2887}}

\begin{document}
\cornersize{1}

\maketitle
%\tableofcontents

%%%%%%%%%%%%%%%%%%%%%%%%%%%%%%%%%%%%%%%%%%%%%%%%%%%%%%%%%%%%%%%%%%%%%

\section{Introduction}
\label{sec:intro}

It is believed that a choice of a 3-manifold $M_3$ and a Lie algebra $\frak g = \text{Lie} (G)$
of ADE type labels a supersymmetric 3d $\CN=2$ theory $T[M_3;G]$,
\be
M_3 \quad \leadsto \quad T[M_3;G] \,,
\label{3d3dbasic}
\ee
defined via compactification of the 6d $(2,0)$ theory of the same (Cartan) type on the 3-manifold $M_3$.
In most of applications, one assumes $G$ to be fixed (typically, $U(N)$ or $SU(N)$)
and views it as a correspondence between 3-manifolds and 3d theories, in which case $T[M_3; G]$ is denoted simply as $T[M_3]$.
Moreover, in such cases, the theory $T[M_3]$ can be thought of as the effective three-dimensional theory on the $\R^3$ part of
the fivebrane world-volume in an M-theory setup:
\be
\begin{matrix}
{\mbox{\rm space-time:}} & \qquad & \R^5 & \times & CY_3 \\
& \qquad & \cup &  & \cup \\
{\mbox{\rm fivebranes:}} & \qquad & \R^3 & \times & M_3
\end{matrix}
\label{surfeng}
\ee
where $M_3$ is embedded in a Calabi-Yau 3-fold $CY_3$ as a special Lagrangian submanifold.
The neighborhood of every special Lagrangian submanifold always looks like the total space of the cotangent bundle, $T^* M_3$,
which is one popular choice of $CY_3$. Another popular choice --- that appears {\it e.g.} in physical realization of knot homologies ---
is the resolved conifold geometry, {\it i.e.} when $CY_3$ is the total space of $\CO (-1) \oplus \CO (-1)$ bundle over $\cp^1$.
Various partition functions of the 3d $\CN=2$ theory $T[M_3]$ have a nice geometric interpretation and can be
realized in the setup \eqref{surfeng} by replacing $\R^3$ with a (squashed) 3-sphere, $S^2 \times_q S^1$, or $\R^2_{\hbar} \times S^1$.

The effective 3d $\CN=2$ theory $T[M_3]$ should exhibit all properties of the fivebrane system \eqref{surfeng},
including symmetries and the space of classical vacua. In particular, the fivebrane system \eqref{surfeng} has at least
three $U(1)$ symmetries which are independent of the choice of $M_3$:
one is a Cartan subgroup of the $SO(3)$ rotation symmetry of $\R^3$,
another is a rotation symmetry in two transverse dimensions of $\R^3 \subset \R^5$,
and the third $U(1)$ is the R-symmetry.
Certain combinations of these symmetries give rise to three conserved charges which are familiar in the study of
surface operators in $\CN=2$ gauge theories, {\it e.g.} the index of such 2d-4d systems depends on three universal
fugacities, whose nature is independent of the details of the theory or choice of a surface operator.

\begin{wrapfigure}{l}{0.5\textwidth}
\begin{center}
\vspace{.2cm}
\includegraphics[width=0.48\textwidth]{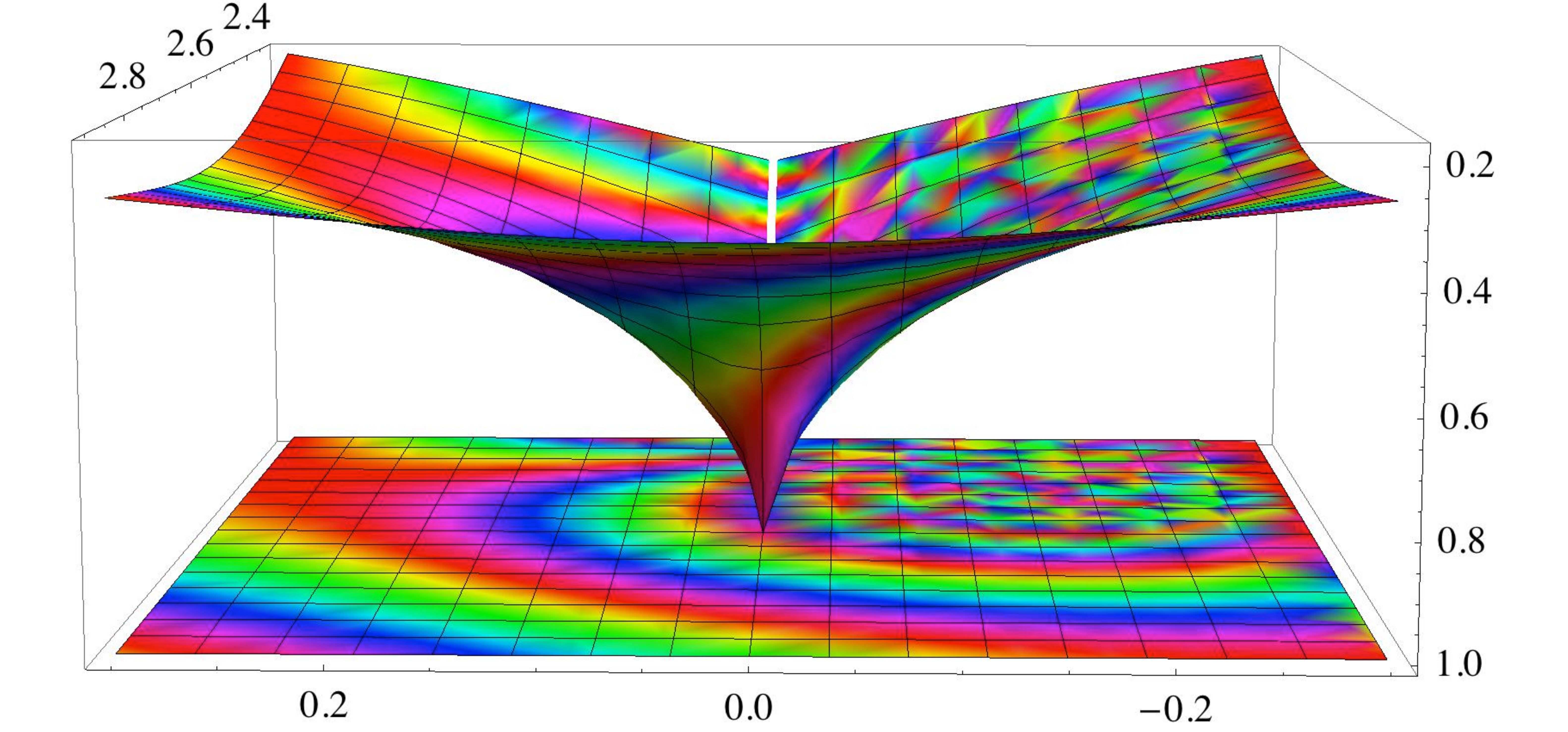}
\end{center}
\caption{The space of SUSY vacua (parameters) of the 3d $\CN=2$ theory $T[M_3]$ has several branches,
which often touch at singular points. Turning on $t\neq -1$ resolves (some of) the singularities and reconnects different branches into a single component.}
\label{fig:A41plot}
\end{wrapfigure}

{}From the viewpoint of the 3d theory $T[M_3]$ on the $\R^3$ part of the fivebrane world-volume, these three $U(1)$ symmetries
have the following interpretation. The rotation along $\R^3$ is part of the Lorentz symmetry, while (certain combinations of)
the other two become the $U(1)_R$ R-symmetry group of 3d $\CN=2$ supersymmetry algebra and a close cousin that we denote $U(1)_t$.
The non-R flavor symmetry $U(1)_t$ is not present in a generic, garden variety 3d $\CN=2$ theory.
But, since it is a symmetry of the system \eqref{surfeng}, theories $T[M_3]$ should have it as well.
This special symmetry illustrates to what extent $T[M_3]$ are non-generic 3d $\CN=2$ theories and plays a key role in the 3d-3d correspondence,
as will be discussed in this paper.

Another basic aspect of the fivebrane system \eqref{surfeng} that should be reflected in the physics of $T[M_3]$
is the relation between $G_{\C}$ flat connections on $M_3$ and the space of supersymmetric vacua
of the 3d $\CN=2$ theory on $S^1 \times \R^2$:
\be
\CM_{\text{SUSY}} (T[M_3;G]) \; = \; \CM_{\text{flat}} (M_3; G_{\C})
\label{mspaces}
\ee
This basic property of the 3d-3d correspondence \eqref{3d3dbasic} was originally taken \cite{DGH} as a definition\footnote{One
might wonder whether such definition specifies an equivalence relation
on 3d $\CN=2$ theories, {\it i.e.} if two theories that obey \eqref{mspaces} are dual in some sense.}
of the theory $T[M_3]$.
Since then many attempts to construct $T[M_3]$ systematically have been undertaken, including the approach \cite{DGG, DGG-Kdec} based on
triangulations of $M_3$. It leads to a 3d $\CN=2$ theory $T_{DGG} [M_3]$ with many desired properties, but also presents some puzzles.
In particular, as noted already in \cite{DGG, DGG-Kdec},
\be
\CM_{\text{SUSY}} (T_{DGG} [M_3;G]) \; \ne \; \CM_{\text{flat}} (M_3; G_{\C})
\label{missingDGG}
\ee
since certain branches of flat connections are always missing.
Examples of such ``lost branches'' include even the simplest flat connections on $M_3$, namely the abelian ones
({\it i.e.} flat connections that can be conjugated to the maximal torus of $G_{\C}$).

At first, it was unclear how severe this problem is.
However, a number of independent recent developments all lead to the same conclusion:
the complete theory $T[M_3;G]$ must realize all $G_{\C}$ flat connections on $M_3$.
We review some of these developments in Appendix \ref{sec:bdy}.
\\

Our first goal in this paper is to construct some theories $T[M_3;G]$ with all expected flavor symmetries and with vacua corresponding to all flat connections on $M_3$, and to investigate their relation to theories $T_{DGG}[M_3;G]$\,.
We will mainly focus on the case $G=SU(2)$, and on knot complements $M_3=S^3\backslash K$. A knot-complement theory $T[M_3]:=T[M_3;SU(2)]$ is defined by compactification of the 6d (2,0) theory on $S^3$ with a codimension-two defect wrapping the knot $K\subset S^3$. In this case $T[M_3]$ should gain a $U(1)_x$ flavor symmetry, part of the $SU(2)_x$ flavor symmetry of the defect, in addition to $U(1)_t$ and $U(1)_R$.
What we find can be then summarized by the following diagram:
\be \label{intro-diag}
\begin{array}{rcl}
  & T [M_3] & \; \\
 {}^{\langle\partial_r\CO_x\rangle\neq 0}\swarrow & \; & \searrow^{\langle \CO_t \rangle \ne 0} \\
 T_{\rm poly}[M_3;r]^{U(1)_x \hspace{-.75cm} \text{-------}} \quad \hspace{0in} & \; & \qquad T_{DGG} [M_3]^{U(1)_t \hspace{-.75cm} \text{-------}}
\end{array}
\ee
In particular, the theory $T_{DGG}[M_3]$ is a particular subsector of $T[M_3]$ obtained by Higgsing the $U(1)_t$ symmetry.

The left-hand side of the diagram \eqref{intro-diag} indicates an expected relation between $T[M_3]$ and a theory $T_{\rm poly}[M_3;r]$ whose partition functions compute the Poincar\'e polynomials of $r$-colored $SU(2)$ knot homology for $K$.
Indeed, our practical approach to constructing $T[M_3]$ will be to identify a 3d $\CN=2$ theory with $U(1)_x\times U(1)_t$ symmetry whose partition functions reduce to the desired Poincar\'e polynomials in a special limit. Physically this limit corresponds to another Higgsing procedure, this time breaking the $U(1)_x$ symmetry of $T[M_3]$ while creating a line defect or vortex, similar to scenarios studied in \cite{GRR-bootstrap, Bullimore:2014nla, Razamat:2014pta}.

An important feature of \eqref{intro-diag} is that the two arrows corresponding to Higgsing do not commute. In particular, while it is easy to obtain Jones polynomials of knots from the Poincar\'e polynomials on the left-hand side by ignoring $U(1)_t$ fugacities, it is (seemingly) impossible to do this from $T_{DGG}[M_3]$ on the right-hand side. Jones polynomials include a crucial contribution from the abelian flat connection on a knot complement $M_3$, and vacua corresponding to the abelian flat connection are lost during the Higgsing of $U(1)_t$.

\bigskip
\begin{figure}[ht]
\centering
\includegraphics[width=5.0in]{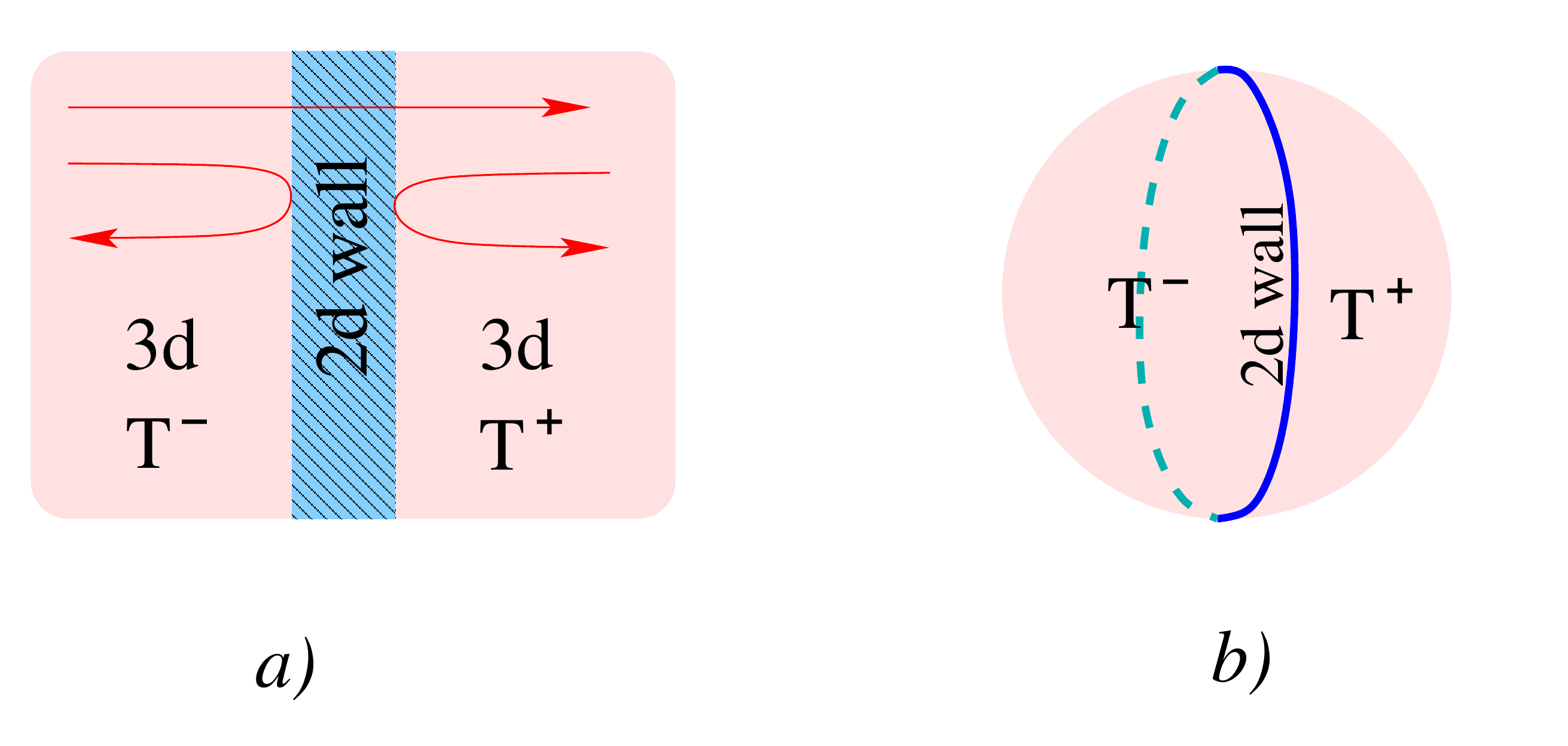}
\caption{The index of 3d $\CN=2$ theories can be generalized to include domain walls and boundary conditions~\cite{GGP-walls}.
It is obtained from two copies of the half-index $\CI_{S^1 \times_q D^{\pm}} (T^{\pm}) \simeq Z_{\text{vortex}} (T^{\pm})$
convoluted via the index (flavored elliptic genus) of the wall supported on $S^1 \times S^1_{\text{eq}}$,
where $D^{\pm}$ is the disk covering right (resp. left) hemisphere of the $S^2$
and $S^1_{\text{eq}}:= \partial D^+ = - \partial D^-$ is the equator of the $S^2$.}
\label{fig:wallindex}
\end{figure}

Later, in Section \ref{sec:surgeries} we discuss gluing of knot and link-complement theories to form closed $M_3$,
in particular 3d $\CN=2$ theories for lens spaces, Seifert manifolds, and Brieskorn spheres.
The importance of such gluing or surgery operations is two-fold.
First, it will give us another clear illustration why all flat connections need to be accounted by
3d $\CN=2$ theories $T[M_3]$ in order for cutting and gluing operations to work.
Moreover, it will help us to understand half-BPS boundary conditions
that one needs to choose in order to compute the half-index of $T[M_3]$.
As explained in \cite{GGP-4d}, a large class (``class $\CH$'') of boundary conditions
can be associated to 4-manifolds bounded by $M_3$,
\be
\boxed{ \phantom{\oint} {\text{4-manifold } M_4 \atop \text{bounded by } M_3} \phantom{\oint} }
\quad \leadsto \quad
\boxed{ \phantom{\oint} {\text{boundary condition for} \atop \text{3d } \CN=2 \text{ theory } T[M_3]} \phantom{\oint} }
\label{4mfldbc}
\ee
therefore making the half-index of $T[M_3]$ naturally labeled by 4-manifolds.

%%%%%%%%%%%%%%%%%%%%%%%%%%%%%%%%%%%%%%%%%%%%%%%%%%%%%%%%%%%%%%%%%%%%%%%%%%%%%%%%%%%%%
%%%%%%%%%%%%%%%%%%%%%%%%%%%%%%	 The Beginning of the Section 2		%%%%%%%%%%%%%%%%%%%%%%%%%%%%%%%%%%
%%%%%%%%%%%%%%%%%%%%%%%%%%%%%%%%%%%%%%%%%%%%%%%%%%%%%%%%%%%%%%%%%%%%%%%%%%%%%%%%%%%%%

\section{Contour integrals for Poincar\'e polynomials}
\label{sec:pintegral}

Even though our main goal is to identify all symmetries and flat connections in the 3d $\CN=2$ theory $T[M_3]$,
one of the intermediate steps is of mathematical value on its own.
Namely, the point of this section will be to show that Poincar\'e polynomials of homological link invariants
can be expressed as contour integrals
\be
P_{K} (q,t \ldots) \; = \; \int_{\Gamma} \frac{ds}{2\pi i s} \Upsilon (s,q,t, \ldots)
\label{Pintbasic}
\ee
in complex space $\C^m$ parametrized by (multi-)variable $s$.
Here, $P_{K} (q,t \ldots)$ stands for the Poincar\'e polynomial of a doubly-graded \cite{Khovanov,KhR1,Yonezawa,Wu}
or triply-graded \cite{DGR,KhR2,GS}
homology theory $\CH (K)$
of a link $K$:
\be
P_{K} (q,t \ldots) \; = \; \sum_{i,j,\ldots} q^i t^j \ldots \text{dim}\, \CH^{i,j} (K)
\label{Poincdef}
\ee
that categorifies either quantum $sl(N)$ invariant \cite{Wit-Jones} or colored HOMFLY polynomial \cite{HOMFLY}, respectively.
Depending on the context and the homology theory in question, the sum runs over all available gradings,
among which two universal ones --- manifest in \eqref{Poincdef} --- are the homological grading and the so-called $q$-grading.
In the case of HOMFLY homology, there is at least one extra grading and, correspondingly,
the Poincar\'e polynomial depends on one extra variable $a$, whose specialization to $a = q^N$ makes contact with $sl(N)$ invariants.
The Poincar\'e polynomials of triply-graded HOMFLY homology theories are often called superpolynomials.
In general, such invariants are also labeled by a representation / Young diagram $R$ and referred to as {\it colored},
unless $R = \Box$ in which case the adjective `colored' is often omitted.

In this section we will write the Poincar\'e polynomials of colored knot homologies in the form \eqref{Pintbasic} of contour integrals,
whose physical interpretation will be discussed in the later sections.
Our basic examples here (and throughout the paper) will be the unknot, trefoil, and figure-eight knot complements.

In general, superpolynomials or Poincar\'e polynomials are expressed as finite sums of products of $q$-Pochhammer symbols
\be (z;q)_n := \prod_{i=0}^{n-1}(1-q^iz)\, \ee
and monomials. For instance, the {\it unnormalized} superpolynomial of the trefoil ${\mathbf 3_1}$ is \cite{FGS-superA} (see also \cite{GS,DMMSS,AS-refinedCS,Cherednik}):
\begin{align}
\overline{\mathcal{P}}^{\mathcal{S}^{r}}_{\mathbf{3_1}}(a,q,t) = \sum _{k=0}^r \frac{(a(-t)^{3};q)_{r}(-a q^{-1} t;q )_{k}}{(q;q)_{k}(q;q)_{r-k}}  a^\frac{r}{2} q^{-\frac{r}{2}}  q^{(r+1) k} (-t)^{2 k -\frac{3r}{2}}\,.  \label{ptrefoil-a}
\end{align}
This is the Poincar\'e polynomial of the HOMFLY homology \eqref{Poincdef}
colored by the $r$-th symmetric power of the fundamental representation of $SU(N)$ or, in the language of Young diagrams,
by a Young tableau with a single row and $r$ boxes.
For our applications here, we specialize to $SU(2)$ homology\footnote{Specialization $a=q^2$ leads to Poincar\'e polynomials of colored $SU(2)$ knot homologies for a certain class of knots, which include unknot, trefoil, and figure-8 knot considered in this paper. In general and for more complicated knots, specialization of superpolynomials to Poincar\'e polynomials of $SU(N)$ knot homologies requires taking into account a nontrivial action of differentials \cite{DGR,GS,FGS-VC}. Note that in \cite{FGSS-AD,FGS-VC, FGS-superA} normalized colored superpolynomials were considered, \emph{i.e.} divided by the superpolynomial of the unknot colored by the same representation. In this paper we do not implement such a normalization.} by setting $a=q^2$.
It is further convenient to renormalize the $SU(2)$ polynomial by a factor $(-1)^r$, defining%
\footnote{In the next section, the rescaling by $(-1)^r$ leads to a convenient choice of fermion-number twist when identifying $P^r_{\mathbf{3_1}}(t;q)$ with a partition function of $T[\mathbf{3_1}]$ on $\R^2\times_q S^1$.} %
\begin{align} P^r_{\mathbf{3_1}}(t;q) &:= (-1)^r \, \overline{\mathcal{P}}^{\mathcal{S}^{r}}_{\mathbf{3_1}}(a=q^2,q,t) \notag \\
 & = \sum_{k=0}^{r} \frac{(q^{2}(-t)^{3};q)_{r} (q (-t);q)_{k}}{(q;q)_{k} (q;q)_{r-k}}(-q^{\frac12})^{2rk+2k+r}  (-t)^{2k-2r}\,. \label{ptrefoil}
\end{align}
We remark that the following steps could also be carried out for generic $a$, though for our applications we specialize from $SU(N)$ to $SU(2)$.

Let us suppose that $|q|>1$ (for reasons that will become clear momentarily), and define
\be (z)_\infty^- := (z;q^{-1})_\infty = \prod_{i=0}^\infty (1-q^{-i}z)\,,\qquad \theta^-(z) := \theta(z;q^{-1}) = (-q^{-\frac 12}z)^-_\infty(-q^{-\frac 12}z^{-1})^-_\infty\,, \ee
as well as
\be \theta^-(z_1,...,z_n) := \theta^-(z_1)\cdots \theta^-(z_n)\,. \ee
Then, by using identities such as $(q^rz)_\infty^-/(z)_\infty^-=(qz;q)_r = (-1)^rq^{\frac{r(r+1)}{2}}z^r(q^{-1}z^{-1};q^{-1})_r$ and $\theta^-(q^nz)/\theta^-(z) = q^{\frac{n^2}{2}}z^n$, we may rewrite
\begin{align}
 P^r_{\mathbf{3_1}}(t;q) &= \frac{(-1/(q^2t^3))^-_\infty(-1/(qt))^-_\infty}{(q^{-1})^-_\infty(-1/(q^2xt^3))^-_\infty} \sum_{k=0}^\infty \frac{(s/(qx))^-_\infty}{(q;q)_k(-1/(qst))^-_\infty}
 \frac{\theta^-(q^{\frac32} sxt^3,-q^{\frac12}x,-q^{\frac32}x(-t)^{\frac32},1)}
 {\theta^-(q^{\frac32}xt^3,-q^{\frac12}x/s,-q^{\frac32}(-t)^{\frac32},x)}\bigg|_{x=q^r\!,\,s=q^k} \notag \\
 &=: \sum_{k=0}^\infty \frac{1}{(q;q)_k(q^{-1})^-_\infty} \Upsilon^{(0)}_{\mathbf{3_1}}(s,x,t;q)\big|_{x=q^r\!,\,s=q^k}\,.  \label{sum31}
\end{align}
Note in particular that upon setting $x=q^r$ and $s=q^k$ the term $(s/(qx))^-_\infty$ in the numerator on the LHS vanishes unless $k\leq r$. Thus the sum naturally truncates to the one in \eqref{ptrefoil}.

\begin{figure}[htb]
\centering
\includegraphics[width=6in]{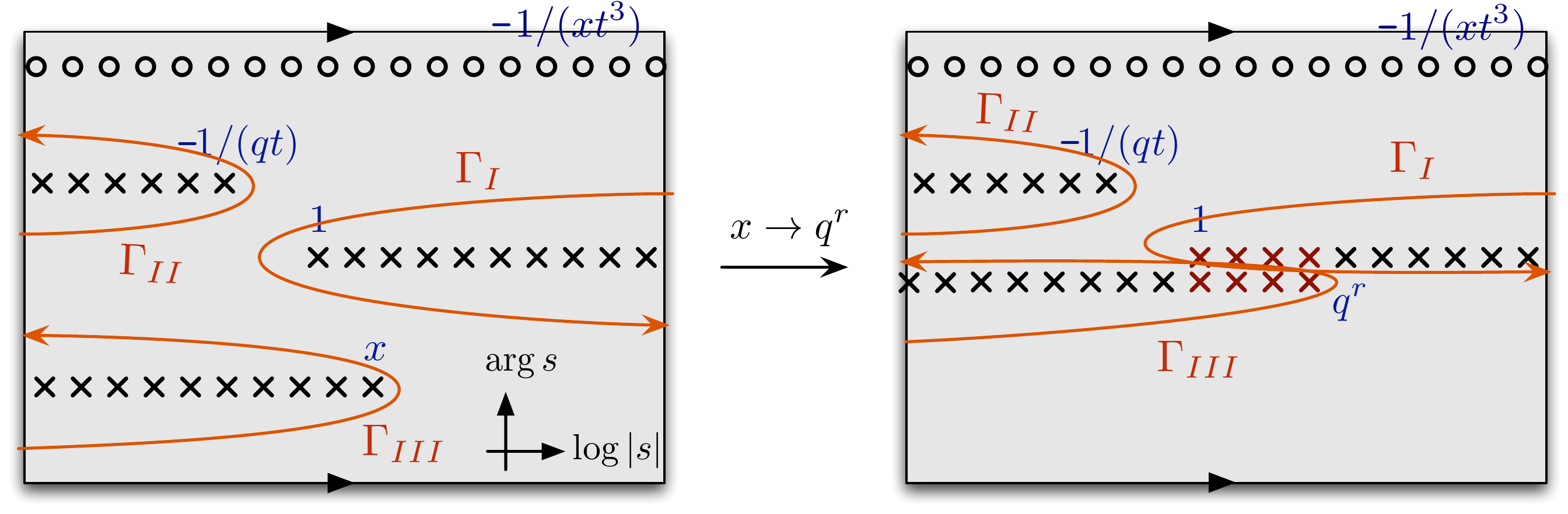}
\caption{Possible integration contours for the trefoil, drawn on the cylinder parametrized by $\log s$. There are three half-lines of poles in the integrand $\Upsilon_{\mathbf{3_1}}(s,x,t;q)$, coming from $(s)^-_\infty,\, (-1/(qst))^-_\infty,\, (x/s)^-_\infty$ in the denominator; and a full line of zeroes from $\theta^-(q^{\frac32} sxt^3)$ in the numerator. On the right, we demonstrate a pinching of contours as $x\to q^r$.}
\label{fig:cont31}
\end{figure}

Going further, we observe that the sum in \eqref{sum31} may be rewritten as a sum of residues
\be
 P^r_{\mathbf{3_1}}(t;q) = \bigg[\sum_{k=0}^\infty {\rm Res}_{s=q^k} \frac{1}{2\pi i s} \frac{1}{(s)^-_\infty} \Upsilon^{(0)}_{\mathbf{3_1}}(s,x,t;q)\bigg]_{x=q^r}\,, \ee
since $\Upsilon^{(0)}_{\mathbf{3_1}}$ is smooth at $s=q^k$, while the residue of $1/[2\pi i s(s)^-_\infty]$ at $s=q^k$ is precisely $1/[(q;q)_k(q^{-1})^-_\infty]$. It was the initial choice $|q|>1$ that allowed us to write the sum as residues like this. Therefore, at least formally,
\be P^r_{\mathbf{3_1}}(t;q) = \int_{\Gamma_I} \frac{ds}{2\pi is} \Upsilon_{\mathbf{3_1}}(s,x,t;q) \bigg|_{x=q^r}\, \ee
with
\begin{align} \label{integ31}
&\Upsilon_{\mathbf{3_1}}(s,x,t;q) := \frac{1}{(s)^-_\infty}\Upsilon^{(0)}_{\mathbf{3_1}}(s,x,t;q)  \\
&\qquad = \frac{\theta^-(-q^{\frac12}x,-q^{\frac32}x(-t)^{\frac32},1)}{\theta^-(q^{\frac32}xt^3,-q^{\frac32}(-t)^{\frac32},x)} \frac{(-1/(q^2t^3))^-_\infty (-1/(qt))^-_\infty}{(-1/(q^2xt^3))^-_\infty} \times
\frac{\theta^-(q^{\frac32} sxt^3)}{(s)^-_\infty (-1/(qst))^-_\infty (x/s)^-_\infty}\,, \notag
\end{align}
where the contour $\Gamma_I$ is shown in Figure \ref{fig:cont31}. (We have put all $s$-dependent terms in $\Upsilon_{\mathbf{3_1}}$ on the right.) This is now the form of a contour integral \eqref{Pintbasic}.

Note that the three terms $(s)^-_\infty$, $(-1/(qst))^-_\infty$, and $(x/s)^-_\infty$ each contribute a half-line of poles to $\Upsilon_{\mathbf{3_1}}$. If we take $q>1$ to be real, then the asymptotics of the integrand are given by
\be \Upsilon_{\mathbf{3_1}} \sim \begin{cases} \exp\,\frac{1}{\log q}\big[(\log x+3\log (-t))\log s+\ldots\big]  & \log|s|\to \infty \\
 \exp\,\frac{1}{\log q}\big[(-\frac12(\log s)^2+\ldots \big] & \log|s|\to -\infty\,,
\end{cases}
\ee
so the integral along $\Gamma_I$ does converge in a suitable range of $x$ and $t$ (namely, if $|xt^3|<1$). In contrast, the integrals along the other obvious cycles here, $\Gamma_{II}$ and $\Gamma_{III}$, \emph{always} converge. Moreover, a little thought shows that upon setting $x=q^r$ the integral along $\Gamma_I$ must equal the integral along $\Gamma_{III}$; indeed, as $x\to q^r$, some $r+1$ pairs of poles in the lines surrounded by $\Gamma_I$ and $\Gamma_{III}$ collide, and all contributions to the integrals along either $\Gamma_I$ and $\Gamma_{III}$ come from the $r+1$ points where the contours get pinched by colliding poles. (Such pinching would usually cause integrals to diverge, but here the divergence is cancelled by one of the $s$-independent theta-functions in $\Upsilon_{\mathbf{3_1}}$.) Therefore, letting
\be  B^{\mathbf{3_1}}_*(x,t;q) :=  \int_{\Gamma_*} \frac{ds}{2\pi is} \Upsilon_{\mathbf{3_1}}(s,x,t;q)\,, \label{31blocks} \ee
(with the obvious relation $B_I+B_{II}+B_{III}=0$), we find
\be P^r_{\mathbf{3_1}}(t;q) = B^{\mathbf{3_1}}_I(x,t;q)\big|_{x=q^r} = B^{\mathbf{3_1}}_{III}(x,t;q)\big|_{x=q^r}\,. \ee \\

We can repeat the analysis for the unknot $U=\mathbf{0_1}$ and figure-eight knot $\mathbf{4_1}$. The superpolynomials of these knots are given by \cite{GIKV,AS-refinedCS,ItoyamaMMM,FGS-superA}:
\begin{align}
\overline{\mathcal{P}}_{\mathbf{0_1}}^{S^{r}}(a,q,t) &=  a^{-\frac{r}{2}} q^{\frac{r}{2}} (-t)^{-\frac{3}{2}r} \frac{(a(-t)^{3};q)_{r}}{(q;q)_{r}}    \label{punknot}   \\
\overline{\mathcal{P}}_{\mathbf{4_1}}^{S^{r}}(a,q,t) &= \sum_{k=0}^{r} \frac{(a(-t)^{3};q)_{r}}{(q;q)_{k}(q;q)_{r-k}} (aq^{-1}(-t);q)_{k} (aq^{r}(-t)^{3};q)_{k}  a^{-k-\frac{r}{2}} q^{\frac{r}{2}+k(1-r)} (-t)^{-2k-\frac{3}{2}r}\,,   \label{pfigure8}
\end{align}
and Poincar\'e polynomials for $G=SU(2)$, \emph{i.e.} specializations to $a=q^2$,  normalized by $(-1)^r$, are given by
\begin{align}
{P}_{\mathbf{0_1}}^{{r}}(t;q) =   (-q^{\frac12})^{-r} (-t)^{-\frac{3}{2}r} \frac{(q^{2}(-t)^{3};q)_{r}}{(q;q)_{r}}
\end{align}

\begin{align}
{P}_{\mathbf{4_1}}^{{r}}(t;q) = \sum_{k=0}^{r} \frac{(q^{2}(-t)^{3};q)_{r}}{(q;q)_{k}(q;q)_{r-k}} (q(-t);q)_{k} (q^{2}q^{r}(-t)^{3};q)_{k} (-q^{\frac12})^{-r-2k(1+r)}(-t)^{-2k-\frac32 r}\,,
\end{align}
respectively.
Repeating the above procedure, we find
\be \label{blockU}
 {P}_{\mathbf{0_1}}^{{r}}(t;q) = B^{\mathbf{0_1}}(x,t;q)\big|_{x=q^r}\,,\qquad B^{\mathbf{0_1}}(x,t;q):=
 \frac{\theta^-(1,-q^{\frac12}x(-t)^{\frac32})}{\theta^-(x,-q^{\frac12}(-t)^{\frac32})} \frac{(q^{-1}/x)^-_\infty(-q^{-2}/t^3)^-_\infty}
  {(q^{-1})^-_\infty (-q^{-2}/(xt^3))^-_\infty}
\ee
for the unknot, and
\begin{align} \label{integ41}
{\Upsilon}_{\mathbf{4_1}}(s,x,t;q) := \frac{\theta^-(-q^{\frac12}x,q^{\frac12}tx,(-t)^{-\frac12})}{\theta^-(q,t^2,q^{\frac12}t,x(-t)^{-\frac12})}\theta^-(qs,t^2s)\frac{(-1/(q^2t^3))^-_\infty (-1/(qt))^-_\infty}
{ (s)^-_\infty (-1/(qts))^-_\infty (x/s)^-_\infty (-1/(q^2xt^3s))^-_\infty}\,.
\end{align}
for the figure-eight knot.
In the latter case, the integrand ${\Upsilon}_{\mathbf{4_1}}$ has four half-lines of poles in the $s$-plane, coming from the four factors $(s)^-_\infty,\, (-1/(qts))^-_\infty,\, (x/s)^-_\infty,\, (-1/(q^2xt^3s))^-_\infty$ in the denominator of \eqref{integ41}. Let $\Gamma_I,\Gamma_{II},\Gamma_{III},\Gamma_{IV}$ be contours encircling these respective half-lines of poles. A formal sum of residues along poles in the first half-line, evaluated at $x=q^r$, most directly gives ${P}_{\mathbf{4_1}}^{{r}}(t;q)$; but the actual integral along $\Gamma_I$ does not converge for generic $x$. In contrast, the integrals along $\Gamma_{II},\Gamma_{III},\Gamma_{IV}$ always converge, and
\be  {P}_{\mathbf{4_1}}^{{r}}(t;q) = B^{\mathbf{4_1}}_{III}(x,t;q)\big|_{x=q^r} =  \text{``}B^{\mathbf{4_1}}_{I}(x,t;q)\big|_{x=q^r}\text{''}\,,
\ee
where
\be \label{blocks41}
B_*^{\mathbf{4_1}}(x,t;q) := \int_{\Gamma_*}\frac{ds}{2\pi is}{\Upsilon}_{\mathbf{4_1}}(s,x,t;q)\,.
\ee

These examples indicate how the analysis may be extended to other knots and links (\emph{e.g.} those whose superpolynomials are found in \cite{FGSS-AD,Nawata}), and to Poincar\'e polynomials of other homological invariants. In general, the required integrals will not be one-dimensional, but will require higher-dimensional integration cycles. Generalizations of some results of this paper to other knots and links are also discussed in section \ref{sec:gen}.

%%%%%%%%%%%%%%%%%%%%%%%%%%%%%%%%%%%%%%%%%%%%%%%%%%%%%%%%%%%%%%%%%%%%%%%%%%%%%%%%%%%%%%%%%%%%%%%%%%%%%%%%%%%%%%

\section{Knot polynomials as partition functions of $T[M_3]$}
\label{sec:TM}

In this section, we construct some examples of 3d $\CN=2$ theories $T[M_3]$ for knot complements $M_3$ (and $G=SU(2)$) with the properties outlined above. In particular, we would like the vacua of $T[M_3]$ on $\R^2\times S^1$ to match all flat connections on $M_3$.

Although our strategy will be a little indirect, it is based on a simple key observation: the contour integral \eqref{Pintbasic} for colored Poincar\'e polynomials has the form of localization integrals in supersymmetric 3d $\CN=2$ theories as well as in Chern-Simons theory on certain 3-manifolds. Indeed, powerful localization techniques reduce the computation of Chern-Simons partition functions to finite dimensional integrals of the form \eqref{Pintbasic}, where the choice of the contour is related to the choice of the classical vacuum
\cite{Marino0207,deHaro2004uz,Beasley2005vf,deHaro2005rn,Blau2013oha},
as we briefly review in section~\ref{sec:surgeries}.

Similar --- and, in fact, closer to our immediate interest --- contour integrals of the form \eqref{Pintbasic} appear as a result
of localization in supersymmetric partition functions of 3d $\CN=2$ theories,
such as the (squashed) sphere partition function \cite{Kapustin-3dloc,HHL}, the index \cite{Kim-index,IY-index,KW-index},
and the vortex partition function \cite{DGG} or the half-index \cite{DGG-index}.
Since in the last case the space-time is non-compact it requires a choice of the asymptotic boundary condition
or vacuum of the theory on $\R^2\times_q S^1$, which manifests as a choice of the integration contour
in the localization calculation. (The integrand is completely determined by the Lagrangian of 3d $\CN=2$ theory.)
This has to be compared with the first two cases, where localization of 3-sphere partition function and index
lead to a contour integral with canonical choices of the integration contour.

Therefore, in order to interpret \eqref{Pintbasic} as a suitable partition function of 3d $\CN=2$ theory
in this paper we mainly focus on half-indices and vortex partition functions.
This gives us enough flexibility to interpret \eqref{Pintbasic}
and we generically expect that the \emph{full} set of partition functions for $T[M_3]$, labelled by a full set of vacua, corresponds to a complete basis of independent convergent contours for the integrals of Section \ref{sec:pintegral}. On the other hand, we also expect that a basis of convergent contours is in 1--1 correspondence with flat connections on $M_3$:
\be \label{vacuaconn}   \text{vacua of $T[M_3]$} \;\;\leftrightarrow\;\; \text{convergent contours} \;\;\leftrightarrow\;\; \text{flat conn's}\,. \ee

The reason we expect these correspondences to hold is outlined more carefully in Section \ref{sec:Ahat}. In order to capture \emph{all} flat connections, it turns out to be crucial that we start with Poincar\'e polynomials for knot homology rather than unrefined Jones polynomials. In Section \ref{sec:theory} we then demonstrate the construction of $T[M_3]$ in a few examples.

In Section \ref{sec:ind} we examine the physical meaning of the limit $x\to q^r$ that recovers Poincar\'e polynomials from $T[M_3]$. We argue that it is a combination of Higgsing and creation of a line operator in $T[M_3]$, as on the left-hand side of \eqref{intro-diag}. We also show that Poincar\'e polynomials can be obtained by directly taking residues of $S^2\times_q S^1$ indices and $S^3_b$ partition functions of $T[M_3]$.

\subsection{Recursion relations}
\label{sec:Ahat}

One understanding of why contour integrals as in Section \ref{sec:pintegral} should capture all flat connections on a knot complement follows from looking at the $q$-difference relations that the integrals satisfy.

Let us start with the Poincar\'e polynomials $P_K^r(t;q)$ for colored $SU(2)$ knot homology of a knot $K$. As found in \cite{FGS-VC,FGS-superA,FGSS-AD}, the sequence of Poincar\'e polynomials obeys a recursion relation of the form
\be \hat A^{\rm ref}(\hat x,\hat y;t;q) \cdot P^r_K(t;q) = 0\,,\ee
where $\hat A^{\rm ref}(\hat x,\hat y;t;q)$ is a polynomial operator in which $\hat x,\hat y$ act as $\hat x P^r_K = q^r P^r_K$ and $\hat y P^r_K = P^{r+1}_K$.
The limit $q\to 1$ of $\hat A^{\rm ref}(\hat x,\hat y;t;q)$ is a classical polynomial $A^{\rm ref}(x,y;t)$, whose subsequent $t\to -1$ limit contains the classical A-polynomial of $K$ \cite{cooper-1994} as a factor,
\be \hat A^{\rm ref}(\hat x,\hat y;t;q) \;\overset{q\to 1}\to\; A^{\rm ref}(x,y;t) \;\overset{t\to -1}\to\; A(x,y)\,.\ee
The physical interpretation of the classical A-polynomial $A(x,y)$ goes back to \cite{gukov-2003}.
Its roots at fixed $x$ are in 1-1 correspondence with all flat connections on $M$ (with fixed boundary conditions at $K$); but the root corresponding to the abelian flat connection is distinguished because it comes from a universal factor $(y-1)$ in $A(x,y)$. However, the $t$-deformed polynomial $A^{\rm ref}(x,y;t)$ is irreducible (at least in simple examples%
\footnote{To be more precise, both $A^{\rm ref}(x,y;t)$ and $A(x,y)$, obtained as appropriate limits of super-A-polynomials, may contain some additional factors.
As explained in \cite{Dimofte-QRS, GS-quant} (for $t=-1$) and \cite{FGS-VC,FGS-superA} (for general $t$), these factors are necessary for quantization but are not associated to classical flat connections.
For knots considered in this paper these factors are independent of $y$, they do not affect the structure of roots of $A^{\rm ref}(x,y;t)$ or $A(x,y)$ at generic fixed $x$, and therefore they do not modify our discussion here.
}%
), and none of its roots is more or less important than the others.

Alternatively, note that the $t\to -1$ limit of $A^{\rm ref}(\hat x,\hat y;t;q)$ leads to a shift operator known as the quantum A-polynomial, $\widehat{A}(\hat x,\hat y;q)$, which annihilates colored Jones polynomials \cite{gukov-2003, garoufalidis-2004}.
One can also consider $a$--deformations of these shift operators. Such a deformation of the quantum A-polynomial  was called $Q$-deformed A-polynomial in \cite{AVqdef}, and it agrees with the mathematically defined augmentation polynomial of \cite{NgFramed,Ng}. More generally, one can consider shift operators $\widehat{A}^{\rm super}(\hat x,\hat y;a;t;q)$ depending on both $a$ and $t$, which annihilate colored superpolynomials, and which were called super-A-polynomials in \cite{FGS-superA} (for a concise review see \cite{FS-superA}). However, as mentioned above, we are only interested here in $a=q^2$ specializations.

Now, in Section \ref{sec:pintegral} we expressed
\be P_K^r(t;q) = \bigg[\int_{\Gamma_P} \frac{ds}{2\pi is} \Upsilon_K(s,x,t;q)\bigg]_{x=q^r} = B_P(x,t;q)\big|_{x=q^r} \label{intP} \ee
for a suitable integrand $\Upsilon_K$ and a choice of integration contour $\Gamma_P$. It is easy to see that $B_P(x,t;q)$ satisfies a $q$-difference equation
\be \hat A^{\rm ref}(\hat x,\hat y;t;q)\cdot B_P(x,t;q)=0 \ee
even before setting $x=q^r$, with $\hat x,\hat y$ acting as $\hat x B_P(x,...) =x B_P(x,...)$ and $\hat y B_P(x,...) = B_P(qx,...)$. More so, the integral $B_\alpha = \int_{\Gamma_\alpha} ds/s\,\Upsilon_K$ for \emph{any} convergent integration contour $\Gamma_\alpha$ (that stays sufficiently far away from poles) should provide a solution to the $q$-difference equation $\hat A^{\rm ref}\cdot B=0$, and one generally expects that a maximal independent set of integration contours generates the full vector space of solutions.

The situation is entirely analogous to the solution of Picard-Fuchs equations for periods of a holomorphic form on a complex manifold. Here $\hat A^{\rm ref}$ plays the role of a $q$-deformed Picard-Fuchs operator, and $B_P$ is a fundamental period; the general integrals $B_\alpha$ compute the remaining periods.

If we fix the values of $x$, $t$, and $q$, the convergent integration cycles $\Gamma_\alpha$ can be labelled by the roots $y^{(\alpha)}(x,t)$ of the classical equation $A^{\rm ref}(x,y;t)=0$ --- \emph{i.e.} by the flat connections on $M_3$ with boundary conditions (meridian holonomy) fixed by $x$. The correspondence follows roughly by identifying the solutions to $A^{\rm ref}(x,y;t)=0$ with critical points of the integrand $\Upsilon_K(s,x,t;q)$ at $q\approx 1$, then using downward gradient flow with respect to $\log|\Upsilon_K(s,x,t;q)|$ to extend the critical points into integration cycles $\Gamma_\alpha$. This is the standard construction of so-called ``Lefschetz thimbles,'' modulo some subtleties that were discussed in \cite{HIV,Wit-anal}.

We have claimed that by writing one solution of $\hat A^{\rm ref}\cdot B=0$ as a contour integral \eqref{intP}, we can actually reproduce all other solutions from integrals on a full basis of contours $\Gamma_\alpha$. This reasoning relies on an important assumption: that the quantum $\hat A^{\rm ref}$ (and hence the classical $A^{\rm ref})$ is irreducible. Otherwise, we may only get solutions corresponding to one irreducible component. For this reason, it is crucial that we use \emph{refined} knot polynomials and recursion relations rather than Jones polynomials and the quantum A-polynomial. See \cite{GS-rev,FS-superA} for further details as well as pedagogical introduction.

To complete the chain of correspondences \eqref{vacuaconn}, we simply use \cite{DGH,Yamazaki-3d,Yamazaki-layered,DG-Sdual,DGG,CCV,FGS-VC,FGSS-AD,Cordova-tangles,DGG-Kdec} to translate the above observations to the language of gauge theory. Momentarily we will engineer gauge theories $T[M_3]$ for which the integrals $\int_* ds/s\, \Upsilon_K$ compute various partition functions on $\R^2\times S^1$ annihilated by $\hat A^{\rm ref}$ and labelled by vacua of $T[M_3]$ on $\R^2\times S^1$, \emph{i.e.} classical solutions of $A(x,y;t)=0$.

\subsection{$3d$ $\mathcal{N}=2$ gauge theories for unknot, trefoil knot, and figure-eight knot}
\label{sec:theory}

Having rewritten the Poincar\'e polynomials of colored $SU(2)$ knot homologies as special values of a contour integral, we try to engineer $T[M_3]$ so that the contour integral computes its partition function.
In particular, by examining the integrand $\Upsilon_K$ and associating
\be \label{blockCSM}
\begin{array}{c@{\quad}c@{\quad}c}
\text{fugacities $x,(-t),q$} &\leadsto & \text{flavor and $R$ symmetries} \\
\text{fugacity $s$} &\leadsto & \text{$U(1)_s$ gauge symmetry} \\
\text{$(*)^-_\infty$ factors} &\leadsto & \text{chiral multiplets} \\
\text{$\theta^-$ functions} &\leadsto & \text{(mixed) Chern-Simons couplings}
\end{array}
\ee
we can construct a putative UV description for $T[M_3]$ as an abelian Chern-Simons-matter theory.

This approach is almost successful, and good enough for our present purposes, though we should mention an important caveat. In general, one must also specify relevant superpotential couplings for a UV description of $T[M_3]$, which are crucial for attaining the right superconformal theory in the IR; but it is very difficult to specify such couplings just by looking at partition functions. At the very least one would like to find superpotential couplings that break all ``extraneous'' flavor symmetries whose fugacities don't appear in supersymmetric partition functions, and are not expected for the true $T[M_3]$. Even this is difficult, because the naive prescription \eqref{blockCSM} leads to theories that simply don't have chiral operators charged only under the extraneous symmetries. This problem was discussed in \cite[Section 4]{DGG}, and solved by finding ``resolved'' theories with the same partition functions as the naive ones, but with all necessary symmetry-breaking operators present.

Presently, we will follow the naive approach to obtain simple UV descriptions for putative $T[M_3]$'s, where some but not all symmetry-breaking superpotential couplings are present. We expect that these theories are limits of the ``true'' superconformal knot-complement theories $T[M_3]$, where some marginal couplings have been sent to infinity. Thus, any observables of $T[M_3]$ that are insensitive to marginal deformations --- such as supersymmetric indices, massive vacua on $S^1$, {\it etc.} --- can be calculated just as well in our naive descriptions as in the true theories, as long as masses or fugacities corresponding to extraneous flavor symmetries are turned off by hand. This is sufficient for testing many of the properties we are interested in.

\subsubsection*{Theory for unknot, $T[\mathbf{0_1}]$}

The theory for the unknot that gives \eqref{blockU} was already discussed in \cite{FGS-superA} and has four chirals $\Phi_i$, corresponding to the terms $(q^{-1}/x)^-_\infty,\,(q^{-2}/t^3)^-_\infty,\,(q^{-1})^-_\infty,\,(q^{-2}/(xt^3))^-_\infty$. Letting $x$ and $(-t)$ be fugacities for flavor symmetries $U(1)_x$ and $U(1)_t$, we use the rules of \cite{DGH,Yamazaki-3d,Yamazaki-layered,DG-Sdual,DGG,CCV,FGS-VC,FGSS-AD,Cordova-tangles,DGG-Kdec} to read off the precise charge assignments and levels of (mixed) background Chern-Simons couplings
\be \label{Tunknot}
T[\mathbf{0_1}]:\qquad
\begin{array}{c|cccc}
 & \Phi_{1}   & \Phi_{2}   & \Phi_{3} & \Phi_{4} \\
    \hline
U(1)_{x}    &-1  &0   &0   & 1   \\
U(1)_{t}    &0   &-3   &0  & 3  \\ \hline
U(1)_{R}    &0   &-2   &2  & 4
\end{array} \qquad\;\;
\text{CS:}\; \begin{array}{c|cc|c}
 & U(1)_x & U(1)_t & U(1)_R \\\hline
U(1)_x & 0 & 0 & 0 \\
U(1)_t & 0 & 0 & 0 \\\hline
U(1)_R & 0 & 0 & 0 \end{array}
\ee
(Here all background Chern-Simons couplings simply vanish.\footnote{We can multiply an extra normalization factor to $SU(2)$ Poincar\'e polynomials to make the mixed IR CS levels for $U(1)_t$ to be integers, but we will work formally without such an extra normalization.}). In this case, we can add an obvious superpotential
\begin{align}
W_{\mathbf{0_1}} = \mu\, \Phi_{1} \Phi_{2} \Phi_{4} \label{Wunknot}
\end{align}
that breaks most extraneous flavor symmetries and preserves $U(1)_x,U(1)_t$, and $U(1)_R$ (note that the operator in \eqref{Wunknot} has R-charge two).
The chiral $\Phi_3$ is completely decoupled from the rest of the theory and rotated by an extraneous $U(1)$ symmetry. We could break this $U(1)$ by adding $\Phi_3$ to the superpotential \eqref{Wunknot}, but prefer not to do this as it would forbid $T[\mathbf{0_1}]$ from having a supersymmetric vacuum. Ignoring the $\Phi_3$ sector, the putative unknot theory looks just like the 3d $\CN=2$ XYZ model.

Similarly, if we follow \cite{FGS-VC} and
compactify $T[\mathbf{0_1}]$ on a circle turning on masses (\emph{i.e.} complexified scalars in background gauge multiplets) for $U(1)_x$ and $U(1)_t$, we find that the theory is governed by an effective twisted superpotential
\be \wt W_{\mathbf{0_1}} = \Li_2(x)+\Li_2(-t^3)+\Li_2(-x^{-1}t^{-3})+ \frac12\big[(\log x)^2+3\log x\log(-t)+9(\log (-t))^2\big]\,.
\ee
(We have removed from $\wt W_{\mathbf{0_1}}$ an infinite contribution from the massless $\Phi_3$; this could be regularized by turning on a mass for the $U(1)$ symmetry rotating $\Phi_3$.) The equation for the supersymmetric parameter space,%
\footnote{On the RHS we define an effective FI parameter as $-y$ rather than $y$ in order to match the knot-theoretic A-polynomial below. This is correlated with the renormalization of Poincar\'e polynomials above by $(-1)^r$.}
\be \exp \bigg(x \frac{\partial \wt W_{\mathbf{0_1}}}{\partial x}\bigg) = -y\,,\ee
becomes the refined A-polynomial equation
\be\label{ArefU} (-t)^{\frac32} y = \frac{1+t^3x}{1-x}\,, \ee
which further reduces to the unknot A-polynomial $y-1=0$ at $t\to -1$. Equation \eqref{ArefU} has a unique solution in $y$ at generic fixed $x,t$, corresponding to the unique, abelian flat connection on the unknot complement (with fixed holonomy eigenvalue $x$ on a cycle linking the unknot).

\subsubsection*{Theory for trefoil knot, $T[\mathbf{3_1}]$}

In this case, the integrand \eqref{integ31} suggests a theory with six chirals, with charges and Chern-Simons levels
\be \label{T31}
T[\mathbf{3_1}]:\quad
\begin{array}{c|cccccc|c}
      &\Phi_{1} &\Phi_{2} &\Phi_{3} &\Phi_{4} &\Phi_{5} &\Phi_{6} & V_-\\ \hline
U(1)_s  &-1     &1     &1     &0      &0        &0                        &0      \\
U(1)_x  &0      &-1    &0     &0      &0        &1                        &-1     \\
U(1)_t  &0      &0     &1     &-1     &-3       &3                        &-3     \\\hline
U(1)_R  &0      &0     &2     &0	      &-2       &4                        &-2     \end{array}
\,,\qquad
\text{CS}:\; \begin{array}{c|c|cc|c}
& U(1)_s & U(1)_x & U(1)_t & U(1)_R \\\hline
U(1)_s & - 1/2 & 3/2 & 5/2 & 5/2 \\\hline
U(1)_x & 3/2 & 0 & 0 & 2 \\
U(1)_t & 5/2 & 0 & 0 & 0 \\\hline
U(1)_R & 5/2 & 2 & 0 & 0
\end{array}\,.
\ee
This is now a gauge symmetry with a dynamical $U(1)_s$ symmetry in addition to $U(1)_x$ and $U(1)_t$ flavor symmetries. Standard analysis of \cite{AHISS} shows that this theory has a gauge-invariant anti-monopole operator $V_-$ formed from the dual photon, with charges as indicated in the table. Altogether we can write a superpotential
\begin{align}
W_{\mathbf{3_1}} = \mu_{1}\, \Phi_{1}\Phi_{2}\Phi_{5}\Phi_{6} + \mu_{2}\, \Phi_{1}\Phi_{3}\Phi_{4} + \mu_3\,\Phi_{6} V_{-}
\end{align}
that preserves all symmetries we want to keep, and breaks almost all other flavor symmetry. There remains a single extraneous $U(1)$, just like in the unknot theory, which plays (roughly) the role of a topological symmetry dual to $U(1)_s$.

When compactifying the theory on a circle with generic twisted masses $x$ and $(-t)$ for $U(1)_x$ and $U(1)_t$, and scalar $s$ in the $U(1)_s$ gauge multiplet, we obtain the effective twisted superpotential
\begin{align} \label{Wt31}
 \wt W_{\mathbf{3_1}} &= \Li_2(s)+\Li_2(-1/(st))+\Li_2(x/s)+\Li_2(-t^{3})+\Li_2(-t)+\Li_2(-1/(t^3x)) \\
&\quad + \tfrac12\big( (\log s)^2+\log s(6\log t+2\log x)+\log x(\log x+3\log(- t))+10(\log t)^2\big)\,. \notag
\end{align}
The critical-point equation $\exp\big(s\, \partial \wt W_{\mathbf{3_1}} /\partial s)=1$, namely
\be \frac{t^2(1+st)(s-x)x}{s(1-s)}=1 \label{crit31} \ee
determines two solutions in $s$ at generic values of $x$ and $t$; plugging these into the SUSY-parameter-space equation
\be
-y = \exp\big(x\, \partial \wt W_{\mathbf{3_1}} /\partial x) = -s^2(-t)^{-3/2} \frac{1+t^3 x}{s-x}  \label{crit31bis}
\ee
then determines two values of $y$. More directly, they are solutions of the quadratic
\be A_{\mathbf{3_1}}^{\rm ref}(x,y;t) = (1-x)t^2y^2 - (1-t^2x+2t^2x^2+2t^3x^2+t^5x^3+t^6x^4)(-t)^{\frac12}y + t^3(x^3+t^3x^4) =0\,, \notag\ee
which collapses to the $A_{\mathbf{3_1}}^{\rm ref}(x,y;-1) = (x-1)(y-1)(y+x^3)$, the trefoil's A-polynomial (with an extra $(x-1)$ factor) as $t\to -1$. Thus $T[\mathbf{3_1}]$ has vacua corresponding to \emph{both} of the flat $SL(2,\C)$ connections on the trefoil complement, one irreducible, and one abelian.
The two independent contour integrals $B_{II}$ and $B_{III}$ of \eqref{31blocks} are in 1-1 correspondence with the two flat connections.

\subsubsection*{Theory for figure-eight knot, $T[\mathbf{4_1}]$}

Finally, for the figure-eight knot, the integrand \eqref{integ41} suggests a theory with $U(1)_s$ gauge symmetry, $U(1)_x\times U(1)_t$ flavor symmetry, and six chirals of charges
\be \label{T41}
T[\mathbf{4_1}]:\quad
\begin{array}{ c | c  c  c  c  c  c }
          &\Phi_{1} &\Phi_{2} &\Phi_{3} &\Phi_{4} &\Phi_{5} &\Phi_{6}  \\  \hline
U(1)_{s}  &-1 &1  &1	&0	  &1  &0       \\
U(1)_{x}  &0  &-1 &0	&0	  &1  &0     \\
U(1)_{t}  &0  &0  &1	&-1 &3  &-3     \\\hline
U(1)_{R}  &0  &0  &2	&0	  &4  &-2
\end{array}
\ee
The net Chern-Simons couplings all turn out to vanish. This particular theory does not admit gauge-invariant monopole or anti-monopole operators. We can introduce a superpotential
\begin{align}
W_{4_1} = \mu_{1}\, \Phi_{1}\Phi_{3}\Phi_{4}+ \mu_{2}\, \Phi_{1}^{2} \Phi_{2} \Phi_{5} \Phi_{6}\,,
\end{align}
which breaks flavor symmetry to $U(1)^4$, including $U(1)_x\times U(1)_t$. Thus there are two extraneous $U(1)$'s, including the topological symmetry of the theory.

As before, we can find an effective twisted superpotential on $\R^2\times S^1$ of the form
\be \label{Wt41}
 \wt W_{\mathbf{4_1}} = \Li_2(s)+\Li_2(x/s)+\Li_2(-1/(st))+\Li_2(-1/(sxt^3))+\Li_2(-t)+\Li_2(-t^3) + \text{log's}\,, \ee
whose critical point equation
\be
\exp\bigg(s \frac{\partial \wt W_{\mathbf{4_1}}}{\partial s}\bigg) =
\frac{(1+st)(s-x)(1+st^3x)}{(1-s)st^2x} = 1    \label{crit41}
\ee
generically has three solutions in $s$ --- which in turn determine
\be
y = -\exp\big(x\,\partial \wt W_{\mathbf{4_1}}/\partial x\big) \sim \frac{1+s t^3 x}{s-x}.   \label{crit41bis}
\ee
More directly, the solutions in $y$ are roots of the cubic
\begin{align} A^{\rm ref}_{\mathbf{4_1}} &= (x^3-x^2)(-t)^{\frac92}y^3 - (1+tx-t^2x+2t^2x^2+2t^3x^2+2t^4x^3+2t^5x^3-t^5x^4+t^6x^4+t^7x^5)ty^2 \notag \\
&+(-1-tx+t^2x-2t^3x^2-2t^4x^2+2t^4x^3+2t^5x^3-t^6x^4+t^7x^4+t^8x^5)(-t)^{\frac12}y-(x^2+t^3x)t^3\,, \notag
\end{align}
which deforms the standard figure-eight A-polynomial $A^{\rm ref}_{\mathbf{4_1}}(x,y;t=-1)=(x-1)(y-1)(x^2-(1-x-2x^2-x^3+x^4)y+x^2y^2)$\,. Thus $T[\mathbf{4_1}]$ has massive vacua on $S^1$ corresponding to all three flat $SL(2,\C)$ connections on the figure-eight knot complement, two irreducible and one abelian. Again, these flat connections label linear combinations of the three independent contour integrals $B_{II}^{\mathbf{4_1}}, B_{III}^{\mathbf{4_1}}, B_{IV}^{\mathbf{4_1}}$ in \eqref{blocks41}.

\subsection{Vortices in $S^2\times_q S^1$ and $S^3_b$}
\label{sec:ind}

Having obtained a theory $T[M_3]$ whose vacua on $\R^2\times S^1$ match flat connections on the knot complement $M_3$, it is interesting to probe its other protected observables. Here we focus on the $S^2\times_q S^1$ indices of $T[M_3]$, and make some preliminary observations as to the nature of the ``Poincar\'e polynomial theories'' $T_{\rm poly}[M_3;r]$ on the left-hand side of the flow diagram \eqref{intro-diag}.

The 3d index \cite{Kim-index, IY-index, KW-index} of a knot-complement theory, or equivalently a partition function on $S^2\times_q S^1$, depends on three fugacities $q,\xi,\tau$ and two integer monopole numbers $n,p$\,:
\be  \begin{array}{c@{\quad}c@{\quad}c}
   \text{\underline{fugacity}} & \text{\underline{monopole \#}} & \text{\underline{symmetry}} \\
   q & - & \text{combo of $U(1)_J\subset SO(3)_{\rm Lorentz}$ and $U(1)_R$} \\
   \xi & n & U(1)_x \\
   \tau & p & U(1)_t
   \end{array}
\ee
We'll consider ``twisted'' indices $\CI(\zeta,n;\tau,p;q)=\Tr_{\CH_{n,p}(S^2)}e^{i\pi R}q^{\frac R2-J} \zeta^{e_x}\tau^{e_p}$ as in \cite{DGG-index}, in which case it's convenient to regroup fugacities into pairs of holomorphic and anti-holomorphic variables
\be \label{indfug}  q=q\,,\quad \wt q=q^{-1}\,;\qquad x=q^{\frac n2}\xi\,,\quad \wt x=q^{\frac n2}\xi^{-1}\,;\qquad -t=q^{\frac p2}\tau\,,\quad -\wt t=q^{\frac p2}\tau^{-1}\,.\ee
Then we find in examples below that the indices $\CI[M_3]$ of $T[M_3]$ develop poles at $n=r$ and $\xi=q^{\frac r2}$, or $(x,\wt x) = (q^r,1)$, whose (logarithmic) residue is the $r$-th Poincar\'e polynomial of the colored $SU(2)$ knot homology,
\be \boxed{{\rm Res}_{(x,\WT x)\to (q^r,1)} \CI[M_3]
\,=\, \lim_{\xi\to q^{r/2}} (1-q^{\frac r2}\xi^{-1})\cdot \CI[M_3](\xi,n;\tau,p;q)\Big|_{n=r} \,=\, P_K^r(t;q)}\,. \label{indxr} \ee

A similar statement holds for $S^3_b$ partition functions. The $S^3_b$ partition function $\CZ_b$ \cite{Kapustin-3dloc, HHL} of a knot-complement theory depends on the ellipsoid deformation $b$ as well as two dimensionless complexified masses $m_x$, $m_t$ for $U(1)_x,U(1)_t$, which are conveniently grouped into holomorphic and anti-holomorphic parameters
\be q = e^{2\pi i b^2}\,,\; \wt q=e^{2\pi i/b^2}\,;\quad
    x= e^{2\pi b m_x}\,,\; \wt x=e^{2\pi m_x/b}\,;\quad
    -t=e^{2\pi b m_t}\,,\; -\wt t=e^{2\pi m_t/b}\,.
\ee
Then the $S^3_b$ partition function has poles at $m_x=ibr$, or $(x,\wt x)=(q^r,1)$, with
\be \boxed{{\rm Res}_{(x,\WT x)\to (q^r,1)} \CZ_b[M_3]
\,=\, \lim_{m_x\to ibr} (m_x-ibr)\cdot \CZ_b[M_3](m_x,m_t;b) \,=\, P_K^r(t;q)}\,. \label{Zbxr} \ee

These relations are not altogether surprising, since both $\CI[M_3]$ and $\CZ_b[M_3]$ should take the form of a sum of products of vortex partition functions,
\be \label{IZblocks} \CI[M_3],\,\CZ_b[M_3] \sim \sum_\alpha B_\alpha(x,t;q) \wt B_\alpha(\wt x,\wt t;\wt q)\,, \ee
and our theory $T[M_3]$ was engineered so that the $x\to q^r$ specialization of a specific linear combination of $B_P$ would reproduce Poincar\'e polynomials. Below we will choose a convenient basis of contours so that $B_P$ is one of the $B_\alpha$'s, and manifestly gives the only contribution to the residues \eqref{indxr}, \eqref{Zbxr}. (Nevertheless, in the natural basis of contours labelled by flat connections at fixed $(x,t\approx -1,q=1)$, $B_P$ may easily correspond to a sum over multiple flat connections, including the abelian one.)

Taking the residue of a pole in an index such as \eqref{indxr} has an important physical interpretation, which was discussed in \cite{GRR-bootstrap} in the context of 4d indices and, closer to our present subject, in \cite{Bullimore:2014nla, Razamat:2014pta} in the context of 3d indices.
Let us suppose that $\CI[M_3]$ is a superconformal index --- \emph{i.e.} that we have adjusted R-charges to take their superconformal values. Then the index counts chiral operators at the origin in $\R^3$, and a pole signals the presence of an unconstrained operator $\CO$ whose vev can parametrize a flat direction in the moduli space of $T[M_3]$. Taking the residue of the pole is equivalent to giving a large vev to $\CO$, thus Higgsing any flavor symmetries under which $\CO$ transforms, and decoupling massless excitations of $T[M_3]$ around this vev.

Consider, for example, the pole at $(x,\wt x)=(1,1)$, or $(\xi,n)=(1,0)$. The pole suggests the presence of an operator $\CO_x$, of charge $+1$ under $U(1)_x$, in the zero-th $U(1)_x$ monopole sector. The contribution of this operator and its powers to the index is
\be (1+\xi + \xi^2 + \ldots)\times \CI'  = \frac{1}{1-\xi}\times\CI'\,.\ee
Taking the residue $\CI'$ amounts to giving a vev to $\CO_x$ and decoupling massless excitations around it, thereby Higgsing $U(1)_x$ symmetry. One can interpret $\CI'$ as the index of a new superconformal theory, the IR fixed point of a flow triggered by the vev $\langle \CO_x\rangle$.

More generally, taking a residue at $(x,\wt x)=(q^r,1)$ or $(\xi,n)=(q^{\frac r2},r)$ gives a space-dependent vev (with nontrivial spin) to an operator in the $r$-th monopole sector. This not only Higgses the $U(1)_x$ symmetry of $T[M_3]$ but creates a vortex defect. We therefore expect that the residue of $\CI[M_3]$ at $(x,\wt x)=(q^r,1)$ is the index of a new 3d theory $T_{\rm poly}[M_3,r]$ in the presence of a (complicated!) line operator.

In the context of 4d theories $T[C;G]$ coming from compactification of the 6d $(2,0)$ theory on a punctured Riemann surface $C$, taking the residue at a pole in the index amounted to removing a puncture from $C$ --- or more generally replacing the codimension-two defect at the puncture by a dimension-two defect in a finite-dimensional representation of $G$. Similarly, we expect here that taking a residue replaces the codimension-two defect along a knot $K\subset M$ by a dimension-two defect in the $(r+1)$-dimensional representation of $SU(2)$. We hope to elucidate this interpretation in future work. \\

We proceed to examples of \eqref{indxr}.
Our conventions for indices follow \cite{DGG-index}. Below, all indices depend on fugacities from \eqref{indfug} as well as the pair
\be
 s = q^{\frac k2}\sigma\,,\quad \wt s = q^{\frac k2}\sigma^{-1}\,,\ee
which is used for summations/integrations. We assume $|q|<1$, as is physically sensible for the index.
Thus, the convergent $q$-Pochhammer symbols are
\be \textstyle (z)_\infty := (z;q)_\infty = \prod_{i=1}^\infty (1-q^iz)\,, \ee
and theta-functions are
\be \theta(z_1,...,z_n) := \theta(z_1;q)\cdots \theta(z_n;q)\,,\qquad \theta(z;q):= (-q^{\frac12}z)_\infty (-q^{\frac12}z^{-1})_\infty\,. \ee

\subsection*{Unknot}

The index of the unknot theory $T[\mathbf{0_1}]$ from \eqref{Tunknot} is given equivalently by
\begin{align}
\label{indU}
 \CI[\mathbf{0_1}] &= (-q^{\frac12})^{n}\xi^{\frac32 p}\tau^{\frac32 n}
 \frac{(q/\wt x)_\infty (-q^2/\wt t^3)_\infty (-1/(qxt^3))_\infty}
 {(x^{-1})_\infty (-q^{-1}/t^3)_\infty (-q^2/(\wt x\wt t^3))_\infty} \\
  & = \bigg|\!\bigg|\frac{\theta(x,-q^{\frac12}(-t)^{\frac32})}{\theta(1,-q^{\frac12}x(-t)^{\frac32})} \frac{(-1/(qxt^3))_\infty}{(x^{-1})_\infty (-1/(qt^3))_\infty} \bigg|\!\bigg|^2_{\rm id} \notag \\
  &=
  \frac{\theta(x,-q^{\frac12}(-t)^{\frac32},-q^{-\frac12}\wt x(-\wt t)^{\frac32})}
 {\theta(\wt x,-q^{-\frac12}(-\wt t)^{\frac32},-q^{\frac12}x(-t)^{\frac32})}
 \times \frac{(q/\wt x)_\infty (-q^2/\wt t^3)_\infty (-1/(qxt^3))_\infty}
 {(x^{-1})_\infty (-q^{-1}/t^3)_\infty (-q^2/(\wt x\wt t^3))_\infty}\,. \notag
\end{align}
In the first line, we simply write down the index as defined by the theory --- with the massless chiral $\Phi_3$ decoupled
in order to remove an otherwise infinite factor.
In the second line, we show that this index comes from a fusion norm $\big|\!\big|B^{\mathbf{0_1}}(x,t;q)\big|\!\big|^2_{\rm id}$ of \eqref{blockU}, with $(q^{-1})^-_\infty$ removed. Since we are working at $|q|<1$, we replace all $q$-Pochhammer symbols and theta-functions
\be (z)^-_\infty \to \frac{1}{(q^{-1}z)_\infty}\,,\qquad \theta^-(z) \to \frac{1}{\theta(z)} \label{qq} \ee

We could take the limit $(\xi, n)\to (q^{\frac r2},r)$ in the first line of \eqref{indU}; after setting $n=r$, we would find a pole at $\xi\to q^{\frac r2}$ whose residue is the Poincare polynomial $P_U^r(t,q)$. But it is more illustrative to take the equivalent limit $(x,\wt x)\to (q^r,1)$ in the factorized expression on the last line. Setting $\wt x=1$ produces no divergence. The pole we are looking for comes from $(x^{-1})_\infty$ in the denominator. We get
\begin{align} \lim_{(x,\overset{\sim}{x})\to(q^r,1)}  (1-q^{-r} x)\,\CI[U]
&= \frac{\theta(q^r,-q^{\frac12}(-t)^{\frac32},-q^{-\frac12}(-\wt t)^{\frac32})}
 {\theta(1,-q^{-\frac12}(-\wt t)^{\frac32},-q^{\frac12+r}(-t)^{\frac32})}
 \times \frac{ (q)_\infty(-q^2/\wt t^3)_\infty (-q^{-r-1}/t^3)_\infty}
 {(q^{-1};q^{-1})_r(q)_\infty (-q^{-1}/t^3)_\infty (-q^2/\wt t^3)_\infty} \notag \\
 &= (-q^{\frac12})^{-r}(-t)^{-\frac{3r}{2}} \frac{(-q^2t^3)_r}{(q)_r} = P_U^{r}(t;q)\,.
\end{align}
Note how the $\wt t$ dependence completely cancelled out of the problem. If we had taken a more general limit $(x,\wt x)\to (q^r,q^{r'})$, we would have found a similar pole, with residue $P_U^r(t;q)P_U^{r'}(\wt t;q^{-1})$. The fact that the $\wt t$ dependence cancels out follows from the simple identity $P_U^{r'=0}(\wt t;q^{-1})=1$.

\subsection*{Trefoil}

For the trefoil, the theory $T[\mathbf{3_1}]$ of \eqref{T31} leads to an integral formula for the index,
\be \label{ind031}
 \CI[\mathbf{3_1}] = \CI_0 \sum_{k\in \Z}\oint \frac{d\sigma}{2\pi i\sigma} \frac{\theta(-q^{-\frac32}\wt s\wt x(-\wt t)^3)}{\theta(-q^{\frac32}sx(-t)^3)}
  \frac{(qs)_\infty(1/(-st))_\infty (qx/s)_\infty}{
  (\wt s)_\infty (q/(-\wt s\wt t))_\infty (\wt x/\wt s)_\infty}\,, \ee
where
\be \label{pref31}
 \CI_0 = \frac{\theta(-q^{-\frac12}\wt x,-q^{-\frac32}\wt x(-\wt t)^{\frac32},-q^{\frac32}x(-t)^3,-q^{\frac32}(-t)^{\frac32},x)}
{\theta(-q^{\frac12} x,-q^{\frac32}x(-t)^{\frac32},-q^{-\frac32}\wt x(-\wt t)^3,-q^{-\frac32}(-\wt t)^{\frac32},\wt x)} \times \frac{(-1/(qxt^3))_\infty (-q^2/\wt t^3)_\infty (-q/\wt t)_\infty}{(-q^2/(\wt x\wt t^3))_\infty (-1/(qt^3))_\infty (-1/t)_\infty}
\ee
Again, we have chosen to regroup Chern-Simons contributions into ratios of theta-functions, separating out the $x$ and $\wt x$ dependence. The integrand in \eqref{ind031} has three pairs of half-lines of zeroes and poles in the $\sigma$-plane, coming from the three terms $(\,)_\infty/(\,)_\infty$. They lie at
\be \label{poles31}
\begin{array}{c@{\quad}c@{\quad}c@{\quad}c}
& {\rm I}\;\;(qs)_\infty/(\wt s)_\infty & {\rm II}\;\; (-1/st)_\infty/(-q/\wt s\wt t)_\infty & {\rm III} \;\; (qx/s)_\infty/(\wt x/\wt s)_\infty \\\hline
{\rm zeroes} & \sigma  = q^{-\frac k2-1-m} & \sigma = q^{-\frac{k+p}{2}+m}\tau^{-1}& \sigma = q^{-\frac{k-n}{2}+1+m}\xi \\
{\rm poles} & \sigma = q^{\frac k2+m} & \sigma = q^{\frac{k+p}{2}-1-m}\tau^{-1} & \sigma = q^{\frac{k-n}{2}-m}\xi \\
& m\geq {\rm max}(-k,0) & m \geq {\rm max}(k+p,0) & m \geq {\rm max}(k-n,0)
\end{array}
\ee
The real, physical contour in \eqref{ind031} should lie on or around the unit circle, separating each half-line of zeroes from its corresponding half-line of poles.

We also observe that the integrand of \eqref{ind031} vanishes as $|\sigma|\to \infty$, if we stay away from half-lines of poles. Thus we can attempt to deform the contour outwards, closing it around $\sigma=\infty$. We pick up the poles in lines II and III, obtaining an expression of the form
\be \label{ind31} \CI[\mathbf{3_1}] = \CI_0\,\big( |\!|B_{II}|\!|^2_{\rm id} + |\!|B_{III}|\!|^2_{\rm id}\big)\,, \ee
where%
\footnote{A redefinition of summation indices turns the sum over $k\in \Z$ into sums $k\geq 0$.}
\begin{subequations} \label{ind31sub}
\begin{align}
|\!|B_{II}|\!|^2_{\rm id} &=
 \sum_{k,m\geq 0} \frac{\theta(-q^{-\frac12+m}\wt x\wt t^2)}{\theta(-q^{\frac12-k}xt^2)} \frac{1}{(q)_k(q^{-1};q^{-1})_m} \frac{(-q^{-k}t^{-1})_\infty(-q^{2+k}tx)_\infty}{(-q^{m+1}\wt t^{-1})_\infty(-q^{-1-m}\wt t\wt x)_\infty}\,, \\
|\!|B_{III}|\!|^2_{\rm id} &= \sum_{k,m\geq 0} \frac{\theta(q^{-\frac32+m}\wt x^2\wt t^3)}{\theta(q^{\frac32-k}x^2t^3)}
 \frac{1}{(q)_k(q^{-1};q^{-1})_m}
 \frac{(q^{1-k}x)_\infty (-q^k/(xt))_\infty}{(q^m\wt x)_\infty(-q^{1-m}/(\wt x\wt t))_\infty}\,.
\end{align}
\end{subequations}
The integrals $B_{II}$ and $B_{III}$ here correspond to contours $\Gamma_{II}$ and $\Gamma_{III}$ in Figure \ref{fig:cont31}, with substitutions of the form $(x)_\infty^-\to 1/(qx)_\infty$ to account for $|q|<1$.

Now, if we send $(x,\wt x)\to (q^r,1)$, the leading pole in line I can collide with the leading pole in line III, pinching the integration contour in the $\sigma$-plane, and leading to a divergence of the the index. We see this explicitly in the evaluated expression \eqref{ind31}: while the prefactor $\CI_0$ and the integrals $|\!|B_{II}|\!|^2_{\rm id}$ are finite in this limit, the integrals $|\!|B_{III}|\!|^2_{\rm id}$ have the expected divergence. It comes from the denominator $(q^m\wt x)_\infty$ in (\ref{ind31sub}b), and occurs only for $m=0$. The related factor $(q^{1-k}x)_\infty$ in the numerator vanishes as $x=q^r$ unless $k\leq r$. Therefore, we find a residue
\begin{align}  \lim_{(x,\overset{\sim}{x})\to(q^r,1)}  (1-\wt x)\,\CI[\mathbf{3_1}] &=
  \lim_{(x,\overset{\sim}{x})\to(q^r,1)}  (1-\wt x)\,\CI_0\, |\!|B_{III}|\!|^2_{\rm id} \notag \\
  &=  \CI_0(x=q^r,\wt x=1;t,\wt t;q) \sum_{k=0}^r  \frac{\theta(q^{-\frac32} \wt t^3)}{\theta(q^{\frac 32-k}t^3)}\frac{(q^{1-k+r})_\infty(-q^{k-r}t^{-1})_\infty}{(q)_k(q)_\infty (-q/\wt t)_\infty} \notag \\
  &= P_{3_1}^r(t;q) \; P_{3_1}^0(\wt t;q^{-1}) = P_{3_1}^r(t;q)\,,
\end{align}
reproducing the superpolynomial after some straightforward manipulations.

\subsection*{Figure-eight knot}

The setup for the figure-eight knot is almost identical to that for the trefoil. Now the index is given by
\be \label{ind410} \CI[\mathbf{4_1}] = \CI_0 \sum_{k\in \Z} \oint \frac{d\sigma}{2\pi i\sigma}
\frac{\theta(q^{-1}\wt s,\wt t^2\wt s)}{\theta(qs,t^2s)} \frac{(qs)_\infty  (-1/(ts))_\infty (qx/s)_\infty (-1/(qxt^3s))_\infty}
   {(\wt s)_\infty (-q/(\wt t\wt s))_\infty (\wt x/\wt s)_\infty (-q^2/(\wt x\wt t^3\wt s))_\infty }\,,
\ee
with
\be \CI_0 = \frac{\theta(t^2,q^{\frac12}t,x(-t)^{-\frac12},-q^{-\frac12}\wt x,q^{-\frac12}\wt t\wt x,(-\wt t)^{-\frac12})}
{\theta(\wt t^2,q^{-\frac12}\wt t,\wt x(-\wt t)^{-\frac12},-q^{\frac12}x,q^{\frac12}tx,(-t)^{-\frac12})} \times \frac{(-q^2/\wt t^3)_\infty(-q/\wt t)_\infty}{(-1/qt^3)_\infty(-1/t)_\infty}\,.
\ee
There are four pairs of half-lines of zeroes and poles in the integrand; three are identical to those in the trefoil integrand above, which we denote I, II, III as in \eqref{poles31}, and there is one new pair
\be \label{poles41}
{\rm IV}:\qquad \begin{array}{c}
\text{zeroes $\sigma = q^{-\frac{k+n+3p}{2}-1+s}\xi^{-1}\tau^{-3}$} \\
\text{poles $\sigma = q^{\frac{k+n+3p}{2}-2+s}\xi^{-1}\tau^{-3}$}
\end{array},
\quad \text{for $m\geq {\rm max}(k+n+3p,0)\,.$}
\ee
We close the contour around $\sigma=\infty$ (where the integrand generically vanishes), picking up the poles in lines II, III, and IV, to give
\be \CI[\mathbf{4_1}] = \CI_0\,\big( |\!|B_{II}|\!|^2_{\rm id} + |\!|B_{III}|\!|^2_{\rm id}+ |\!|B_{IV}|\!|^2_{\rm id}\big)\,, \label{ind41} \ee
with
\begin{subequations} \label{ind41sub}
\begin{align}
|\!|B_{II}|\!|^2_{\rm id} &= \sum_{k,m\geq 0} \frac{\theta(-q^m/\wt t,-q^{m+1}\wt t)}{\theta(-q^{-k}/t,-q^{-k-1}t)} \frac{1}{(q)_k(q^{-1},q^{-1})_m}
\frac{(-q^{-k}/t)_\infty (-q^{2+k}xt)_\infty (q^k/(xt^2))_\infty}
 {(-q^{m+1}/\wt t)_\infty (-q^{-m-1}\wt x\wt t)_\infty (q^{1-m}/(\wt x \wt t^2))_\infty}\,, \\
|\!|B_{III}|\!|^2_{\rm id} &= \sum_{k,m\geq 0} \frac{\theta(q^{m-1}\wt x,q^m\wt t^2\wt x)}{\theta(q^{1-k}x,q^{-k}t^2x)} \frac{1}{(q)_k(q^{-1},q^{-1})_m}
\frac{(q^{1-k}x)_\infty (-q^k/(xt))_\infty (-q^{k-1}/(x^2t^3))_\infty}
{(q^m\wt x)_\infty (-q^{1-m}/(\wt t\wt x))_\infty (-q^{2-m}/(\wt x^2\wt t^3))_\infty}\,, \\
|\!|B_{IV}|\!|^2_{\rm id} &= \sum_{k,m\geq 0} \frac{\theta\Big(\frac{-q^{m+1}}{\WT x\WT t^3}, \frac{-q^{m+2}}{\WT x\WT t}\Big)} {\theta\Big(\frac{-q^{-k-1}}{xt^3},\frac{-q^{-k-2}}{xt}\Big)}
\frac{1}{(q)_k(q^{-1},q^{-1})_m}
\frac{(-q^{-k-1}/(xt^3))_\infty (q^{2+k}xt^2)_\infty (-q^{k+3}x^2 t^3)_\infty }
{(-q^{m+2}/(\wt x\wt t^3))_\infty (q^{-m-1}\wt x\wt t^2)_\infty (-q^{-m-2}\wt x^2\wt t^3)_\infty }\,,
\end{align}
\end{subequations}
The integrals here correspond to contours discussed above \eqref{blocks41} (with the usual translation from $|q|>1$ to $|q|<1$).

Now as $(x,\wt x)\to (q^r,1)$, the prefactor $\CI_0$ along with $|\!|B_{II}|\!|^2_{\rm id}$ and $|\!|B_{VI}|\!|^2_{\rm id}$ all have finite limits; while $|\!|B_{III}|\!|^2_{\rm id}$ has a pole due $1/(q^m\wt x)_\infty$ at $m=0$, and is nonvanishing for $k\leq r$. As in the case of the trefoil, the divergence can be attributed to the poles of lines I and III pinching the contour of the integrand \eqref{ind410}. We then find
\begin{align}  \lim_{(x,\overset{\sim}{x})\to(q^r,1)}  (1-\wt x)\,\CI[\mathbf{4_1}] &=
  \lim_{(x,\overset{\sim}{x})\to(q^r,1)}  (1-\wt x)\,\CI_0\, |\!|B_{III}|\!|^2_{\rm id} \notag \\
  &= P_{4_1}^r(t;q) \; P_{4_1}^0(\wt t;q^{-1}) = P_{4_1}^r(t;q)\,.
\end{align}

\section{The $t=-1$ limit and DGG theories}
\label{sec:t1}

Above, we saw that sending $x\to q^r$ in partition functions of $T[M_3]$ (and perhaps discarding an overall divergence) produced finite Poincar\'e polynomials of colored $SU(2)$ knot homologies. Once the Poincar\'e polynomials are obtained, we are free to send $t\to -1$ to directly recover the colored Jones polynomials.
No further divergences are encountered.
Physically, we  proposed an identification of the regularized $x\to q^r$ limit with a physical ``Higgsing'' process, by which an operator in $T[M_3]$ charged under $U(1)_x$ is given a space-dependent vev, initiating an RG flow to a new theory in the presence of a line defect.
Subsequently sending $t\to -1$ should not correspond to any further flow.

One may wonder what would happen if we sent $t\to -1$ \emph{before} $x\to q^r$. We present evidence in this section that this initiates a \emph{different} RG flow in $T[M_3]$, which ends at a DGG theory $T_{DGG}[M_3]$. In particular, an operator $\CO_t$ is given a (constant) vev, breaking the $U(1)_t$ symmetry characteristic of $T[M_3]$. Moreover, vacua of $T[M_3]$ on $\R^2 \times S^1$ that correspond to abelian or reducible flat connections on $M_3$ are lost.

As above, our analysis will be largely example-driven. In Section~\ref{sec:DGG} we examine how the trefoil and figure-eight knot theories of Section~\ref{sec:theory} flow to DGG theories. We verify in Section~\ref{sec:t1ind} that  $t\to -1$ limits induce divergences in $S^2\times_q S^1$ indices, indicative of Higgsing.
Then in Section \ref{sec:t1blocks} we use effective twisted superpotentials on $\R^2\times S^1$ to better understand how vacua corresponding to abelian flat connections decouple.

\subsection{The DGG theories}
\label{sec:DGG}

We can see an explicit example of the proposed DGG flow by considering the trefoil theory $T[\mathbf{3_1}]$ of \eqref{T31}. If we turn off the real mass for the flavor symmetry $U(1)_t$, then the chiral operator $\CO_t = \Phi_4$ can get a vev,
\be \langle \Phi_4 \rangle = \Lambda\,. \ee
The vev breaks $U(1)_t$, but no other symmetries. Moreover, it induces a complex mass for $\Phi_1$ and $\Phi_3$ due to the superpotential
\be W_{\mathbf{3_1}} = \mu_1\,\Phi_1\Phi_2\Phi_5\Phi_6 + \mu_2\Lambda\, \Phi_1\Phi_3 + \mu_3 \Phi_6V_-\,. \ee
Therefore, taking $\Lambda\to \infty$, we may decouple fluctuations of $\Phi_4$ and integrate out $\Phi_1$ and $\Phi_3$, arriving at
\be \label{T31'}
T'[\mathbf{3_1}]:\quad
\begin{array}{c|ccc|c}
       &\Phi_{2}  &\Phi_{5} &\Phi_{6} & V_-\\ \hline
U(1)_s      &1         &0        &0                        &0      \\
U(1)_x     &-1         &0        &1                        &-1     \\\hline
U(1)_R      &0         &-2       &4                        &-2     \end{array}
\,,\qquad
\text{CS}:\; \begin{array}{c|c|c|c}
& U(1)_s & U(1)_x & U(1)_R \\\hline
U(1)_s & - 1/2 & 3/2 & 5/2 \\\hline
U(1)_x & 3/2 & 0 & 2 \\\hline
U(1)_R & 5/2 & 2 & 0
\end{array}\,
\ee
with superpotential
\be W_{\mathbf{3_1}}' = \mu_3'\, \Phi_6 V_-\,. \ee

At this point, we observe that $T[\mathbf{3_1}]$ has a sector containing a $U(1)_s$ gauge theory with a single charged chiral $\Phi_2$, together with minus half a unit of background Chern-Simons coupling.  This sector can be dualized to an ungauged chiral $\varphi$ as in \cite[Sec 3.3]{DGG}, a consequence of a basic 3d mirror symmetry \cite{IS, dBHOY}. Indeed, the dual ungauged chiral is identified with the (anti-)monopole operator $\varphi=V_-$ of $U(1)_s$\,! Thus, $T'[\mathbf{3_1}]$ is dual to
\be \label{T31''}
T''[\mathbf{3_1}]:\quad
\begin{array}{c|ccc}
         &\Phi_{5} &\Phi_{6} & \varphi \\ \hline
U(1)_s            &0        &0                        &0      \\
U(1)_x            &0        &1                        &-1     \\\hline
U(1)_R             &-2       &4                        &-2     \end{array}
\,,\qquad
\text{CS}:\; \begin{array}{c|c|c}
 & U(1)_x & U(1)_R \\\hline
U(1)_x   & 3 & 6 \\\hline
U(1)_R  & 6 & *
\end{array}\,.
\ee
with $W_{\mathbf{3_1}}'' = \mu_3''\,\Phi_6\varphi$. The superpotential lets us integrate out $\Phi_6$ and $\varphi$, leaving behind
\be T''[\mathbf{3_1}] \;\leadsto\; T_{DGG}[\mathbf{3_1}]\otimes T_{\Phi_5}\,. \ee
Here $\Phi_5$ is a fully decoupled free chiral, while $T_{DGG}[\mathbf{3_1}]$ is a slightly degenerate description of the DGG trefoil theory.

Namely, $T_{DGG}[\mathbf{3_1}]$ here is a ``theory'' consisting only of a background Chern-Simons coupling at level 3 for the flavor symmetry $U(1)_x$, and some flavor-R contact terms given by the matrix on the RHS of \eqref{T31''}.
A similar ``theory'' was obtained by DGG methods in \cite[Section 4.3]{DGG-index}, using a degenerate triangulation of the trefoil knot complement into two ideal tetrahedra. It was interpreted as an extreme limit of the true DGG theory $T_{DGG}[\mathbf{3_1}]$ in marginal parameter space. It is not surprising that we have hit such a limit, since, as discussed at the beginning of Section \ref{sec:theory}, we are ignoring some marginal deformations.

Our $T_{DGG}[\mathbf{3_1}]$ becomes identical to that in \cite[Section 4.3]{DGG-index} upon shifting R-charges by minus two units of $U(1)_x$ charge. The shift is due to difference of conventions: we initially set $x=q^r$ in Poincar\'e polynomials whereas the equivalent choice for \cite{DGG,DGG-index} would be $x=q^{r+1}$. \medskip

We can repeat this exercise for the figure-eight knot. The theory $T[\mathbf{4_1}]$ of \eqref{T41} again has a chiral operator $\CO_t = \Phi_4$ that is charged only under $U(1)_t$, and can get a vev when the real mass corresponding to $U(1)_t$ is turned off,
\be \langle \Phi_4\rangle = \Lambda\,. \ee
Then the effective superpotential
\be W_{\mathbf{4_1}} = \mu_1\Lambda\,\Phi_1\Phi_3 + \mu_2\,\Phi_1^2\Phi_2\Phi_5\Phi_6 \ee
lets us integrate out $\Phi_1$ and $\Phi_3$. We flow directly to a theory
\be T[\mathbf{4_1}] \;\leadsto\; T_{DGG}[\mathbf{4_1}]\otimes T_{\Phi_6}\,, \ee
where $\Phi_6$ is a decoupled chiral and
\be \label{TDGG41}
T_{DGG}[\mathbf{4_1}]:\quad
\begin{array}{ c |  c  c }
           &\Phi_{2}  &\Phi_{5}   \\  \hline
U(1)_{s}   &1  	  &1        \\\hline
U(1)_{x}   &-1 	  &1      \\\hline
U(1)_{R}   &0 	  &4
\end{array}\;,\qquad\quad \text{(CS vanishing)}
\ee
is basically the GLSM description of the $\mathbb{CP}^1$ sigma-model. It is equivalent (after shifting R-charges by minus two units of $U(1)_x$ charge) to the DGG theory obtained from a triangulation of the figure-eight knot complement into two tetrahedra.
Again, this triangulation is a little degenerate (as discussed explicitly in \cite[Section 4.6]{DGG}), so \eqref{TDGG41} should be viewed as a limit of the true $T_{DGG}[\mathbf{4_1}]$, which has the same protected partition functions (index, half-indices, etc.).

\subsection{Indices and residues}
\label{sec:t1ind}

The $S^2\times_qS^1$ indices of theories $T[M_3]$ help us to further illustrate the breaking of $U(1)_t$ by ``Higgsing'' and the flow to $T_{DGG}[M_3]$. As discussed in Section \ref{sec:ind}, Higgsing corresponds to taking residues in an index. In particular, we expect here to find the indices $\CI_{DGG}[M_3]$ of DGG theories as residues of $\CI[M_3]$ at $(t,\wt t)\to (-1,-1)$.

Consider, for example, the index $\CI[\mathbf{3_1}]$ of the trefoil theory as given by \eqref{ind31}. Sending $t\to -1$, the prefactor $\CI_0$ develops a pole due to the factor $1/(-1/t)_\infty$. This factor comes directly from the chiral $\Phi_4$ in $T[\mathbf{3_1}]$. (The factor $1/(-1/(qt^3))_\infty$ in $\CI_0$, coming from the chiral $\Phi_4$, also develops a pole, but it is not relevant for the Higgsing we want to do.)
In addition, we see that $|\!|B_{III}|\!|^2_{\rm id}$ has a finite limit as $(t,\wt t)\to (-1,-1)$, whereas $|\!|B_{III}|\!|^2_{\rm id}$ \emph{vanishes} due to $(-q^{-k}t^{-1})_\infty$ in the numerator. One way to understand this vanishing is to observe that the zeroes in line I of the index integrand perfectly cancel all poles in line II when $(t,\wt t)= (-1,-1)$. Therefore,
\begin{align} \label{indt31}
 &\lim_{t,\WT t\to -1} (1-t) \CI[\mathbf{3_1}] \;= \lim_{t,\WT t\to -1} (1-t)\,\CI_0\,|\!|B_{III}|\!|^2_{\rm id} \\
 &\qquad = \raisebox{.3cm}{\text{``}} \frac{(-q^2/\wt t^3)_\infty}{(-1/(qt^3))_\infty} \raisebox{.3cm}{\text{''}} \frac{\theta(-q^{-\frac12}\wt x,x) (1/(qx))_\infty}{\theta(-q^{\frac12}x,\wt x) (q^2/\wt x)_\infty} \sum_{k,m\geq 0} \frac{1}{(q)_k(q^{-1};q^{-1})_m}\frac{\theta(-q^{m-\frac32}\wt x^2,-q^{\frac12-k}x)}{\theta(-q^{\frac32-k}x^2,-q^{m-\frac12}\wt x)} \notag \\
 &\qquad =  \raisebox{.3cm}{\text{``}} \frac{(-q^2/\wt t^3)_\infty}{(-1/(qt^3))_\infty} \raisebox{.3cm}{\text{''}}
  \frac{\theta(x,q^{-\frac32}\wt x^2)}{\theta(\wt x,-q^{\frac32}x^2)}
   \;=\; \raisebox{.3cm}{\text{``}} \frac{(-q^2/\wt t^3)_\infty}{(-1/(qt^3))_\infty} \raisebox{.3cm}{\text{''}} q^{3n}\xi^{3n} \notag \\
 &\qquad  =  \raisebox{.3cm}{\text{``}} \frac{(-q^2/\wt t^3)_\infty}{(-1/(qt^3))_\infty} \raisebox{.3cm}{\text{''}} \CI_{DGG}[\mathbf{3_1}]\,. \notag
\end{align}
The resummation in the third line captures the duality between a charged chiral $(\Phi_2)$ and a free chiral $(\varphi=V_-)$ discussed in Section \ref{sec:DGG}. Then the expression $q^{3n}\xi^{3n}$ matches the DGG trefoil index of \cite{DGG-index}, modulo a redefinition of R-charges $\xi \to q^{-1}\xi$.
The infinite prefactor $(-q^2/\wt t^3)_\infty/(-1/(qt^3))_\infty\to (q^2)_\infty/(q^{-1})_\infty$ is the contribution of the decoupled chiral $\Phi_5$. \medskip

When considering the $t,\wt t\to -1$ limit of the figure-eight index $\CI[\mathbf{4_1}]$ from \eqref{ind41}, the prefactor $\CI_0$ has the same divergent term $(-1/t)_\infty^{-1}$ that appeared for the trefoil. Moreover, the contribution $|\!|B_{II}|\!|^2_{\rm id}$ to the figure-eight index vanishes, because poles of the index integrand in line II are cancelled by zeroes in line I. Thus, following a short calculation, the figure-eight index takes the form
\begin{align} \label{indt41}
 &\lim_{t,\WT t\to -1} (1-t) \CI[\mathbf{4_1}] \;= \lim_{t,\WT t\to -1} (1-t)\,\CI_0\,\big(|\!|B_{III}|\!|^2_{\rm id}+|\!|B_{IV}|\!|^2_{\rm id}\big) \\
  &\qquad = \raisebox{.3cm}{\text{``}} \frac{(-q^2/\wt t^3)_\infty}{(-1/(qt^3))_\infty} \raisebox{.3cm}{\text{''}}
  (q\xi)^{2n}\bigg[ (-q^{\frac12})^{n} \sum_{k,m\geq 0} \frac{(qx)^{k}(q^{-1}\wt x)^{m}}{(q^{-1};q^{-1})_k(q)_m}\frac{(q^{k+1}(qx)^2)_\infty}{(q^{-m}(q^{-1}\wt x)^2)_\infty} \notag \\
  &\hspace{2.5in} + (n,q\xi)\leftrightarrow(-n,1/(q\xi)) \;\bigg] \notag \\
  &\qquad  =  \raisebox{.3cm}{\text{``}} \frac{(-q^2/\wt t^3)_\infty}{(-1/(qt^3))_\infty} \raisebox{.3cm}{\text{''}} \CI_{DGG}[\mathbf{4_1}]\,. \notag
\end{align}
We recognize in this the DGG index of the figure-eight knot, where we should again rescale $\xi \to q^{-1}\xi$, or $(x,\wt x)\to (q^{-1}x,q\wt x)$.

\subsection{Critical points and missing vacua}
\label{sec:t1blocks}

We saw in Section \ref{sec:t1ind} that in the limit $t,\wt t \to -1$, some parts of indices $\CI[M_3]$ vanished, while others contributed to $\CI_{DGG}[M_3]$. This is a reflection of the fact that the DGG theories $T_{DGG}[M_3]$ don't capture all information about flat connections on $M_3$, and in particular don't have massive vacua on $\R^2\times S^1$ corresponding to abelian or reducible flat $SL(2,\C)$ connections.

We can make this idea much more precise by considering the effective twisted superpotentials that govern theories $T[M_3]$ on $\R^2\times S^1$. For example, for the trefoil, this was given by \eqref{Wt31}:
\begin{align}  \wt W_{\mathbf{3_1}}(s;x,t) &= \Li_2(s)+\Li_2(-1/(st))+\Li_2(x/s)+\Li_2(-t^{3})+\Li_2(-t)+\Li_2(-1/(t^3x)) \notag\\
&\hspace{-1cm} + \tfrac12\big( (\log s)^2+\log s(6\log t+2\log x)+\log x(\log x+3\log(- t))+10(\log t)^2\big)\,. \label{W31rep}
\end{align}
It is important to note that this function on $\C^*$ (parametrized by the dynamic variable $s$) has branch cuts coming from integrating out chiral matter that at some points in the $s$-plane becomes massless. In particular, each term $\Li_2(f(s))$ has a cut along a half-line starting at the branch point $f(s)=1$ and running to zero or infinity.
Such cuts and their consequences have been discussed from various perspectives in \emph{e.g.} \cite{Witten-phases, NS-I, GW-Jones, GGP-walls}.
Often one writes the vacuum or critical-point equations as
\be \label{expW} \exp\big(s\, \partial \wt W_{\mathbf{3_1}}/\partial s\big) = 1\,, \ee
because in this form they are algebraic in $s$. However, when analyzing vacua of $T[M_3]$ on $\R^2\times S^1$, one must remember to lift solutions of \eqref{expW} back to the cover of the $s$-plane defined by $\wt W$ --- and to make sure they are actual critical points on some sheets of the cover.

Now consider what happens if we send $t\to -1$. The branch points of $\Li_2(s)$ and $\Li_2(-1/(st))$, located at $s=1$ and $s=-1/t$, collide. (These branch points came directly from the chirals $\Phi_1$ and $\Phi_3$, which we integrated out of $T[\mathbf{3_1}]$ in \eqref{T31'}.) In the process, the half-line cuts originating at these branch points coalesce into a full cut running from $s=0$ to $s=\infty$; this is easy to see from the inversion formula
\be
\Li_2(s) + \Li_2(1/s) = -\tfrac{\pi^2}{6}-\tfrac12\log(-s)^2   \qquad (s\notin [0,1)\;)\,.     \label{cancel31}
\ee
Moreover, one of the solutions $s_*$ to \eqref{expW}, or rather its lift(s) to the covering of the $s$-plane, gets trapped between the colliding branch points and ceases to be a critical point as $t\to -1$. One can see this from the explicit form of the critical-point equations \eqref{crit31}, which are reduced from quadratic to linear order in $s$ by a cancellation at $t=-1$. However, to properly interpret this limit, it is helpful to think about the branched cover of the $s$-plane as we have done.

Physically, each solution of \eqref{expW} is a vacuum of $T[M_3]$ on $\R^2\times S^1$. As $t\to -1$, the vacuum at $s_*$ is lost. This is possible precisely because the $t\to -1$ limit is singular. Indeed, we know that $t\to -1$ corresponds to making $T[M_3]$ massless, so that the reduction on $\R^2\times S^1$ is no longer fully described by an effective twisted superpotential. The specialized superpotential $\wt W(s;x,t=-1)$ does not describe $T[M_3]$ itself at the massless point, but rather the Higgsed $T_{DGG}[M]$ as found in Section \ref{sec:DGG}.

In the case of the trefoil, the vacuum at $s_*$ close to $t=-1$ is labelled (via the 3d-3d correspondence) by the abelian flat connection on $M_3=S^3\backslash K$. Indeed, if we substitute the limiting $t\to -1$ value of $s_*$ (namely $s_*=1$) into the SUSY-parameter-space equation $\exp\big(x\,\partial \wt W/\partial x\big) = y$, we find
\be y_* := \exp\big(x\,\partial \wt W/\partial x\big)\big|_{s_*} = 1 \qquad \text{at $t=-1$}\,, \ee
corresponding to the abelian factor $y-1=0$ of the trefoil's classical A-polynomial. Thus we see explicitly that the DGG theory $T_{DGG}[\mathbf{3_1}]$ loses a vacuum corresponding to the abelian flat connection.

We may also perform this analysis at the level of vortex partition functions with suitable boundary conditions. Boundary conditions are labelled by ($q$-deformed) critical points of $\wt W$ --- or more precisely by integration cycles $\Gamma_\alpha$ obtained by starting at a critical point of $\wt W$ and approximately following gradient flow with respect to $\text{Re}\,\frac{1}{\log q}\wt W$. For the trefoil we can choose a basis of integration cycles given by $\Gamma_{II}$ and $\Gamma_{III}$ in Figure \ref{fig:cont31}. The precise correspondence with critical points depends on $x,t,q$. Close to $t=-1$, however, it is clear that $\Gamma_{II}$ corresponds to the ``abelian'' critical point $s_*$. As $t\to-1$, the contour $\Gamma_{II}$ gets trapped crossing a full line of poles (resolutions of the classical branch cuts described above), and ceases to be a good integration cycle.%
\footnote{Of course, $\Gamma_{II}$ \emph{is} still a reasonable integration cycle, mathematically, at $t=-1$ and any finite $q$. The integral along it does reproduce a Jones polynomial as $x\to q^r$. It is tempting to wonder whether one could engineer such a ``Jones'' cycle starting directly with $T_{DGG}[M_3]$, with no prior knowledge of the full $T[M_3]$ --- and what the physical meaning of this cycle might be.} %
Most importantly, it no longer flows from any classical critical point. Beautifully, the remaining contour $\Gamma_{III}$ is isolated away from the point $s_*$ where half-lines of poles merge. The $t\to-1$ limit of the corresponding integral $B_{III}(x,t;q)$ is precisely the contour integral of $T_{DGG}[\mathbf{3_1}]$, labelled by the irreducible flat $SL(2,\C)$ connection, and contributing to the index (\ref{indt31}).

Analogous remarks apply to the figure-eight example. The 3d Higgsing and integrating out of $\Phi_1,\Phi_3$ in $T[\mathbf{4_1}]$ translates on $\R^2\times S^1$ to branch points of $\Li_2(s)$ and $\Li_2(-1/(st))$ colliding in \eqref{Wt41}, and trapping a critical point between them. Thus, as $t\to -1$, $T[\mathbf{4_1}]$ looses one of its three massive vacua on $\R^2\times S^1$ --- the one labeled by the abelian connection on the figure-eight knot complement. The $T_{DGG}[\mathbf{4_1}]$ only has two massive vacua, labelled by irreducible flat connections. The remaining vacua correspond to the contour integrals $B_{III}$ and $B_{IV}$, which at $t\to -1$ become those of $T_{DGG}[\mathbf{4_1}]$.

\subsection{Relation to colored differentials}
\label{sec:gen}

We expect that the Higgsing procedure found to relate $T[M_3]$ to $T_{DGG}[M_3]$ in the examples above holds much more generally.
We can actually recognize some key signatures of the reduction in a much larger family of examples, which include so-called thin knots. The phenomena described above follow from the structure of colored Poincar\'e polynomials for these knots. The structure of the Poincar\'e polynomials  is highly constrained by the properties of colored differentials whose existence in $S^r$-colored homologies was postulated in \cite{GS,Gorsky:2013jxa}, as well as by the so-called exponential growth. Using these properties, in \cite{FGSS-AD} colored Poincar\'e polynomials  of many thin knots, including the infinite series of $(2,2p+1)$ torus knots and twist knots with $2n+2$ crossings, were uniquely determined.

More precisely, colored differentials enable transitions between homology theories labeled by the $r$-th and $k$-th symmetric-power representations $S^r$ and $S^k$. The  existence of these differentials implies that Poincar\'e polynomials  take the form of a summation (over $k=0,\ldots,r$), with the summand involving a factor  $(-a q^{-1} t;q)_k$. On the other hand, the exponential growth  is the statement that for $q=1$ (normalized) colored Poincar\'e polynomials (superpolynomials) satisfy the relation
\be
\CP^{S^r}_K(a,q=1,t)=\left( \CP^{S^1}_K(a,q=1,t) \right)^r.
\ee
If the uncolored superpolynomial on the right hand side is a sum of a few terms, its $r$'th power can be written as a (multiple) summation involving Newton binomials, which for arbitrary $q$ turn out to be replaced by $q$-binomials \cite{FGSS-AD,Nawata}. This structure can be clearly seen in the example of $(2,2p+1)$ torus knots considered in \cite{FGSS-AD,Nawata}, whose (normalized) colored superpolynomials take the form
\begin{eqnarray}\label{fort2k}
\!\!\!\!\!\!\!\!{\CP}^{S^r}_{T^{2,2p+1}}(a,q,t) &=& a^{pr} q^{-pr}  \sum_{0\le k_p \le \ldots \le k_2 \le k_1 \le r}
\left[\!\begin{array}{c} r\\k_1 \end{array}\!\right]\left[\!\begin{array}{c} k_1\\k_2 \end{array}\!\right]\cdots\left[\!\begin{array}{c} k_{p-1}\\k_p \end{array}\!\right]  \times\\
& & \nonumber \!\!\!\!\!\!\!\!\!\!\!\!\!\!\!\!\!\!\!\! \times \,\,\, q^{(2r+1)(k_1+k_2+\ldots+k_p)-\Sigma_{i=1}^p k_{i-1}k_i} t^{2(k_1+k_2+\ldots+k_p)} \prod_{i=1}^{k_1}(1+aq^{i-2}t).
\end{eqnarray}
Here the last product originating from the structure of differentials, as well as a series of $q$-binomials originating from the exponential growth, are manifest (in this formula $k_0=r$). Poincar\'e polynomials for infinite families of twist knots derived in \cite{FGSS-AD,Nawata} share analogous features.

It becomes clear now that various properties of trefoil and figure-8 knots,  discussed earlier, should also be present for other knots, such as thin knots discussed above. For example, as discussed in section \ref{sec:t1ind}, the divergence at $t\to -1$ in the trefoil and figure-8 indices, $\CI[\mathbf{3_1}]$ and $\CI[\mathbf{4_1}]$, is a manifestation of a pole due to the factor $1/(-1/t)_\infty$. This factor originates from the $q$-Pochhammer symbol $(-a q^{-1} t;q)_k$ in corresponding Poincar\'e polynomials (\ref{ptrefoil-a}) and (\ref{pfigure8}), after setting $a=q^2$ and rewriting this term in the denominator. As follows from the discussion above, such a factor is present in general for other thin knots (and represents the action of colored differentials), so for such knots an analogous pole at $t\to -1$ should develop. We postulate that the residue at this pole in general reproduces indices $\CI_{DGG}[M]$ for theories dual to other (thin) knots.

Similarly, a decoupling of the abelian branch for more general knots is a consequence of the structure of superpolynomials described above. From this perspective, let us recall once more how this works for trefoil and figure-8 knot, just on the level of critical point equations (\ref{crit31}) and (\ref{crit31bis}), or (\ref{crit41}) and (\ref{crit41bis}).
If we set $t=-1$ in (\ref{crit31}) or (\ref{crit41}), the ratio $\frac{1+st}{1-s}$ on the left hand side drops out of the equation (this is a manifestation of the cancelation (\ref{cancel31}) at the level of twisted superpotential ). In this ratio the numerator $1+st$ has its origin in the $(-a q^{-1} t;q)_k$ term in superpolynomials  (\ref{ptrefoil-a}) and (\ref{pfigure8}), while the denominator $1-s$ originates from $q$-Pochhamer $(q;q)_k$ being a part of the $q$-binomial in those superpolynomials. As explained above, such terms appear universally in superpolynomials for thin knots. Similarly, for $t=-1$ the equations (\ref{crit31bis}) and (\ref{crit41bis}) reduce to $y=1$ (which represents the abelian branch that drops out when $t\to -1$ is set first) due to a cancellation between the term in their numerator and $s-x$ in denominator. The terms in numerator have the origin in $(a(-t)^3;q)_r$ from unknot normalization (\ref{punknot}), possibly combined with another term $(aq^r(-t)^3;q)_k$ representing colored differentials for figure-8 knot (\ref{pfigure8}). The term $s-x$ in denominator has its origin in $(q;q)_{r-k}$ ingredient of $q$-binomial. Analogous terms, responsible for cancellations, are also universally present in superpolynomials for other knots. The analysis is slightly more involved if Poincar\'e polynomials include multiple summations --- \emph{e.g.} for $(2,2p+1)$ torus knots (\ref{fort2k}) --- however one can check that similar cancellations between ``universal'' terms decrease the degree of saddle equations and result in the decoupling of the abelian branch.

%%%%%%%%%%%%%%%%%%%%%%%%%%%%%%%%%%%%%%%%%%%%%%%%%%%%%%%%%%%%%%%%%%%%%%%%%%%%%%%%%%%%%%%%%%%%%%

\section{Boundaries in three dimensions}
\label{sec:surgeries}

In this section we discuss the gluing along boundaries of $M_3$ and the boundary conditions in 3d $\CN=2$ theories $T[M_3]$.

In particular, understanding the operations of cutting and gluing $M_3$ along a Riemann surface $C$
opens a new window into the world of closed 3-manifolds.
The basic idea of how such operations should manifest in 3d $\CN=2$ theory $T[M_3]$ was already
discussed {\it e.g.} in \cite{Yamazaki-3d,DG-Sdual} and will be reviewed below.
The details, however, cannot work unless $T[M_3]$ accounts for all flat connections on $M_3$.
This was recently emphasized in \cite{GGP-4d} where the general method of building $T[M_3]$ via gluing was carried out for certain homology spheres.

After constructing 3d $\CN=2$ theories $T[M_3]$ for certain homology spheres, we turn our attention to boundary conditions in such theories.
Incorporating boundary conditions and domain walls in general 3d $\CN=2$ theories was discussed in \cite{GGP-walls} and involves
the contribution of the 2d index of the theory on the boundary / wall that is a ``flavored'' generalization
of the elliptic genus. For theories of class $\CR$ that come from 3-manifolds, many such boundary conditions come from 4-manifolds as illustrated in~\eqref{4mfldbc}. In this case, the flavored elliptic genus of a boundary condition / domain wall is equal to the Vafa-Witten partition function of the corresponding 4-manifold \cite{GGP-4d}.

\subsection{Cutting and gluing along boundaries of $M_3$}

It is believed that a 3-manifold with boundary $C$ gives rise to a boundary condition in 4d $\CN=2$ theory of class $\CS$,
see Figure 2 in \cite{DGG} or Figure 6 in \cite{GGP-4d}.
This system can be understood as a result of 6d $(2,0)$ theory compactified on a 3-manifold with cylindrical end $\R_+ \times C$
and to some extent was studied previously.\footnote{See {\it e.g.} \cite{Yamazaki-3d,Yamazaki-layered,DG-Sdual,DGG,DGV-hybrid}
for a sample of earlier work; unfortunately the methods of these papers cannot be used to recover all flat connections for general 3-manifolds, even in the simplest cases of knot complements.}  %
For example, when $C = T^2$ is a 2-torus (with puncture) the corresponding 4d $\CN=2$ theory is actually $\CN=4$ super-Yang Mills
(resp. $\CN=2^*$ theory).

A simple class of 3-manifolds bounded by $C$ includes {\it handlebodies}, which for a genus-$g$ Riemann surface $C$ is determined by a choice
of $g$ pairwise disjoint simple closed curves on $C$ (that are contractible in the handlebody 3-manifold). For example, if $C = T^2$, then the corresponding handlebody is a solid torus:
\be
M_3 \; \cong \; S^1 \times D^2 \,.
\label{solidtorus}
\ee
It is labeled by a choice $(p,q)$ of the 1-cycle that becomes contractible in $M_3$. In the basic case of $(p,q)=(0,1)$ the Chern-Simons path integral on $M_3$ defines a state (in the Hilbert space $\CH_{T^2}$) that is usually denoted $\vert 0 \rangle$, so that we conclude
\be
\vert 0 \rangle \; = \; \vert \text{solid torus} \rangle
\label{soltorus}
\ee
It was proposed in \cite{GGP-4d} that the corresponding boundary condition in 4d theory $T[C]$
is Nahm pole boundary condition \cite{Diaconescu:1996rk,GW-boundary}
that can be described by a system of D3-branes ending on D5-branes\footnote{Whether we identify the state $\vert 0 \rangle$
with D5 or NS5 is a matter of conventions. Here we follow the conventions of \cite{GGP-4d,GGP-walls}.}
\be
\vert 0 \rangle \; = \; \vert \text{Nahm} \rangle \; = \; \vert \text{D5} \rangle
\label{D5bc}
\ee
More generally, for $M_3 \cong S^1 \times D^2$ obtained by filling in the cycle in homology class $(p,q)$
the corresponding boundary condition is defined by a system of D3-branes ending on IIB five-branes of type $(p,q)$.

This class of boundary conditions can be easily generalized to other Riemann surfaces $C$ and 3-manifolds with several
boundary components. The latter correspond to domain walls in 4d $\CN=2$ theories $T[C]$, see {\it e.g.} \cite{DGG,GGP-4d, DGV-hybrid} for details.
For example, each element $\phi$ of the mapping class group of $C$ corresponds, on the one hand, to a mapping cylinder $M_3$
(with two boundary components identified via $\phi$) and, on the other hand, to a duality wall of type $\phi$ in the 4d theory $T[C]$.
In the case $C = T^2$ we have the familiar walls that correspond to the generators $\phi = S$ and $\phi = T$
of the $SL(2,\Z)$ duality group of $\CN = 4$ super-Yang-Mills, and the general ``solid torus boundary condition'' described above
can be viewed as the IR limit of a concatenation of $S$- and $T$-walls with the basic Nahm pole boundary condition,
see \cite[pp.20-21]{GGP-4d} for details. For instance,
\be
S \vert 0 \rangle \; = \; \vert \text{Neumann} \rangle \; = \; \vert \text{NS5} \rangle
\label{NS5bc}
\ee

Clearly, there are still many details to work out, but we have outlined the key elements necessary
to glue 3-manifolds along a common boundary and, in particular,
to illustrate why \eqref{mspaces} must hold in a proper 3d $\CN=2$ theory $T [M_3]$.
Suppose $C = \pm \partial M_3^{\pm}$ is a common boundary component of 3-manifolds $M_3^+$ and $M_3^-$,
which in general may have other boundary components, besides $C$. As we reviewed earlier,
appropriately defined 3d $\CN=2$ theories $T[M_3^+]$ and $T[M_3^-]$ naturally couple to a 4d $\CN=2$ theory $T[C]$,
which becomes dynamical upon the gluing process
\be
M_3 \; = \; M_3^- \cup_{\phi} M_3^+
\label{MMMgluing}
\ee
Note, in the identification of the two boundaries here we included an element $\phi$ of the mapping class group of $C$
that corresponds to duality wall in $T[C]$. Hence, the resulting theory $T[M_3]$ consists of a $\phi$-duality wall in
4d $\CN=2$ theory $T[C]$ sandwiched between $T[M_3^+]$ and $T[M_3^-]$.
At the level of partition functions,
\be
Z_{T[M_3]} \; = \; Z_{CS} (M_3) \; = \; \langle M_3^- \vert \phi \vert M_3^+ \rangle
\label{ZZZgl}
\ee

A particularly simple and useful operation that involves (re)gluing solid tori {\it a la} \eqref{solidtorus}--\eqref{MMMgluing}
is called surgery. In fact, it is also the most general one in a sense that, according to a theorem of Lickorish and Wallace, every closed oriented 3-manifold can be represented by (integral) surgery along a link $K \subset S^3$. Since the operation is defined in the same way on any component of the link $L$ it suffices to explain it in the case when $K$ has only one component, {\it i.e.} when $K$ is a knot.
Then, for a pair of relatively prime integers $p,q \in \Z$, the result of $q/p$ Dehn surgery along $K$ is the 3-manifold:
\be
S^3_{q/p} (K) \; := \; (S^3 - N(K)) \cup_{\phi} (S^1 \times D^2)
\label{knotsurgery}
\ee
where $N(K)$ is the tubular neighborhood of the knot, and $S^1 \times D^2$ is attached to its boundary by a diffeomorphism $\phi : S^1 \times \partial D^2 \to \partial N(K)$ that takes the meridian $\mu$ of the knot to a curve in the homology class
\be
q [\mu] + p [\lambda]
\ee
The ratio $q/p \in \Q \cup \{ \infty \}$ is called the surgery coefficient.

In what follows we discuss various aspects of cutting, gluing, and surgery operations.
In particular, we shall see how the operations \eqref{ZZZgl} and \eqref{knotsurgery} manifest at various
levels in 3d $\CN=2$ theory $T[M_3]$ --- at the level of SUSY vacua, at the level of twisted superpotential,
and at the level of quantum partition functions --- thereby illustrating the important role of abelian flat connections.
Needless to say, there are many directions in which one could extend this analysis, {\it e.g.} to various classes of 3-manifolds
not considered in this paper, as well as more detailed analysis of the ones presented here, to higher rank groups $G$
and to relation with known properties of homological knot invariants.

\subsubsection{Compactification on $S^1$ and branes on the Hitchin moduli space}

A useful perspective on our 3d-4d system can be obtained by compactification on $S^1$ and studying the space of SUSY vacua.
Thus, a compactification of 4d $\CN=2$ theory $T[C]$ on a circle yields a 3d $\CN=4$ sigma-model whose target is the hyper-K\"ahler manifold
\be
\CM_{SUSY} (T[C],G) = \CM_H (G,C)
\label{MTCMH}
\ee
while a 3-manifold bounded by $C$ defines a half-BPS boundary condition, {\it i.e.} a brane in the sigma-model language.

More precisely, a 3-manifold $M_3$ with $C = \partial M_3$ gives rise to a brane of type $(A,B,A)$
with respect to the hyper-K\"ahler structure on $\CM_H (G,C)$.
It is supported on a mid-dimensional submanifold of $\CM_H (G,C)$
which can be identified with the moduli space of flat $G_{\C}$ connections on $M_3$:
\be
\CM_{\text{flat}} (M_3,G_{\C}) \; \subset \; \CM_H (G,C)
\label{M3MC}
\ee
Note, according to \eqref{mspaces}, the space of flat $G_{\C}$ connections on $M_3$ is precisely the space of SUSY vacua (parameters)
of the 3d $\CN=2$ theory $T[M_3]$ on a circle. When combined with \eqref{MTCMH} this gives
\be
\CM_{\text{SUSY}} (T[M_3],G) \subset \CM_{\text{SUSY}} (T[C],G)
\ee

In this description, the mapping class group of the Riemann surface $C$
(which we already identified with the duality group of $T[C]$)
acts by autoequivalences on branes in the sigma-model with the target space $\CM_H (G,C)$.
See \cite{Gukov:2007ck,DG-Sdual} for various examples of the mapping class group action on $(A,B,A)$ branes in the Hitchin moduli space.

In particular, when $G=SU(2)$ and $C = T^2$ is a 2-torus, the Hitchin moduli space is a flat hyper-K\"ahler space
$\CM_H (G,C) \cong (\C^* \times \C^*) / \Z_2$ parametrized by $\C^*$-valued holonomy eigenvalues $x$ and $y$ modulo the Weyl group action.
This is also the space of vacua of $T[C,G]$ after dimensional reduction on a circle.
Each 3-manifold with a toral boundary defines a middle-dimensional submanifold or an $(A,B,A)$ brane.
Thus, when translated to language of geometry, the boundary conditions \eqref{D5bc} and \eqref{NS5bc}
correspond to $(A,B,A)$ branes supported on $x=1$ and $y=1$, respectively:
\bea
\vert x=1 \rangle & = & \vert \text{D5} \rangle \label{xyABAlines} \\
\vert y=1 \rangle & = & \vert \text{NS5} \rangle \nonumber
\eea
Similarly, the duality wall of type $\phi = S$ is a ``correspondence''
$\CM_{\text{flat}} (M_3,G_{\C}) \subset \CM_H (G,C) \times \CM_H (G,C)$
associated with the mapping cylinder $M_3 \cong C \times I$,
\be
x + \frac{1}{x} \; = \; y' + \frac{1}{y'}
\qquad , \qquad
y + \frac{1}{y} \; = \; x' + \frac{1}{x'}
\ee
that exchanges the $SL(2,\C)$ holonomies on $a$- and $b$-cycles of $C = T^2$.
Note, these relations are deformed in
$\CM_{\text{SUSY}} (T[M_3],G) \subset \CM_{\text{SUSY}} (T[C],G) \times \CM_{\text{SUSY}} (T[C],G)$
for a generic value of the fugacity $t$.

\subsubsection{Lens space theories and matrix models}

In the above discussion we used the solid torus \eqref{solidtorus}--\eqref{soltorus} as a simple example of a handlebody,
in this case bounded by $C = T^2$. Likewise, the simplest example of a closed 3-manifold obtained by gluing two solid tori
is the Lens space
\be
L(p,1) \; = \; \langle 0 \vert ST^pS \vert 0 \rangle \; \cong \; S^3 / \Z_p
\label{LensSTS}
\ee
Using the dictionary \eqref{D5bc} and \eqref{NS5bc}, we can identify the corresponding 3d $\CN=2$ theory $T [L(p,1)]$
as the theory on D3-branes suspended between a NS5-brane and a $(p,1)$-fivebrane:
\be
T[L(p,1); G] \; = \; \text{ SUSY } G_{-p} \text{ Chern-Simons theory + adjoint chiral}
\label{TLensG}
\ee
Following \cite{GGP-4d}, here we assumed that the gauge group $G$ is of Cartan type A, {\it i.e.} $G=U(N)$ or $G=SU(N)$.
It would be interesting, however, to test the conjecture \eqref{TLensG} for other groups $G$.

Now, let us discuss this gluing more carefully, first from the viewpoint of flat connections (= SUSY vacua)
and then from the viewpoint of partition functions.
According to \eqref{NS5bc} and \eqref{xyABAlines}, the solid torus boundary condition $S \vert 0 \rangle$
in $\CN=4$ super-Yang-Mills $T[C]$ imposes a Neumann boundary condition on $x$ and a Dirichlet boundary condition on $y$.
In fact, the solid torus theory here is basically the theory of the unknot, $T[\mathbf{0_1}]$, discussed in section \ref{sec:theory}.
Its supersymmetric parameter space \eqref{ArefU} is a linear subspace of $\CM_{\text{SUSY}} (T[C],G)$ defined by $y=1$.
Note, the equation $y-1=0$ is precisely the defining equation of the abelian branch, which in our present example is
the entire moduli space $\CM_{\text{flat}} (M_3,G_{\C}) = \CM_{\text{SUSY}} (T[M_3],G)$.
Therefore, had we ignored this component, the space of SUSY vacua would be completely empty,
both for the solid torus theory $T[S^1 \times D^2]$ and for everything else that can be obtained from it by gluing!

A concatenation of the $T^p$ duality wall with this boundary condition adds a supersymmetric Chern-Simons term at level $p$
for the global $U(1)_x$ symmetry of the theory $T[\mathbf{0_1}]$. If we are only interested in SUSY vacua and parameters of
a theory $T[M_3]$ (= flat connections on $M_3$) we need to know how this operation affects the effective twisted superpotential.
For a general theory $T[M_3]$, this has a simple form:
\be
T^p : \quad \tilde W \to \tilde W + \frac{p}{2} \Tr (\log x)^2
\ee
where $\log x\in \mathfrak t_\C$ denotes the complexified mass parameters valued in the Cartan of the symmetry group $G$, and $x\in T_\C \subset G_\C$ their exponential.
For the case at hand, the result of this operation modifies the space of SUSY parameters from $y=1$ to $y=x^p$.
Finally, gluing $\langle 0 \vert S$ and $T^pS \vert 0 \rangle$ in \eqref{LensSTS} means sandwiching $\CN=4$ super-Yang-Mills
between the corresponding boundary conditions. In our IR description of the boundary conditions, this has two effects: it makes $U(1)_x$ dynamical, and it contributes a chiral multiplet in the adjoint representation of $G$ to the effective 3d theory $T[M_3]$. Altogether, the critical points of the effective twisted superpotential (\emph{cf.}\cite{GGP-4d}):
\be
\tilde \CW_{T[M_3]} \; = \; \tilde \CW_{T[M_3^-]} - \tilde \CW_{\phi \, \circ \, T[M_3^+]} + \tilde \CW_{\text{adj. chiral}}
\label{WWWgluing}
\ee
become SUSY vacua (= flat connections) of the theory $T[M_3]$ associated with the gluing~\eqref{MMMgluing}.

Let us analyze the critical points of \eqref{WWWgluing} in the case of a Lens space $L(p,1)$. There is an extra $U(1)_b$ flavor symmetry that rotates the adjoint chiral multiplet, whose associated mass we denote as $b$. The contribution to the superpotential from the adjoint chiral is
\be \tilde W_{\text{adj. chiral}}(x,b) = \sum_{\text{$\alpha\in$ roots($G$)}}\Big[ \Li_2(-b^{-1}x^{-\alpha}) - \frac14(\log b+\log x^\alpha)^2\Big]\,. \ee
For example, for $G=SU(2)$, we find that that $\tilde W_{T[M_3]} = -p(\log x)^2 + \Li_2(\frac{-x^2}{b}) +\Li_2(\frac{-1}{bx^2}) + \Li_2(\frac{-1}{b})+\tfrac34(\log b)^2+2(\log x)^2$. The critical points are intersections of $\{y=1\}$ and $\{y= x^p\, (x^2+b)/(bx^2+1)\}$, counted modulo the $\Z_2$ Weyl symmetry $x\to x^{-1}$. Moreover, critical points lying on the orbifold locus $x=\pm1$ should be excluded, as they do not correspond to massive vacua. Altogether, we find $[p/2]+1$ solutions, in 1--1 correspondence with the flat $SU(2)$ connections on $L(p,1)$.

More generally, for $G = U(N)$ the flat connections on $L(p,1)$ or, equivalently, the SUSY vacua of \eqref{TLensG}
are labeled by Young diagrams $\rho$ with at most $p-1$ rows and $N$ columns,
{\it i.e.} Young diagrams that fit in a rectangle of size $N \times (p-1)$.
Note, these are in one-to-one correspondence with
integrable representations of $\hat{\mathfrak{su}}(p)_N$ (equivalently, of $\hat{\mathfrak{u}}(N)_p$),
a fact that plays an important role \cite{Nakajima,Vafa:1994tf,dhsv,Dijkgraaf:2007fe}
in the study of Vafa-Witten partition function on ALE spaces bounded by $L(p,1)$.

Finally, we propose a ``lift'' of the gluing formula \eqref{WWWgluing} to a similar formula at the level
of partition functions, {\it cf.} \eqref{ZZZgl}:
\be
Z_{T[M_3]} \; = \; \int [dU(x)] \; Z_{T[M_3^-]} (x) \cdot Z_{\phi \, \circ \, T[M_3^+]} (x^{-1})
\label{ZZZgluing}
\ee
where the integration measure $[dU] = Z_{T[C]} dx$ is determined by the 4d $\CN=2$ theory $T[C;G]$
associated with the Riemann surface $C = \partial M_3^+ = - \partial M_3^-$, and accounts for the adjoint-chiral contribution to \eqref{WWWgluing}.
It would be interesting to test this gluing formula in concrete examples, including the Lens spaces
and Seifert manifolds discussed here.
Note that with the $t$-variable that keeps track of homological grading, \eqref{ZZZgluing} basically is a surgery
formula for homological knot invariants. Such formulas are indeed known in the context of knot Floer homology and
its version for general 3-manifolds, the Heegaard Floer homology.

As explained around \eqref{MMMgluing},
we can construct closed 3-manifolds by gluing open 3-manifolds along their boundaries.
The Chern-Simons partition functions on manifolds with torus boundary depend on a parameter $x$,
which should be integrated out upon gluing. For a particular class of 3-manifolds,
the resulting Chern-Simons partition functions can be represented as matrix integrals,
where the integration measure is responsible for integrating out the parameters $x$.
The integrands of such matrix models take the form
\be
\exp\Big(- \frac{1}{\hbar}V(x) \Big),  \label{Vmatrix}
\ee
where $V(x)$ is usually called potential and $\frac{2\pi i}{\hbar} = \frac{2 \pi i}{\log q}$ is called the ``level''.
Let us note that in the case of 3-manifolds with boundary, when the parameters $x$ are not integrated out,
the same representation of partition functions $Z_{CS} \sim \exp(\frac{1}{\hbar}\widetilde{W}+\ldots)$
was used to read off the twisted superpotentials of dual $\mathcal{N}=2$ theories, such as (\ref{Wt31}) or (\ref{Wt41}).
One is therefore tempted to postulate, that a matrix model potential $V(x)$ might encode information about
the twisted superpotential and field content of the dual $\mathcal{N}=2$ theory $T[M_3]$ associated to a closed 3-manifold $M_3$.
Let us demonstrate that this is indeed the case.

For non-abelian Chern-Simons theories it is convenient to a write matrix model representation of their partition functions in terms of eigenvalues $\sigma_i=\log x_i$. A very well known example is a matrix model representation of the $U(N)$ Chern-Simons partition function on $M_3 = S^3$ \cite{Marino0207,AKMV-matrix}, whose measure takes the form of a trigonometric deformation of the Vandermonde determinant, and the potential $V(\sigma)=\sigma^2/2$ is Gaussian in $\sigma=\log x$. More generally, the matrix model potential for $M_3 = L(p,1)$ and $G = U(N)$ takes the form $V(\sigma)=p \sigma^2/2$. More involved integral representations of Chern-Simons partition functions on other Lens spaces and Seifert homology spheres can be found in \cite{LawRoz,Marino0207,AKMV-matrix}. Various other matrix integral representations of Chern-Simons or related topological string partition functions, including the refined setting, were constructed in \cite{deHaro:2005rz,Klemm:2008yu,Eynard:2010dh,Ooguri:2010yk,Sulkowski:2010ux,AS-refinedCS,Szabo:2013vva,Kokenyesi:2013nxa}.

Let us now consider more seriously the proposal that the potential of a Chern-Simons matrix model determines the dual 3d $\mathcal{N}=2$ theory $T[M_3]$. For example, as reviewed above, the potential for a theory of the Lens space $L(p,1)$ takes the form $V(\sigma)=p \sigma^2/2$. Taking into account a minus sign in (\ref{Vmatrix}), and using by now familiar 3d-3d dictionary, we might conclude that the dual theory is $\mathcal{N}=2$ theory at level $-p$, at least in the abelian case. Due to the universal form of the matrix integral, we might also be tempted to declare that in the nonabelian case the dual theory is $U(N)$ theory at level $-p$. This is precisely the dual theory \eqref{TLensG} which was originally constructed by other means. We also emphasize that the form of the matrix model reflects the structure of the gluing \eqref{MMMgluing}, namely the fact that the resulting Lens space \eqref{LensSTS} is constructed from two solid tori (unknot complements), glued with a suitable $SL(2,\mathbb{Z})$ twist $\phi$. Indeed, in this case the potential factor (\ref{Vmatrix}) represents the gluing $SL(2,\mathbb{Z})$ element $\phi$, while the information about two solid tori is encoded in the matrix model measure. This construction is discussed in detail {\it e.g.} in \cite{AKMV-matrix}.

%***********************************

\subsubsection{Seifert manifolds and D4-D6 systems}

The matrix model potential suggests a dual 3d $\mathcal{N}=2$ theory $T[M_3;G]$ also for other Lens spaces and more general Seifert homology spheres.
In this section, we start with a brief review of the most general Seifert fibered 3-manifolds and then discuss how various ways to look at their geometry find application in 3d-3d correspondence.
For a nice exposition of Seifert manifolds see {\it e.g.} \cite{JankinsNeumann}.\footnote{And don't forget about the exercises!}

\begin{figure}[ht]
\centering
\includegraphics[width=5.0in]{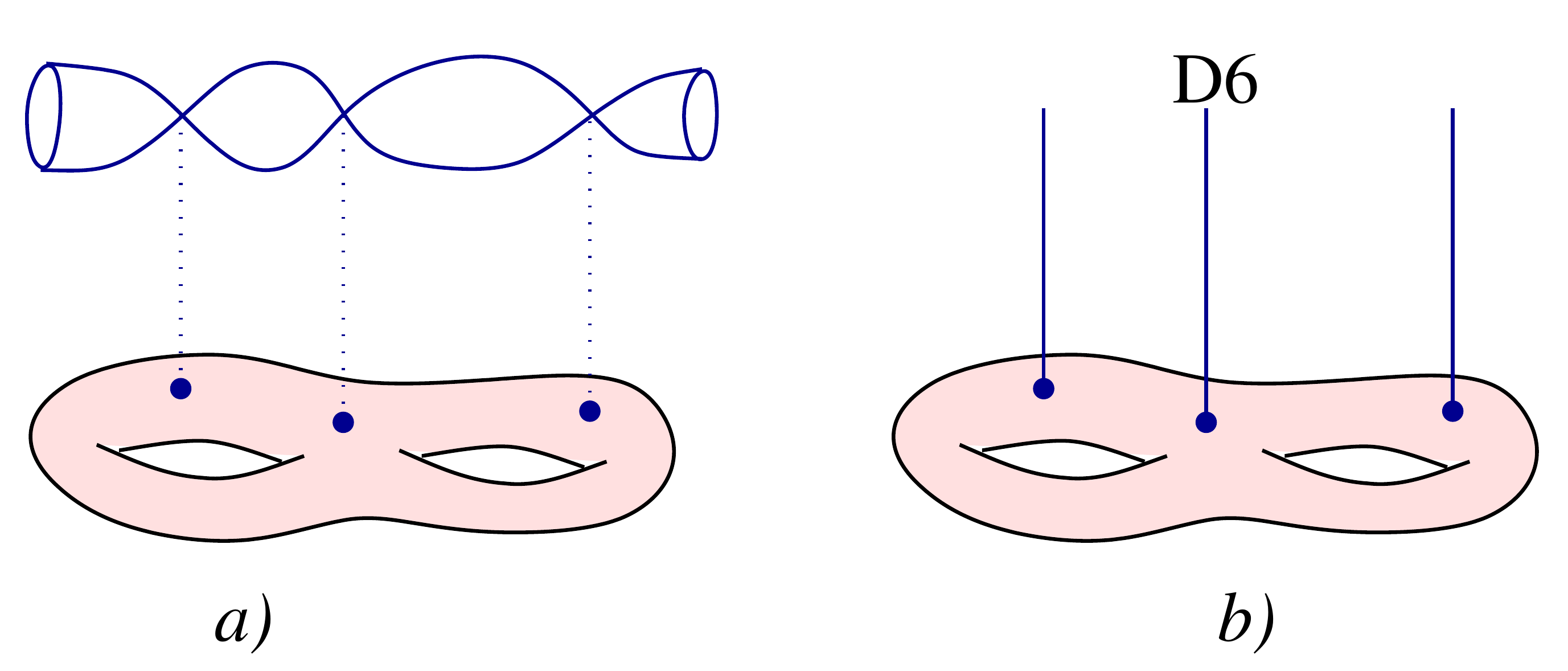}
\caption{$a)$ M-theory on a Seifert fibered 3-manifold $M_3$, and $b)$ its reduction to type IIA string theory with D6-branes.
Upon reduction on the $S^1$ fiber the fivebrane system \eqref{surfeng} turns into a system of D4-branes wrapped
on the (orbifold) surface $\Sigma$ intersecting D6-branes at finitely many points on $\Sigma$.}
\label{fig:SeifertC}
\end{figure}

Seifert manifolds were introduced 80 years ago and can be described in a number of equivalent ways:

\begin{itemize}

\item
A Seifert fibered manifold $M_3$ is a circle bundle over a 2-dimensional orbifold $\Sigma$.

\item
A 3-manifold $M_3$ is Seifert fibered if and only if it is finitely covered by an $S^1$-bundle over a surface.

\item
Finally, a Seifert manifold $M_3$ can be constructed by a sequence of surgeries on a trivial circle fibration
over a Riemann surface.

\end{itemize}

\noindent
Each closed Seifert fibration with $n$ exceptional fibers is classified by the following set of Seifert invariants
(also known as the symbol of the Seifert manifold):
\be
\{ b, (\epsilon,g) ; (p_1,q_1) , \ldots , (p_n , q_n) \}
\,, \qquad \text{gcd} (p_i,q_i) = 1
\label{Seifertinvts}
\ee
where $\epsilon$ tells us whether $M_3$ and $\Sigma$ are orientable. Since both will be assumed to be orientable, $\epsilon$ will not play an important role in our discussion. The integer-valued invariant $b$ is (minus) the Euler number of the $S^1$-bundle, while the non-negative integer $g$ is the genus of the underlying base orbifold $\Sigma$, whose orbifold Euler characteristic is
\be
\chi (\Sigma) = \chi (\Sigma_g) - \sum_i \left( 1 - \frac{1}{p_i} \right)
\ee
The pair $(p_j,q_j)$ of relatively prime integers are the Seifert invariants of the $j$-th exceptional fiber,
locally modeled on the $\Z_{p_j}$ orbifold:
\be
(z, \theta) \; \mapsto \; \left( e^{\frac{2\pi i}{p_j}} z ,\, \theta + \frac{2 \pi i q_j}{p_j} \right)
\ee

For $n=0,1,2$ and $g=0$, the Seifert fibration produces a Lens space $L(p,q)$ with
\be \label{Seifertg0}
(p,q) \; = \;
\begin{cases}
(b,1) \,, & \text{if } n=0 \\
(b p_1 + q_1, p_1) \,, & \text{if } n=1 \\
(b p_1 p_2 + p_1 q_2 + p_2 q_1, c p_2 + d q_2) \,, & \text{if } n=2
\end{cases}
\ee
where $c p_1 - d (b p_1 + q_1) = 1$.
For $n=3$ it gives a Brieskorn sphere $\Sigma (p_1,p_2,p_3)$ for any choice of $q_1$, $q_2$, $q_3$.

One can add integers to each of the rational numbers $b$, $\frac{q_1}{p_1}$, $\ldots$, $\frac{q_n}{p_n}$
provided that their sum remains constant. In other words,
\be
b + \sum_i \frac{q_i}{p_i}
\ee
is an invariant of oriented fibrations. Usually, the symmetries
\be
b \to b+1 \,, \qquad q_i \to q_i - p_i \qquad (\text{fixed } i)
\label{bashiftsym}
\ee
are used to achieve $1 \le q_i < p_i$ for all $i = 1, \ldots, n$.
Another popular choice of (partial) ``gauge fixing'' is to use the symmetry \eqref{bashiftsym} to set $b=0$.
This choice gives the so-called {\it non-normalized} Seifert invariants and clearly is very non-unique~\cite{JankinsNeumann}.

The description of a Seifert manifold $M_3$ as an $S^1$-bundle over a 2-dimensional orbifold $\Sigma$
is very helpful in understanding the fivebrane system \eqref{surfeng} and, therefore, the corresponding 3d $\CN=2$ theory \eqref{3d3dbasic}.
Indeed, by interpreting the circle fiber of $M_3$ as the ``M-theory circle'' we can equivalently describe \eqref{surfeng}
in type IIA string theory.
Upon this reduction, the fivebranes supported on $\R^3 \times M_3$ become D4-branes with world-volume $\R^3 \times \Sigma$.
In addition, singular fibers of the $S^1$ fibration in general give rise to D6-branes supported at (orbifold) points on $\Sigma$
and intersecting D4-branes along the $\R^3$ part of their world-volume, as illustrated in Figure~\ref{fig:SeifertC}$(b)$.
Since D4-branes carry maximally supersymmetric Yang-Mills theory,
in this approach 3d $\CN=2$ theory $T[M_3;G]$ is the result of the reduction of 5d super-Yang-Mills on $\Sigma$.
Among other things, it has 3d $\CN=2$ Chern-Simons coupling at level $b$ induced by $b$ units of Ramond-Ramond 2-form flux
and additional matter multiplets that come from D4-D6 string states.
A detailed derivation of $T[M_3;G]$ from this D4-D6 systems will appear elsewhere~\cite{GPSeifert}.

\begin{figure}[htb]
\centering
\includegraphics[width=0.43\textwidth]{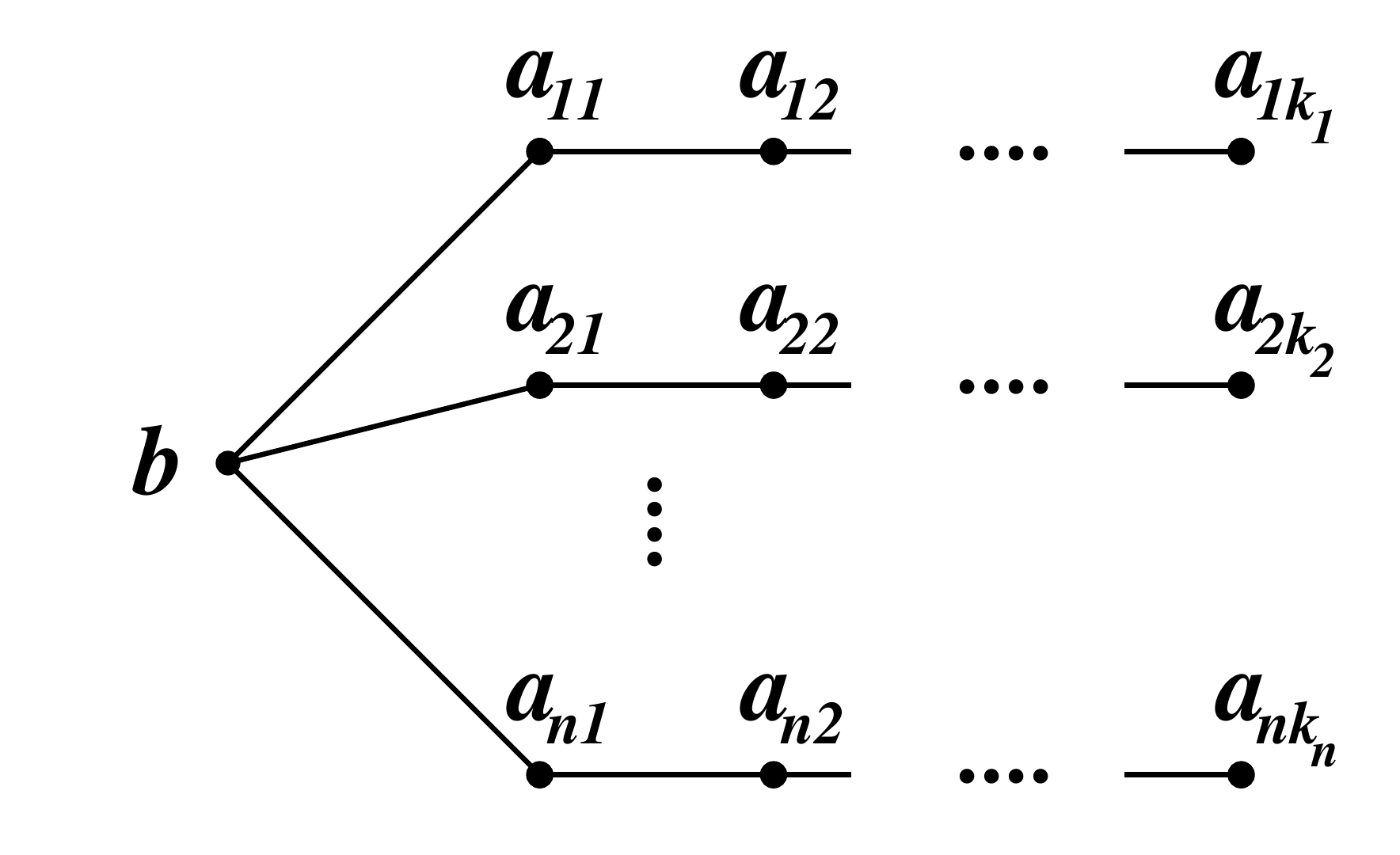}
\caption{Plumbing graph of a Seifert fibered homology 3-sphere with $n$ exceptional fibers.}
\label{fig:Splumb}
\end{figure}

Equivalently, a Seifert manifold $M_3$ can be produced by a sequence of Dehn surgery operations along the fibers of the trivial $S^1$ bundle over $\Sigma_g$.
Indeed, since the tubular neighborhood of every fiber is bounded by a 2-torus, each surgery operation is specified by the image of the meridian circle or, more precisely, by its homology class
\be
q [\mu] + p [\lambda] \in H_1 (T^2,\Z)
\ee
where $p \in \Z_+$ and $q \in \Z$ are coprime integers. The integral surgery (with $p=1$) is special and can be represented by a four-dimensional cobordism of attaching a 2-handle. It does not introduce a singular fiber and merely changes the degree of the $S^1$ bundle by $q$.

Hence, a Seifert manifold with the symbol \eqref{Seifertinvts} can be constructed by a sequence
of $n+1$ surgeries on $S^1 \times \Sigma_g$ with the surgery coefficients $b$, $\frac{q_1}{p_1}$, $\ldots$, $\frac{q_n}{p_n}$.
In the surgery presentation, the symmetries \eqref{bashiftsym} correspond to basic Kirby moves~\cite{GGP-4d}
represented by dualities of the 3d $\CN=2$ theory $T[M_3]$.
Surgery presentation is especially useful for applications to Chern-Simons theory, quantum group invariants, and their categorification.

For future use, let us note that the fundamental group of $M_3$ fits into the following exact sequence
\be
\pi_1(S^1) \to \pi_1 (M_3) \to \pi_1 (\Sigma) \to 1
\ee
and
\be
H^2 (M_3, \Z) \; \cong \; \Z^{2g} \oplus \text{Pic} (\Sigma) / \Z [\CL]
\ee
where $\CL$ is a line bundle over $\Sigma$ associated with the circle bundle $M_3 = S (\CL)$.
In particular, if we want to work with a homology sphere, we need to take $g=0$.
Then, the resulting space is a plumbing 3-manifold given by the graph in Figure~\ref{fig:Splumb}
and its (co)homology can be computed using the algorithm described in~\cite[sec.2.2]{GGP-4d}.
One of the results of this calculation is that
\be
|H_1(M_3,\Z)| \; = \; \left( b + \sum_{i=1}^n \frac{q_i}{p_i} \right) \cdot \prod_{i=1}^n p_i
\ee
A Seifert homology sphere $M_3$ can be
constructed by a surgery on a link in $S^3$ with $n+1$ components, which consists of $n$ parallel mutually unlinked unknots,
all linked (with linking number one) to one additional copy of the unknot.
The surgery coefficients for $n$ parallel unknots are $\frac{q_i}{p_i}$, $i=1,\ldots n$.

An integral representation for $U(N)$ Chern-Simons partition function on a Seifert homology sphere $M_3$
was found in \cite{Marino0207} and it takes the form
\be
Z_{CS} = \int  D\sigma \, e^{-\sum_k  \frac{\sigma_k^2}{2\hat{\hbar}} - \sum_k l t_k\sigma_k  } ,\qquad
D\sigma = \Big(\prod_{k=1}^N d\sigma_k \Big)
\frac{\prod_{i=1}^n \prod_{k<l}  2\sinh\frac{\sigma_k-\sigma_l}{2p_i}  }{ \prod_{k<l} \big( 2\sinh\frac{\sigma_k-\sigma_l}{2}  \big)^{n-2}} .    \label{seifert-matrix}
\ee
For the Lens space case $n=1,2$, with $p_i=1$, the integration measure $D\sigma$ in this expression reduces to the standard (trigonometric) Vandermonde determinant $\prod_{k<l}(2\sinh\frac{\sigma_k-\sigma_l}{2})^2$, which has straightforward interpretation as a unitary matrix integral; for other cases we get more general integral representation, with modified measure. More precisely, the above integral represents contribution from some particular flat connection, whose choice is specified by the choice of $t_i$ and the linear term $\sum_k l t_k\sigma_k$ in the potential.
Such linear terms in the potential and, therefore, the choice of the flat connection $\rho$
corresponds to the choice of the FI term in the partition function of the dual 3d $\CN=2$ theory $T[M_3]$.
The full Chern-Simons partition function is given by a sum of such contributions, taking into account all flat connections.
Finally, it is important to remember that the coefficient of the Gaussian term in the potential is rescaled and takes form
\be
\hat{\hbar} = \Big(  \sum_{j=1}^n \frac{q_j}{p_j} \Big)^{-1} \hbar.    \label{seifert-hbar}
\ee

Again, by applying the standard 3d-3d dictionary, at least for $G=U(1)$, one might conclude that the theory dual to a Seifert homology sphere is 3d $\mathcal{N}=2$ Chern-Simons theory at a fractional level. This is again consistent with the predictions of~\cite{GGP-4d}. Moreover, it is well known that Chern-Simons theory at fractional level is equivalent to a quiver Chern-Simons theory with integer levels.

To be more specific, focusing on the Lens space $M_3 = L(p,q)$ and $G=U(1)$ let us demonstrate how this data determines the dual quiver theory $T[M_3;G]$ and show the equivalence of this quiver theory to a 3d $\CN=2$ abelian Chern-Simons theory at a fractional level. According to \cite{GGP-4d}, at least in the abelian case, the theory dual to $L(p,q)$ Lens space is a $U(1)^k$ quiver Chern-Simons theory with interactions between various $U(1)$ gauge fields specified by a quadratic matrix
\be
Q_{ij} = \left[ \begin{array}{ccccc}
a_1 & 1 & 0 & 0 & \cdots\\
1 & a_2 & 1 & 0 & \cdots\\
0 & 1 & a_3 & 1 & \cdots\\
0 & 0 & 1 & a_4 & \cdots \\
&& \vdots && \ddots \end{array} \right]    \label{seifert-Qij}
\ee
where $a_1,\ldots,a_k$ arise from the continuous fraction expansion of $p/q$:
\be
\frac{p}{q} = a_1 - \frac{1}{a_2-\frac{1}{a_3-\frac{1}{a_4-\ldots}}}.
\ee
Schematically, denoting $U(1)$ Chern-Simons gauge fields by $u_i$, the twisted superpotential of this quiver theory takes the form
\be
\widetilde{W} = -\frac{1}{2} \sum_{i,j=1}^k Q_{ij} u_i u_j =  -\frac{1}{2}\Big(\big( \sum_i a_i u_i^2\big)   +  (u_1u_2 + u_{k-1}u_k) + \sum_{i=2}^{k-1} u_i(u_{i-1}+u_{i+1})  \Big).  \label{WQij}
\ee
We can now integrate out $u_2,\ldots,u_k$ fields using their equations of motion
\bea
a_2 u_2 + (u_1+u_3) & = & 0 \nonumber\\
a_3 u_3 + (u_2+u_4) & = & 0 \nonumber\\
&\vdots& \nonumber\\
a_{k-1} u_{k-1} + (u_{k-2}+u_k) & = & 0 \nonumber\\
a_k u_k + u_{k-1} & = & 0 \nonumber
\eea
 Solving these equations, starting from the last one and proceeding to the first one, we find
\be
u_{k} = -\frac{u_{k-1}}{a_k}, \qquad u_{k-1} = -\frac{u_{k-2}}{a_{k-1}-\frac{1}{a_k}},\quad \dots \quad,  \qquad u_2 = -\frac{u_1}{a_2 - \frac{1}{a_3-\frac{1}{a_4-\ldots}}}.
\ee
We can also use the equations of motion to get rid of all $u_i(u_{i-1}+u_{i+1})$ and $u_k u_{k-1}$ terms in (\ref{WQij}). Finally, substituting the above result for $u_2$, the twisted superpotential takes the form
\be
\widetilde{W} = -\frac{1}{2}\big(  a_1 u_1^2 + u_1 u_2 \big) = -\frac{u_1^2}{2} \Big( a_1 -  \frac{1}{a_2 - \frac{1}{a_3-\frac{1}{a_4-\ldots}}}  \Big) = -\frac{p}{q}\frac{u_1^2}{2},
\ee
which indeed represents the abelian 3d $\CN=2$ super-Chern-Simons theory at fractional level $-p/q$. This $p/q$ factor precisely agrees with the rescaling (\ref{seifert-hbar}) of the Gaussian potential in (\ref{seifert-matrix}) for the Lens space $X(q/p)=L(p,q)$. As a special case, let us also note that for $a_i=2$ we find $u_i=\frac{k+1-i}{k}u_1$ and $p/q=(k+1)/k$, which corresponds to $L(k+1,k)$ Lens space.

Therefore, by 3d-3d dictionary, the potential rescaled by $p/q$ in (\ref{seifert-matrix}) suggests that the dual 3d theory is $\mathcal{N}=2$ Chern-Simons at level $-p/q$, or equivalently the quiver Chern-Simons theory determined by the interaction matrix (\ref{seifert-Qij}). Since matrix integrals result from ``abelianization'' of non-abelian theories, it is tempting to speculate that similar correspondence holds for non-abelian $G$ as well.

Using this dictionary one could also consider other matrix models representation of Chern-Simons partition functions \cite{LawRoz,Marino0207,AKMV-matrix,deHaro:2005rz,Klemm:2008yu,Eynard:2010dh,Ooguri:2010yk,Sulkowski:2010ux,AS-refinedCS,Szabo:2013vva,Kokenyesi:2013nxa}, either for various interesting manifolds or in the refined setting, and predict the form of dual 3d $\mathcal{N}=2$ theories $T[M_3;G]$. Interestingly, the models derived in {\it loc. cit.} have potentials that consist of quadratic and dilogarithmic terms, which indeed are the basic ingredients in modeling the content of dual 3d $\CN=2$ theories. Also, in some cases inequivalent matrix model representations of the same Chern-Simons partition function are known and, hence, might lead to interesting new dualities of 3d $\mathcal{N}=2$ theories.

\subsubsection{Dehn surgery}
\label{sec:Dehn}

As a final simple illustration of the necessity of accounting for all flat connections, we return to the basic Dehn surgery operation \eqref{knotsurgery}. Suppose that the knot $K=\mathbf {3_1}$ is the trefoil. As we know well from Section \ref{sec:theory}, the A-polynomial%
\footnote{In contrast to the rest of the paper, we take care in this section to write A-polynomials in terms of actual $SL(2,\C)$ meridian and longitude eigenvalues rather than their squares. Thus, for the trefoil, the non-abelian A-polynomial is written as $y+x^6$ rather than $y+x^3$. The distinction is important for consistently counting $SL(2,\C)$ (as opposed to $PSL(2,\C)$, etc.) flat connections resulting from surgery.} %
for the trefoil, parametrizing $\CM_{\rm SUSY}(T[\mathbf{3_1}],SU(2))$ for the full trefoil-complement theory $T[\mathbf{3_1},SU(2)]$, is
\be A(x,y) = (y-1)(y+x^6) \;\subset (\C^*\times \C^*)/\Z_2\,. \ee
Here $x$ and $y$ are the $\C^*$-valued eigenvalues of longitude and meridian $SL(2,\C)$ holonomies on the torus boundary of the knot complement, well defined up to the Weyl-group action $(x,y)\mapsto (x^{-1},y^{-1})$.
We recall that the $(y-1)$ component of the A-polynomial corresponds to an abelian flat connection on the knot complement, while the $(y+x^6)$ component corresponds to an irreducible flat connection.

Suppose that we perform Dehn surgery with $q/p=\pm 1$ on the trefoil knot complement. The result is a closed 3-manifold --- in fact one of the Brieskorn spheres of \eqref{Seifertg0}
\be S^3_{p/q}(\mathbf{3_1}) = \begin{cases} \Sigma[2,3,5] & p/q=+1 \\
 \Sigma[2,3,7] & p/q=-1\end{cases}
\ee
In each case, the moduli space of flat $SL(2,\C)$ connections on $S^3_{p/q}(\mathbf{3_1})$, consists of isolated points. It is easy to count them directly from a presentation of the fundamental groups of the Brieskorn spheres,
\be \pi_1\Sigma[2,3,5] = \langle a,b\,|\, a^3 = b^5 = (ab)^2 \rangle\,,\qquad
 \pi_1\Sigma[2,3,7] = \langle a,b\,|\, a^3 = b^7 = (ab)^2 \rangle\,.\ee
We find $|\CM_{\rm flat}(S^3_{+1}(\mathbf{3_1}),SL(2,\C))| = 3$ and $|\CM_{\rm flat}(S^3_{-1}(\mathbf{3_1}),SL(2,\C))| = 4$. These counts must equal the numbers of isolated vacua of the theories $T[S^3_{\pm 1}(\mathbf{3_1}),SU(2)]$ on $\R^2\times S^1$.

Now compare the count of flat connections on the Brieskorn spheres with the intersection points of the varieties
\be \label{countA}
 (x^py^q = 1) \,\cap\, (A(x,y)=0) \;= \begin{cases} \text{4 points} & p/q=1 \\ \text{5 points} & p/q=-1\,.
\end{cases}\ee
This does not quite match the count of flat connections on the Brieskorn spheres: in each case, there is one extra intersection point in \eqref{countA}. In particular, in each case, the intersection point $(x,y)=(-1,-1)$ corresponds to flat connections on the knot complement $S^3\backslash \mathbf{3_1}$ and the solid surgery torus whose \emph{eigenvalues} match at the $T^2$ surgery interface, but whose full \emph{holonomies} do not. Namely, the flat connection on the solid surgery torus with eigenvalues $(-1,-1)$ is trivial, while the flat connection on the trefoil knot complement with eigenvalues $(-1,-1)$ is parabolic, meaning the full holonomy matrix is $\left(\begin{smallmatrix} -1 & 1 \\ 0 & -1 \end{smallmatrix}\right)$. This is not unexpected, since $(x,y)=(-1,-1)$ lies on the nonabelian branch $y+x^6=0$ of the trefoil's A-polynomial. After subtracting the ``false'' intersection point from the counts in \eqref{countA}, we recover the expected number of flat connections on $S^3_{+1}(\mathbf{3_1})$ and $S^3_{-1}(\mathbf{3_1})$.

Physically, \eqref{countA} is the (naively) expected count of vacua when gluing the trefoil theory to an unknot theory with the appropriate element $\phi\in SL(2,\Z)$ corresponding to the Dehn surgery. The presence of a ``false'' intersection point $(x,y)=(-1,-1)$ suggests that the corresponding vacuum in the glued theory must be lifted. It would be interesting to uncover the mechanism behind this. The remaining vacua match the count of flat connections on the Brieskorn spheres (\emph{i.e.} vacua of $T[S^3_{\pm 1}(\mathbf{3_1}),SU(2)]$), as they should. Crucially the vacuum corresponding to the intersection point $(x,y)=(1,1)$ must be included in order for the count to work out; this intersection point sits on the abelian branch $(y-1)$ of the trefoil A-polynomial, and labels the trivial flat connection on $S^3_{\pm 1}$.

A similar phenomenon occurs when considering simple surgeries on the figure-eight knot complement $S^3\backslash \mathbf{4_1}$. For example, the Brieskorn sphere $\Sigma[2,3,7]$ may be constructed from $+1$ or $-1$ surgeries on $S^3\backslash \mathbf{4_1}$. (The two different surgeries produce opposite orientations on $\Sigma[2,3,7]$.) The intersection of the full figure-eight A-polynomial $A(x,y)=(y-1)\big(x^4-(1-x^2-2x^4-x^6+x^8)y + x^4 y^2\big)$ with the surgery conditions $xy^{\pm 1}=1$ yield
\be \label{countA41}  (xy^{\pm 1}=1)\,\cap\,(A(x,y)=0) \;= \text{5 points}\,. \ee
Four of these five intersection points, including the point on the abelian branch $y-1=0$, correspond to the expected flat $SL(2,\C)$ connections on $\Sigma[2,3,7]$. The fifth intersection point, at $(x,y)=(-1,-1)$, does not correspond to any flat connection on  $\Sigma[2,3,7]$, because the connection with eigenvalues $(x,y)=(-1,-1)$ on the knot complement is parabolic, while on the solid surgery torus it would have to be trivial. Explicitly, the meridian and longitude holonomies of the connections on $S^3\backslash \mathbf{4_1}$ with $(x,y)=(-1,-1)$ are conjugate to $\mu = \left(\begin{smallmatrix} -1 & 1 \\ 0 & -1\end{smallmatrix}\right)$, $\lambda=\left(\begin{smallmatrix} -1 & \pm2i\sqrt{3} \\ 0 & -1\end{smallmatrix}\right)$, which will never satisfy $\mu^p\lambda^q=I$ for any $p,q$.

%***********************************

\subsection{Boundary conditions in 3d $\CN=2$ theories}

So far we discussed what happens when 3-manifolds have boundaries, along which they can be glued, {\it cf.} \eqref{MMMgluing}.
Now let us briefly discuss what happens when the space-time of 3d $\CN=2$ theory $T[M_3;G]$ has a boundary.

\begin{table}[htb]\begin{center}
\begin{tabular}{|c|c|}
\hline
\rule{0pt}{6mm}
{\bf (0,2) multiplet} & {\bf contribution to half-index} \\[5pt]
\hline
\hline
\rule{0pt}{6mm}
chiral & $\theta(-q^{\frac{R-1}{2}} x;q)^{-1}$ \\[5pt]
\hline
\rule{0pt}{6mm}
Fermi & $\theta(- q^{\frac{R}{2}} x;q)$ \\[5pt]
\hline
\rule{0pt}{6mm}
$U(N)$ gauge & $(q;q)_{\infty}^{2N}\prod_{i\neq j}\theta ( - q^{-\frac{1}{2}}\sigma_i/\sigma_j;q)$ \\[5pt]
\hline
\end{tabular}\end{center}
\caption{Building blocks of 2d boundary theories and their contributions to the half-index.}
\label{tablica}
\end{table}

Much like in Chern-Simons theory on $M_3$ the presence of non-trivial boundary requires specifying boundary conditions,
the same is true in the case of 3d $\CN=2$ theories.
One important novelty, though, is that some boundary conditions are now distinguished if they preserve part of supersymmetry,
such as half-BPS boundary conditions that preserve $\CN=(0,2)$ supersymmetry on the boundary.
These ``B-type'' boundary conditions have been studied only recently in \cite{GGP-walls} and then in \cite{Okazaki:2013kaa}.

In the presence of a boundary (or, more generally, a domain wall) one can define a generalization of the index
as a partition function on $S^1 \times_q D$ with a prescribed B-type boundary condition on the boundary torus
$S^1 \times_q S^1 \cong T^2$ of modulus $\tau$, as illustrated in Figure~\ref{fig:wallindex}.
The resulting half-index $\CI_{S^1 \times_q D}$ is essentially a convolution of the flavored elliptic genus
of the 2d $\CN=(0,2)$ boundary theory with the index of a 3d $\CN=2$ theory on $S^1 \times_q D$.
The contribution of $(0,2)$ boundary degrees of freedom is summarized in Table~\ref{tablica} where,
as usual, gauge symmetries result in integrals over the corresponding variables $\sigma_i$.

The half-index $\CI_{S^1 \times_q D}$ labeled by a particular choice of the boundary condition
can be viewed as a UV counterpart of a vortex partition function labeled by a choice of the massive vacuum in the IR.
Moreover, since the half-index is invariant under the RG flow, it makes sense to identify some of
massive vacua and integration contours in the IR theory with specific boundary conditions in the UV.
The latter, in turn, can sometimes be identified with 4-manifolds via \eqref{4mfldbc},
which altogether leads to an interesting correspondence between certain contour integrals discussed here and 4-manifolds.

Note, that for theories $T[M_3;G]$ labeled by closed 3-manifolds, supersymmetric vacua $\rho \in \CM_{\text{SUSY}} (T[M_3;G])$
specify boundary conditions for the Vafa-Witten topological gauge theory on a 4-manifold bounded by $M_3$.
Therefore, had we missed any of the vacua in constructing $T[M_3;G]$ there would be no hope to relate
supersymmetric boundary and 4-manifolds in \eqref{4mfldbc}.

For instance, let us consider one of the simplest 3d $\CN=2$ theories, namely
the super-Chern-Simons theory with gauge group $G=U(N)$ that in \eqref{TLensG} we identified with the Lens space theory.
As we mentioned earlier, the contour integrals for this theory are not known.
However, their UV counterparts $\CI_{S^1 \times_q D}$ are easy to write down
by choosing various B-type boundary conditions constructed in \cite{GGP-walls,GGP-4d,Okazaki:2013kaa}.
Thus, a simple boundary condition involves $pN$ Fermi multiplets on the boundary.
According to the rules in Table~\ref{tablica}, its flavored elliptic genus can be interpreted as
the half-index of 3d $\CN=2$ super-Chern-Simons theory \eqref{TLensG} with gauge group $G = U(N)$:
\be
\CI_{S^1 \times_q D} \; = \; q^{-\frac{pN}{24}} \prod_{i=1}^{p} \prod_{j=1}^{N} \theta(x_{i}z_{j};q)
\ee
Moreover, it can be identified with the Vafa-Witten partition function of the ALE space
\be
A_{p-1}=M_4(\mathfrak{su}(p))=M_4(\underbrace{\raisebox{-0.7ex}{$\overset{-2}{\bullet}\hspace{-0.7em}-\cdots -\hspace{-0.7em}\overset{-2}{\bullet}$}}_{p-1})
\label{ALEplumb}
\ee
written in the ``continuous basis''
\be
Z^{U(N)}_{\text{VW}} [A_{p-1}] (q,x|z) \; := \; \sum_{\rho} \, \chi^{\hat{\mathfrak{u}}(N)_{p}}_{\rho^t}(q,z) \, Z^{U(N)}_{\text{VW}} [A_{p-1}]_{\rho} (q,x)
\ee
where
\be
Z_{\text{VW}}^{U(N)}[A_{p-1}]_\rho (q,x) \; = \; \chi_\rho^{\hat{\mathfrak{su}}(p)_N}(q,x)
\ee
is the well known form of the Vafa-Witten partition function on the ALE space \eqref{ALEplumb}
written in the ``discrete basis'' \cite{Nakajima,Vafa:1994tf,dhsv,Dijkgraaf:2007fe}.
Here, $\rho$ is a Young diagram with at most $p-1$ rows and $N$ columns that in the previous section
we identified with the choice of flat connection on $M_3 = \partial M_4 = L(p,1)$.

%***********************************

\section{Conclusions and open questions}

In this work we have studied 3d-3d correspondence, which relates (fivebranes compactified on) non-trivial 3-manifolds to the 3d $\CN=2$ theories. To much extent we have focused on examples of theories whose partition functions can be identified with homological knot invariants. We discussed their relation to the theories considered previously by Dimofte-Gaiotto-Gukov, stressed the importance of taking into account all flat connections in the construction of the $\CN=2$ theories, and discussed the role of boundary conditions on both sides of the correspondence.

While the approach in the main part of the paper combines the strong points of \cite{DGH} and \cite{DGG,FGS-superA}, there is, however, something deeply puzzling between these two lines of development.
They both morally describe 3d $\CN=2$ theory associated either to a knot $K$ or a 3-manifold $M_3$,
but realize the quantum / categorified invariants of $K$ very differently.
Indeed, the approach of \cite{DGH} leads to $P^r_K (q,t, \ldots)$ as a vortex partition function on $S^1 \times \R^2$
in a sector with vortex number $r$,
\be
Z_{\text{vortex}} (T_{DGH})
= \sum_r z^r P^r_K (q,t, \ldots).  \label{Zvortex}
\ee
On the other hand, in the other approach (and most of the above discussion in this paper) we have
\be
P^r_K (q,t, \ldots) = B^K_* (T[M_3], x, \ldots) \vert_{x = q^r}. \label{Zvortex2}
\ee
It is therefore important to understand the relation between these formulas and between the corresponding theories $T_{DGH}$ and $T[M_3]$. At this stage we can suggest some possible explanations, which however require further studies.

First, it is very suggestive to compare two partition functions, (\ref{Zvortex}) and (\ref{Zvortex2}), to the four-dimensional Nekrasov partition function and its dual partition function discussed in \cite{Nekrasov:2003rj}. The Nekrasov partition function and its dual are indeed related by the Fourier transform, analogous to the one that relates (\ref{Zvortex}) and (\ref{Zvortex2}). In this relation the (original) Nekrasov partition is evaluated on the discrete set of parameters determined by the summation parameters, similar to the form of $P^r_K (q,t, \ldots)$ in (\ref{Zvortex2}). Moreover, in four-dimensional case one can introduce the dual prepotential, which morally describes the same $\mathcal{N}=2$ theory, and is related by the Legendre transform to the (original) Seiberg-Witten prepotential. Similarly, one can associate two twisted superpotentials to both sides of (\ref{Zvortex}), which are then related by the Legendre transform. Nonetheless, one should be cautious in following this analogy; in particular, the dual partition function was introduced in the non-refined limit $\epsilon_1=-\epsilon_2$, and it does not automatically extend to other cases.

As another possibility, one might try to interpret the relation (\ref{Zvortex}) as gauging of the global $U(1)_x$ symmetry in a theory with the (half-)index $B^K_* (x)$:
\be T_{DGH} \;\overset{?}{=}\; \text{$T[M_3]$ with $U(1)_x$ gauged}\,, \label{DGH-TM} \ee
while identifying the $U(1)_z$ flavor symmetry of $T_{DGH}$ as a topological symmetry for $U(1)_x$. The coefficient of $z^r$ in \eqref{Zvortex} is the $\R^2\!\times_q \!S^1$ half-index of $T_{DGH}$ in an $r$-vortex sector. In turn, via the logic of Section \ref{sec:ind}, this half-index could be identified with the \emph{residue} of the index of $T[M_3]$ at $x=q^r$  --- or, equivalently, the specialization of a contour integral of $T[M_3]$ to $x=q^r$. This provides a possible conceptual explanation of \eqref{Zvortex}--\eqref{Zvortex2}, whose details must be worked out with some care.

This brings us back to the main and final question that remains unanswered: Is there a systematic construction of the 3d $\CN=2$ theory $T[M_3]$ that accounts for all classical solutions in $G_{\C}$ Chern-Simons theory on $M_3$? Such construction might come from the triangulation data of $M_3$, extending the work \cite{DGG,DGG-Kdec}, or via representing $M_3$ in some other way.

\appendix

\section{Flat connections in $T[M_3]$}
\label{sec:bdy}

We review here some recent developments that all point to the existence of a complete theory $T[M_3]$ whose vacua on a circle correspond to all flat connections on $M_3$. These developments all have to do, in one way or another, with the study of the basic fivebrane system \eqref{surfeng} and its generalization to knot complements, which we introduce momentarily.

The different ``classes'' of flat connections that we alluded to in the introduction are labelled by their reducibility. The stabilizer $\text{Stab}(\CA)$ of a flat $G_\C$ connection $\CA$ on a 3-manifold $M_3$ is defined to be the group of gauge transformations that leave $\CA$ invariant. These gauge transformations must be constant, and so form a subgroup $\text{Stab}(\CA) \subset G_\C$. Equivalently, noting that a flat connection is uniquely characterized by its holonomy representation $\rho_\CA:\pi_1(M_3)\to G_\C$, one may define $\text{Stab}(\CA)$ as the subgroup of $G_\C$ that preserves the image of $\rho_\CA$.
A given flat connection is called irreducible if the stabilizer is trivial, and reducible otherwise.
On a hyperbolic 3-manifold, there is always at least one irreducible flat $G_\C$ connection for any simple $G_\C$, coming from embedding the hyperbolic holonomy in $G_\C$. Indeed, generic flat connections on generic 3-manifolds are irreducible. However, there also always exist fully reducible or ``abelian'' flat connections, whose holonomy lies in a maximal torus $T_\C\subset G_\C$, and which have $\text{Stab}(\CA) \simeq T_\C$. Since the holonomy of such a representation is abelian, it forms a representation of $H_1(M_3,\Z)$. It is the reducible flat connections, and in particular the abelian ones, that have been missing in previous constructions of $T[M_3]$.

\subsection*{Fivebranes and flat connections}

A simple analysis of the basic fivebrane system \eqref{surfeng} actually indicates the presence of multiple classes of flat connections in the moduli space of $T[M_3]$ (on a circle), and makes some predictions about their physical behavior.

Suppose that we wrap $N$ fivebranes on $M_3\subset T^*M_3$.
At low energies, the branes deform into the cotangent directions, forming a special Lagrangian cycle $\wt M_3\subset T^*M_3$ that is an $N$-sheeted cover of $\wt M_3$, \emph{cf}. \cite{CCV}. This cover is directly analogous to the Seiberg-Witten curve of four-dimensional $\CN=2$ theories, described from an M-theory perspective in \cite{Witten-M}. Notably, both connected and disconnected covers can appear. Configurations of the fivebranes that form a connected cover correspond to vacua of $T[M_3]$ that, via \eqref{mspaces} are irreducible flat $SL(N,\C)$ connections on $M_3$ itself. Disconnected covers correspond to vacua labelled by reducible flat connections. In the extreme case of a fully disconnected (\emph{i.e.} trivial) cover, we find vacua labelled by an abelian flat connection.

Now let us view this same system from the perspective of $\R^5$. 
If $\wt M_3$ is a connected cover, then all $N$ fivebranes must sit at the same point of the transverse $\R^2\subset \R^5$. Indeed, in this case there is really only a single fivebrane in the IR. However, if $\wt M_3$ is disconnected, then various subsets of the fivebranes may separate from each other in $\R^2$. In the case of a fully disconnected cover, all $N$ fivebranes may move independently in $\R^2$.
This suggests that 1) reducible flat connections do appear as vacua of $T[M_3]$; and moreover that 2) for each such ``reducible'' vacuum there are actually continuous flat directions in the moduli space of $T[M_3]$, parameterized by the possible relative deformations of fivebranes in $\R^2$.

The theory $T[M_3]$ has a $U(1)$ symmetry corresponding to rotations of $\R^2$: it is precisely the $U(1)_t$ symmetry that plays a prominent role throughout this paper.%
\footnote{More accurately, rotations of $\R^2$ are an R-symmetry in $T[M_3]$. We are proposing that $U(1)_t$ is a flavor symmetry that arises as a combination of $\R^2$ rotations and a second R-symmetry.} %
The flat directions corresponding to motion of fivebranes in $\R^2$ can be eliminated by turning on a twisted mass for $U(1)_t$ --- thus forcing all fivebranes to lie at the origin. It is only in the presence of this mass deformation that we truly expect a one-to-one correspondence between vacua of $T[M_3]$ (on a circle) and flat $SL(N,\C)$ connections on $M$.
Moreover, partition functions of the full theory $T[M_3]$ on $S^2 \times S^1$ (etc.) should only make sense in the presence of this mass deformation, for otherwise the additional flat directions will lead to divergences.

%%%%%%%%%%%%%%%%%%%%%%%%%%%%%%%%%%%%%%%%%%%%%%%%%%%%

\subsection*{Knot complements}

Throughout most of this paper we focus on examples where $M_3$ is a knot (or link) complement. In this case, the brane setup \eqref{surfeng} requires a slight modification, in order to account for the presence of the knot. The modification ultimately gives rise to the extra symmetry $U(1)_x$ in $T[M_3]$ that we introduced in \eqref{intro-diag}, and is related in Chern-Simons theory to a choice of boundary condition at the knot itself. 

To include a knot in the brane setup, one may introduce two intersecting stacks of fivebranes,
\be \label{brane-knot}
\begin{array}{r@{\quad}ccc}
 \text{space-time:} & \R^5 & \times & T^* \ol M_3 \\[.1cm]
 \text{$N$ M5-branes:} & \R^3 & \times & \;\;\; \ol M_3 \\[.1cm]
 \text{$N'$ M5$'$-branes:} & \R^3 & \times &  L_K\,,
\end{array}
\ee
where $\ol M_3$ is a closed 3-manifold and $L_K := N^*K \subset T^*\ol M_3$ is the conormal bundle of a knot $K\subset \ol M_3$. Configurations of this type were introduced by Ooguri and Vafa in \cite{OV}. The M5$'$ branes that wrap the knot have more noncompact directions than the M5 branes that wrap $\ol M_3$, so one may treat them as non-dynamical probes.

When $\ol M_3=S^3$, the large-$N$ duality of \cite{GV-I, GV-II} can be used to dualize the above configuration to 
\begin{align} \label{largeN}
\begin{array}{r@{\quad}ccc}
 \text{space-time:} & \R^5 & \times & X \\[.1cm]
 \text{$N'$ M5$'$-branes:} & \R^3 & \times & \widetilde L_K\,,
\end{array}
\end{align}
where $X$ is the resolved conifold, {\it i.e.} the total space of $\mathcal{O}(-1) \oplus \mathcal{O}(-1) \rightarrow \mathbb{P}^1$, and $\wt L_K$ is a related special Lagrangian submanifold. The K\"ahler modulus of the base $\cp^1$ is ${Ng_s}$.

The first configuration \eqref{brane-knot} leads, on one hand, to knot-complement theories $T[M_3]$ whose vacua on a circle are flat $SL(N,\C)$ connections on $M_3$. On the other hand, counting BPS states in the first configuration is known to produce doubly-graded $SU(N)$ knot homologies. Rotations of the $\R^2$ transverse to both sets of fivebranes become the homological $U(1)_t$ grading in the knot homologies, while $SO(2)_q\subset SO(3)_q$ rotations of the $\R^3$ common to both sets of fivebranes give the usual internal grading. For $N'\leq N$ probe branes, there is an additional $SU(N')_x$ global symmetry in $T[M_3]$ and in the knot homology theories. Its Cartan $U(1)_x^{N'-1}$ provides a grading related to the ``color'' or representation labelling a knot in the context of knot homology, provides twisted mass deformations of $T[M_3]$, and provides boundary conditions for $SL(N,\C)$ connections in complex Chern-Simons theory on $M_3$.

The second configuration contains yet another grading and leads to triply-graded knot homologies. Namely, $H_2(X,\Z)=\Z$ appears as a HOMFLY-like grading, usually incorporated with a fugacity $Q\sim e^{N g_s}$, sometimes also called ``$a$''. Three-dimensional $\CN=2$ theories arising from the second configuration were studied in \cite{FGSS-AD}.

From various perspectives on the above configurations, which we explain in this section, we find evidence that all flat connections on $M_3$ are inevitably captured in the physics of these systems, and so should appear as vacua of $T[M_3]$.

%%%%%%%%%%%%%%%%%%%%%%%%%%%%%%%%%%%%%%%%%%%%%%%%%%%%

\subsection*{The A-polynomial of a knot}

When $M_3=S^3\backslash K$ is a knot complement and $N=2$, the geometry of the moduli space of flat $SL(2,\C)$ connections on $M_3$ is (partially) captured by the A-polynomial of $K$ \cite{cooper-1994}.
Deformations of the A-polynomial, incorporating $t$ and $Q$ gradings, arise from studying BPS states and vacua of the systems \eqref{brane-knot}, \eqref{largeN}. 
As we will explain, these deformed polynomials provide some of the strongest evidence that the physical systems necessarily incorporate all flat connections. 
(There are obvious generalizations for $N> 2$, but we focus on $N=2$.)

Loosely speaking, the A-polynomial of a knot describes the set of flat $SL(2,\C)$ connections on the boundary of the knot complement $M_3=S^3\backslash K$ that can be extended as flat connections throughout the bulk of $M_3$.
More concretely, each flat connection $SL(2,\C)$ on a knot complement $M_3=S^3\backslash K$ is determined by its holonomy representation $\rho: \pi_1(M_3) \rightarrow SL(2,\C)$, whence
\begin{align}
\mathcal{M}_\text{flat}[M_3, \, SL(2,\mathbb{C})] \,=\, \text{Hom}(\pi_1(M_3), \, SL(2,\mathbb{C}))/\text{conj} \, .
\end{align}
Similarly, the space of flat connections on the boundary $\partial M_3\simeq T^2$ is 
\begin{align}
\mathcal{P} \, := \, \mathcal{M}_\text{flat}[T^2, \, SL(2,\mathbb{C})] \,=\, \text{Hom}(\pi_1(T^2), \, SL(2,\mathbb{C}))/\text{conj}\,. 
\end{align}
As $\pi_1(T^2) = \mathbb{Z} \times \mathbb{Z}$ is abelian, the holonomies along the two independent cycles on the boundary (usually called the meridian $\gamma_x$ and longitude $\gamma_y$) can be simultaneously conjugated to Jordan normal form
\begin{align}
\rho(\gamma_x) \, \sim \, \left( \begin{matrix} x & * \\ 0 & x^{-1}\end{matrix} \right), \qquad \rho(\gamma_y) \, \sim \, \left( \begin{matrix} y & * \\ 0 & y^{-1}\end{matrix} \right),
\end{align}
giving $\mathcal{P} = \{(x,y)\in \C^*\times \C^*\}/\Z_2$. Then the closure of the image of map $\mathcal{M}_\text{flat}[M_3, SL(2,\mathbb{C})]\to \mathcal{P}$ induced by restricting a flat connection to the boundary is the vanishing set of a complex curve
\begin{align}
\mathcal{L} = \{ (x,y) \in \mathcal{P} \, | \, A_K(x,y) \, = \, 0 \}, 
\end{align}
where $A_K$ is the classical A-polynomial.

For example, A-polynomials of the unknot and trefoil knot take form
\begin{align}
A_{\bf{0}_1}(x,y) = y-1, \quad A_{\bf{3}_1}(x,y) = (y-1)(y+x^3)\, .
\end{align}
Note that \emph{every} knot complement admits reducible flat $SL(2,\C)$ connections with abelian holonomy. Since the group $H_1(M_3)=\text{abel}(\pi_1(M_3))=\Z$ is generated by the meridian boundary cycle $\gamma_x$, while the longitude $\gamma_y$ is trivial in homology, these abelian flat connections are fully specified by the meridian eigenvalue $x$ and satisfy $y=1$. They contribute a universal factor $(y-1)$ to every A-polynomial.

The connection between the A-polynomial and Chern-Simons theory was explained in \cite{gukov-2003}: the quantization of $A_K(x,y)$ should provide a recursion relation for both $G=SU(2)$ and $G_\C=SL(2,\C)$ Chern-Simons partition functions on the knot complement. (A similar mathematical conjecture appeared in \cite{garoufalidis-2004}.) Conversely, from the semi-classical asymptotics of Chern-Simons partition functions, one can recover the classical A-polynomial.
 Many other interesting properties of the A-polynomial, which we don't explain further here, are reviewed (\emph{e.g.}) in \cite{GS-rev, FS-superA} and references therein.

%%%%%%%%%%%%%%%%%%%%%%%%%%%%%%%%%%%%%%%%%%%%%%%%%%%%

\subsection*{Physical realization of knot homologies}

There are several different ways to analyze the fivebrane systems \eqref{brane-knot}, \eqref{largeN} in order to extract knot polynomials and knot homologies. All of them are related to the counting of BPS states, either BPS M2 branes in M-theory, or standard BPS states in various field-theory limits. In summary:

\begin{itemize}
\item The BPS degeneracies appear in the A-twisted topological string partition function on $T^*\ol M_3$ or $X$, with topological branes supported on the appropriate Lagrangian cycles $\ol M_3,L_K$, or $\widetilde L_K$ \cite{GV-I, GV-II, OV}. In this context, only a protected index of BPS states appears, rather than an absolute count. The index is missing an independent homological grading $U(1)_t$. Accordingly, one obtains Jones polynomials from \eqref{brane-knot} and HOMFLY polynomials from \eqref{largeN} \cite{OV, LMV}.

Taking into account the additional $U(1)_t$ symmetry leads to ``refined'' counts of BPS states \cite{NekSW, HIVref}, which define the partition functions of ``refined'' topological string theory. It was conjectured in \cite{GSV, DGR} that the Hilbert space of BPS states in the resolved geometry \eqref{largeN} produces triply-graded HOMFLY knot homology
\be \mathcal{H}_\text{BPS}(K) \, = \, \mathcal{H}_\text{knot}(K)\,,
 \ee
and accordingly that a refined generating function of BPS states (including the $U(1)_t$ grading) produces the Poincar\'e polynomial of HOMFLY homology.
The refined topological vertex allowed very concrete computations of these Poincar\'e polynomials in some simple examples \cite{GIKV}.
More recent developments appear in (\emph{e.g.}) \cite{GS, FGS-VC, FGS-superA, Gorsky:2013jxa, FGSS-AD, Nawata, DSV-HOMFLY}.

\item Analyzing the system \eqref{brane-knot} from the perspective of an effective field theory on $M_3$ led to the definition of refined Chern-Simons theory \cite{AS-refinedCS}. This is a deformation of Chern-Simons theory with compact gauge group $G$. Computations of partition functions in refined $SU(N)$ Chern-Simons theory have reproduced the Poincar\'e polynomials of HOMFLY homologies for several Seifert-fibered knot complements.

\item Meanwhile, if one focuses on M5 and M5$'$ on $S^1 \times S^1 \times \mathbb{R}$ instead of $\mathbb{R}^3$ in \eqref{brane-knot}, taking a IIA limit along one circle and T-dualizing along the other, one obtains two stacks of D3 branes that intersect along $\mathbb{R} \times K$. In this context, a B-twisted Landau-Ginzburg theory on $\mathbb{R} \times K$ can be used to obtain doubly-graded $SU(N)$ knot homology \cite{GW-LG, GS}.

\item Finally, one can consider the effective 6d $(2,0)$ theory on the worldvolume of the $N$ M5 branes on $\R^3\times \ol M_3$, with codimension-two defects along $\R^2\times K$ arising from the M5$'$ branes. Deforming $\R^3$ to $\R\times D^2$, where $D^2$ denotes a cigar geometry, and compactifying along the cigar circle leads to a 5d $\CN=2$ theory on $\R\times \R_+\times \ol M_3$ with gauge group $G=SU(N)$. Its BPS equations take the schematic form \cite{Wfiveknots}
\be
F^+ - \frac{1}{4}B \times B - \frac{1}{2} D_y B \, = \, 0\,,\qquad 
F_{y \mu} + D^\nu D_{\nu \mu} \, = \, 0\,,
\ee
where $F$ is the field strength of a $G$-connection, $B$ is an adjoint-valued 2-form that is self-dual along $\R\times \ol M_3$, and $y$ denotes the coordinate on $\R_+$. It was proposed in \cite{Wfiveknots} that counting solutions to these equations again produces doubly-graded $SU(N)$ knot homology.

\end{itemize}

All these examples have to do with compact Chern-Simons theory, say for gauge group $G=SU(N)$, and its refinement/categorification. At this point, we should emphasize that the distinction between compact and complex Chern-Simons theory is controlled by how one treats the $\R^3\subset \R^5$ wrapped by fivebranes in \eqref{brane-knot}. In particular, one expects
\be \label{R3options}
\R^3\;\leadsto\; \begin{array}{c@{\;:\qquad}l}
 \text{$\R^2\times \R$ or $D^2\times \R$} & \text{$SU(N)$ homology} \\
 \text{$\R^2\times S^1$ or $D^2\times S^1$} & \text{$SU(N)$ polynomials/partition functions} \\
S^2\times \R & \text{$SL(N,\C)$ homology} \\
\text{$S^2\times S^1$ or $S^3/\Z_k$} & \text{$SL(N,\C)$ partition functions}\,.
\end{array}
\ee
Thus, it was mainly the geometries $\R^2\times \R$ or $D^2\times \R$ that appeared above.
In \emph{all} of these cases, one can discover contributions from flat $SL(N,\C)$ connections on $M_3$.

The way in which $SL(2,\C)$ flat connections ``contribute'' to $SU(2)$ knot homology was discussed in \cite{FGS-VC, FGSS-AD, FGS-superA}. In brief, the Poincar\'e polynomials $P_n(q,t)$ of $n$-colored $SU(2)$ knot homology satisfy a recursion relation encoded by an operator $\hat A_K(\hat x,\hat y;q,t)$, where $\hat x=q^n$ and $\hat y$ shifts $n\mapsto n+1$. As $q\to 1$, this operator becomes a classical three-variable polynomial $A_K(x,y;t)$ that is a deformation of the A-polynomial, in the sense that $A_K(x,y;t=-1)=A_K(x,y)$. Thus, one can recover the classical moduli space of flat $SL(2,\C)$ connections from $SU(2)$ homology. Crucially, while the classical A-polynomial always has an abelian factor $(y-1)$, it was observed in many examples that the $t$-deformed A-polynomial is irreducible and incorporates all flat connections (reducible or irreducible) on the same footing. This is a strong indication that the full fivebrane system knows about all complex flat connections on $M_3$ (regardless of what is happening on $\R^3$, as in \eqref{R3options}); and thus all flat connections should appear in $T[M_3]$ as well.

In the triply-graded setting, one finds a similar recursion relation for colored HOMFLY homologies, governed by an operator $\hat A_K(\hat x,\hat y;Q;q,t)$ whose $q\to 1$ limit is an ordinary four-variable polynomial $A_K(x,y;Q;t)$, which satisfies $A_K(x,y;Q=1,t=-1)=A_K(x,y)$. Again, it was observed in many examples that introducing the $Q$ deformation (with or without the $t$ deformation) leads to irreducible polynomials that necessarily incorporate all classical flat connections.

%%%%%%%%%%%%%%%%%%%%%%%%%%%%%%%%%%%%%%%%%%%%%%%%%%%%

\subsection*{Topological string theory on mirror geometry and $Q$-deformed A-polynomial}

It was proposed in \cite{AV-Q} that the string theory on the resolved conifold $X$ with a brane wrapping $\widetilde L_K$ \eqref{largeN} is mirror to string theory on a Calabi-Yau manifold $Y_K$ defined by
\begin{align}
\widetilde{X} = \{ x,y,u,v \, | \, A_K(x,y;Q) = uv, \, x, y \in \mathbb{C}^*, \, u, v \in \mathbb{C} \}\,, \label{mirrorCY}
\end{align}
where $A_K(x,y;Q) = A_K(x,y;Q;t=-1)$ is the \emph{same} $Q$-deformed A-polynomial we just described. In this case, the $Q$-deformed A-polynomial was given a mathematical interpretation in terms of knot contact homology \cite{AEV}. Again, this deformed polynomial was found to be irreducible, incorporating all flat connections.

Some examples of $Q$-deformed A-polynomials for the unknot and trefoil knot are
\begin{align}
A^\text{$Q$-def}_{\bf{0}_1}(x,y;a) \, &= \, a^{-1/2}(1-ax) - (1-x) y\, ,  \nonumber \\
A^\text{$Q$-def}_{\bf{3}_1}(x,y;a) \, &= \, a^2(x-1)x^3 - a(1 - x + 2(1-a)x^2 - ax^3 + a^2 x^4)y + (1-a^3 x) y^2 \nonumber
\end{align}
where we use the conventions of \cite{FGS-superA} and set $Q=a$.

%%%%%%%%%%%%%%%%%%%%%%%%%%%%%%%%%%%%%%%%%%%%%%%%%%%%

\subsection*{Flat connections on $M_3$ and the 4d-2d correspondence}

The correspondence between Vafa-Witten theory on 4-manifold $M_4$ and 2d $\mathcal{N}=(0,2)$ theory $T[M_4]$ also indicates that all flat connections should be taken into account in $T[M_3]$ \cite{GGP-4d}.

Consider multiple M5-branes wrapped on $M_4\times \mathbb{R}^2$, where $M_4$ is a 4-manifold with asymptotic boundary $M_3$. (This means that $M_4$ has a noncompact end with topology $M_3\times \R_+$.) Upon compactification on $M_4$, one finds 2d-3d coupled system where 2d $\mathcal{N}=(0,2)$ theory $T[M_4]$ on $\mathbb{R}^2$ provide half-BPS boundary condition for 3d $\mathcal{N}=2$ theory $T[M_3]$ on $\mathbb{R}^2 \times \mathbb{R}_+$ in the vicinity of the boundary. 

One can glue 4-manifolds along their common boundaries. In the corresponding supersymmetric field theory side, this gives a sequence of 2d $\mathcal{N}=(0,2)$ boundary conditions and domain walls in 3d $\mathcal{N}=2$ theories $T[M_3]$. It was observed in several explicit examples that this gluing only makes sense if $T[M_3]$ is a theory whose vacua on a circle account for all the flat complex connections on $M_3$.

For example, consider 4-manifolds $M_4^+$ and $M_4^-$ with boundaries $M_3^+$ and $M_3^-$, respectively, and a cobordism $B$ with boundary $\partial B = -\partial M_3^- \cup \partial M_3^+$ where the minus sign denotes orientation. Suppose that gluing $M_4^-$ to $B$ produces $M_4^+$,
\begin{align}
M_4^+ \, = \, M_4^- \cup_\varphi B\,. \label{4dgluing}
\end{align}
%where we assume $\varphi \, : \, M_3^- \rightarrow M_3^-$ is the identity map.
This translates to a 3d theory $T[M_3^-]$ on $\R^2\times I$ with boundary condition $T[M_4^-]$ on one end of the interval and an interface $T[B]$ at the other end; the interface separates $T[M_3^-]$ from a second 3d theory $T[M_3^+]$.

The BPS equations in Vafa-Witten theory on $M_4$ involve a $G$-connection $A$, an adjoint-valued scalar, and an adjoint-valued 2-form \cite{Vafa:1994tf}. However, under certain conditions, the scalar and the two-form can be set to zero and solutions are simply described by connections on $M_4$ with anti-self-dual curvature
\begin{align}
F^+_A = 0 \, . \label{SD}
\end{align}
For non-compact $M_4$ with asymptotic boundary $M_3$ and (say) $G=U(N)$, one must specify boundary conditions for $A$.
In order for the action to be finite, the connection $A$ should be asymptotically flat 
\begin{align}
A |_{M_3} = A_\rho, \quad F_{A_\rho} = 0\,,
\end{align}
where 
\begin{align}
\rho \, \in \, \mathcal{M}_\text{flat}(M_3; U(N)) \, = \, \text{Hom}(\pi_1(M_3), U(N)) /\text{conj.} \, 
\end{align}

For example, if $M_4$ is an $A$-type ALE space, the resolution of a singularity $M_4 = \widetilde{\R^4/\Z_p}$, the asymptotic boundary is a lens space $M_3 \simeq S^3/\mathbb{Z}_p$. There is a one-to-one correspondence between the $U(N)$ flat connection on $M_3$ and integrable representation of affine Lie algebra $\widehat{\mathfrak{su}}(p)_N$, which is in turn one-to-one correspondence to Young tableaux with at most $N$ columns and $p-1$ rows.
The Vafa-Witten partition function on the ALE space with $G=U(N)$ is given by the character of integrable representation $\rho$ of the affine Lie algebra $\widehat{\mathfrak{su}}(p)_N$ at level $N$,
\begin{align}
\mathcal{Z}^{U(N)}_\text{VW}[M_4]_\rho (q,x) \, = \, \chi^{\widehat{su}(p)_N}_\rho (q,x)\,,
\end{align}
where $q= e^{2\pi i \tau}$
and $x$ is a fugacity associated to the first Chern class of the gauge bundle $F_A\in H_2(M_4,\C)$. 

In order to see why all flat connections should be taken into account in $T[M_3]$, we consider a gluing of the type \eqref{4dgluing}. Take $G=U(N)$ and $M_4^+ = \widetilde{\R^4/\Z_{p+1}}$,  $M_4^- = \widetilde{\R^4/\Z_p}$ to be two different ALE spaces. The partition function on $M_4^+$ is given by
\begin{align}
\mathcal{Z}_\text{VW}^{U(N)}[M_4^+]_\rho (q,x) \, = \, \sum_\lambda \, \mathcal{Z}_\text{VW}^{U(N)}[B]_{\rho, \lambda} (q,x^\perp) \, \mathcal{Z}_\text{VW}^{U(N)}[M_4^-]_\lambda (q, x^\parallel) \label{VWpart}
\end{align}
where $x= (x^\perp, x^\parallel)$ are fugacities associated to the exponential of $H_2(M_4^+,\C) \, = \, H_2(B,\C) \; \oplus \; H_2(M_4^-,\C)$, and $\mathcal{Z}_\text{VW}^{U(N)}[B]^\lambda_\rho (q,x)$ is the branching function of the embedding $\widehat{su}(p)_N \subset \widehat{su}(p+1)_N$ or character of coset $\widehat{su}(p+1)_N \, / \, \widehat{su}(p)_N$,
\begin{align}
\mathcal{Z}_\text{VW}^{U(N)}[B]_{\rho, \lambda} (q,x) = \chi_{\rho, \lambda}^{\widehat{su}(p+1)_N \, / \, \widehat{su}(p)_N} (q,x)\,.
\end{align}
Crucially, the summation in  \eqref{VWpart} runs over $\lambda$ corresponding to all flat connections on $M_3$.\footnote{If we use {\it continuous} fugacity variables instead of discrete label such as $\rho$ in \eqref{VWpart} via appropriate transformation, the resulting expression for the partition function on $M_4^+$ can be expressed as the standard contour integral expression of the partition function (elliptic genus) of 2d $\mathcal{N}=(0,2)$ theory. Here, the fugacities are interpreted as those for flavor symmetry of $T[M_4]$ and also those for gauge symmetry of $T[M_3 = \partial M_4]$. For more detail explanation, see the original paper \cite{GGP-4d}.}

%%%%%%%%%%%%%%%%%%%%%%%%%%%%%%%%%%%%%%%%%%%%%%%%%%%%

\subsection*{Boundary conditions for flat connections}

Finally, we recall the field-theoretic derivation of the correspondence \eqref{mspaces} between flat connections $G_\C$ connections on $M_3$ and vacua of $T[M_3]$ on a circle, in the case that $M_3=S^3\backslash K$ is a knot complement.

As mentioned above in the context of knot homology, the effective low-energy field theory on the stack of $N$ M5 branes in \eqref{brane-knot} is a 6d $(2,0)$ theory of type $A_{N-1}$, with a codimension-two defect wrapping $\R^3\times K\subset \R^3\times \ol  M_3$. The precise type of the defect depends on the configuration of probe M5$'$ branes. We take $N'=N$, which produces a ``maximal'' defect. Compactification of the 6d theory on $S^3$ in the presence of this defect produces our knot-complement theory $T[M_3]$ on $\R^3$.

If we replace $\R^3$ with $\R^2\times S^1$, then we can compactify the 6d theory in two different ways. Reducing first along $\ol M_3$, we find $T[M_3]$ on $\R^2\times S^1$, whose vacua we want to analyze. Alternatively, reducing first along $S^1$ we find 5d $\CN=2$ Yang-Mills theory with gauge group $G=SU(N)$ on $\R^2\times S^3$, in the presence of a defect wrapping $\R^2\times K$. The supersymmetric vacuum equations in the 5d SYM theory \cite{BlauThompson-branes, BlauThompson-SYM, BSV} reduce to a three-dimensional analogue of Hitchin's equations along $M_3$, of the form
\be F_A=\phi\wedge\phi\,,\qquad d_A\phi=d_A *\phi = 0\,, \label{3dHit} \ee
where $A$ is a real $G$-connection and $\phi$ is an adjoint-valued 1-form (see also \cite{Wfiveknots, DGH}). These can be recast as flatness equations for a complex connection $\mathcal A=A+i\phi$ on $M_3$ (counted modulo complex gauge transformations).

Even at this stage, it is clear that any flat connection, reducible or irreducible, gives a solution to \eqref{3dHit} --- and therefore a supersymmetric vacuum for $T[M_3]$ on a circle. Reducible connections are somewhat special, in that they preserve some of the gauge symmetry of the 5d SYM theory, allowing additional Coulomb-branch fields to be turned on as well. These produce the extra flat directions that we found at the beginning of this section from a cursory analysis of the fivebrane system.

The presence of the maximal defect along the knot complicates matters only slightly. The upshot is that the defect comes with parameters valued in the Cartan subalgebra of $\mathfrak g= \mathfrak{su}_N$,
\be x\sim \text{Hol}_\gamma(\mathcal A)\,, \label{x-bdy-hol} \ee
which determines the \emph{eigenvalues} of the holonomy of a flat $G_\C$ connection around a small loop $\gamma$ linking the knot. Thus, the parameters of the defect determine a boundary condition for flat connections on $M_3$. When the complex eigenvalues $x$ are all generic, there is nothing more to say: solutions to the vacuum equations \eqref{3dHit} are all flat connections with diagonalizable boundary holonomy $x$, either reducible or irreducible. For $G_\C=SL(2,\C)$, these are the roots of the full A-polynomial $A_K(x,y)=(y-1)(...)$ at fixed $x$.

If the boundary eigenvalues happen to coincide, the situation is a little more subtle, but the outcome is the same: only eigenvalues of the boundary holonomy are fixed. The boundary holonomy itself may be either diagonal or it may have nontrivial Jordan blocks. A careful analysis of this situation appears in \cite{Ramified} (see especially Section 3.8). The boundary condition at the defect is local, and to study it one can take $M_3\simeq \R\times D^2$, where $D^2$ is a disc, with a ``knot'' wrapping $\R\times\{0\}$. 
If one imposes invariance along $\R$ and radial symmetry along $D^2$, equations \eqref{3dHit} reduce to Nahm's equations. Concretely, one writes
\begin{align}
\begin{split}
A &= a(r) \, d\theta + h(r) \, \frac{dr}{r} \\
\phi &= b(r) \, \frac{dr}{r} - c(r) \, d\theta\,,
\end{split} \label{ansatz}
\end{align}
where $(r,\theta)$ are polar coordinates on $D^2$. After setting $h(r)=0$ with a gauge transformation, equations \eqref{3dHit} become
\begin{align}
\frac{da}{ds} = [b, c], \quad \frac{db}{ds} = [c,a], \quad \frac{dc}{ds} = [a,b]\qquad (s=-\log r)\,. \label{Nahm}
\end{align}

The work of Kronheimer \cite{kronheimer1990hyper, kronheimer1990} relates solutions to Nahm's equations to complex coadjoint orbits in $\mathfrak g_\C$, which in turn determine a boundary condition for complex flat connections by specifying the conjugacy class of the boundary holonomy $\text{Hol}_\gamma(\mathcal A)$. The key point is that to study vacua in the physical 5d SYM theory one must consider the \emph{closure} of any given orbit or conjugacy class. If a conjugacy class is semi-simple, meaning that its eigenvalues $x$ are distinct, the class is automatically closed and we get a simple boundary condition \eqref{x-bdy-hol}. On the other hand, if eigenvalues coincide then a conjugacy class may have nontrivial Jordan-block structure. In this case, its closure always contains fully diagonalizable matrices as well.

For example, for $G_\C=SL(2,\C)$, when the eigenvalues both equal 1, we generically find a conjugacy class with elements of the form
\be \begin{pmatrix} 1 & w \\ 0 & 1 \end{pmatrix}\,. \label{parw} \ee
If $w\neq 0$, then conjugation by $SL(2,\C)$ can change the value of $w$ to any nonzero complex number. The closure of this conjugacy class contains the identity matrix with $w=0$. If $M_3$ is a hyperbolic knot complement, then a typical irreducible flat connection with unit eigenvalues (such as the one coming from the hyperbolic metric) has boundary holonomy of the form \eqref{parw} with $w\neq 0$; whereas the abelian flat connection has trivial holonomy, with $w=0$.

\begin{figure}[htb] \centering \includegraphics[width=5.0in]{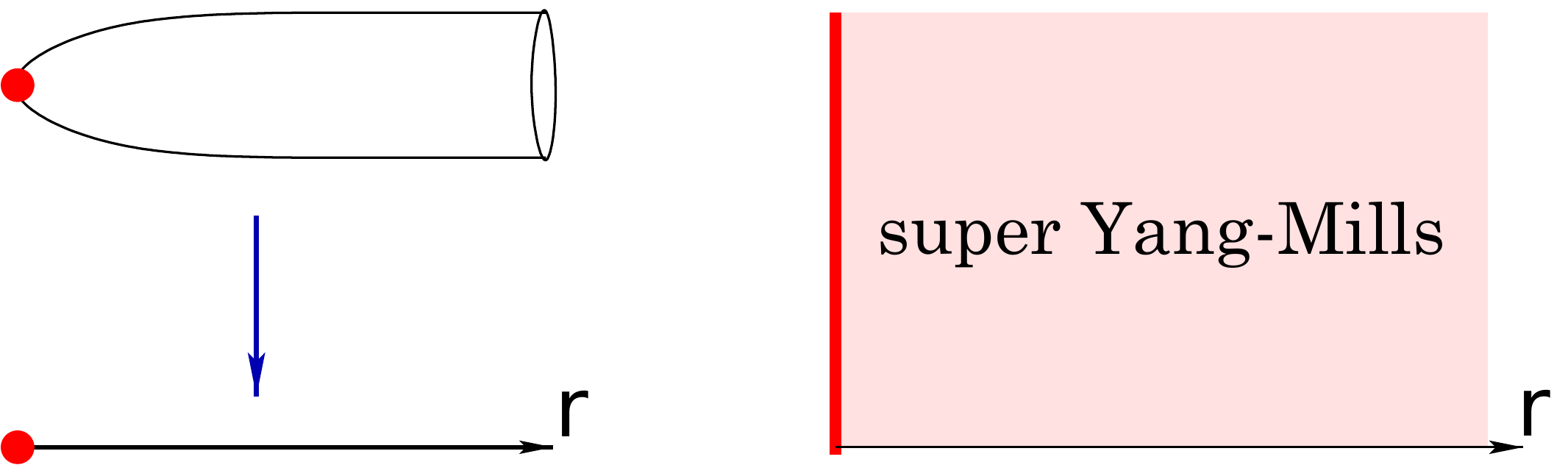}
  \caption{The six-dimensional $(2,0)$ theory with a codimension-2
    defect at the tip of the cigar reduces to 5d super-Yang-Mills
    theory with a non-trivial boundary condition.}
    \label{cigarfig}
\end{figure}

We remark that the standard classification of defects in the 6d (2,0) theory itself follows along similar lines. Namely, if one considers the 6d theory on $\R^4\times D^2$ with a defect supported on $\R^4\times \{0\}$, then compactification along the ``cigar circle'' in $D^2$ produces 5d SYM on $\R^4\times \R_+$ with a boundary condition (Figure \ref{cigarfig}). The relevant boundary conditions were analyzed by Gaiotto and Witten in \cite{GW-Sduality}, and found again to be classified by solutions to Nahm's equations. This is of course no accident; above we started with the 6d (2,0) theory on $\R^3\times S_t^1\times D^2$ and reduced first along $S_t^1$, then found Nahm's equations by asking for radial symmetry, which is equivalent to the cigar compactification. Again, the upshot is that the parameters labeling a defect in the 6d (2,0) theory can specify a local boundary condition at a knot $K$, which determines eigenvalues of a flat $G_\C$ connection (after $S^1_t$ compactification); but otherwise both reducible and irreducible flat connections must be allowed for in the space of vacua.

\subsection*{Flat connections and the partition function of complex Chern-Simons theory}

The above discussion focused on ``classical'' aspects of the 3d-3d correspondence, namely, comparisons of moduli spaces of vacua with flat connections. There is of course a ``quantum" correspondence as well: from conjectures of \cite{Yamazaki-3d, DGG, DGG-index} and the subsequent proofs in \cite{CJ-S3, LY-S2, Yagi-S2}, we expect an equality of partition functions
\be \label{3dCS}
\begin{array}{rl} \CZ(T[M_3;G],S^3_b)  &= \CZ(\text{$G_\C$ Chern-Simons at level $k=1$},M_3) \\[.2cm]
\CZ(T[M_3;G],S^2\times_qS^1)  &= \CZ(\text{$G_\C$ Chern-Simons at level $k=0$},M_3)\,.
\end{array}\ee
For more general level $k$, the partition functions of $G_\C$ Chern-Simons theory are expected to be equal to the partition functions of $T[M_3]$ on squashed Lens spaces $S^3/\mathbb{Z}_k$ \cite{D-levelk}. Again, if $M_3$ is a knot complement, there is additional data on both sides: on the LHS, fixed masses and R-charges/Wilson lines; and on the RHS, a fixed meridian holonomy eigenvalues as a boundary condition in $G_\C$ Chern-Simons theory.

If it is so important that all flat connections appear in the vacua of $T[M_3]$, one may wonder why they have not appeared in previous computations of non-perturbative partition functions of the corresponding $G_\C$ Chern-Simons theory. 
Indeed, there are now systematic definitions of $SL(2,\C)$ partition functions for (say) hyperbolic knot complements \cite{KashAnd, AK-new, GHRS-index, D-levelk} (following \cite{hikami-2006, DGLZ, Dimofte-QRS, DGG-index}). These partition functions can be decomposed into holomorphic blocks \cite{BDP-blocks} that correspond to contributions from different classical flat connections, and so far reducible flat connections have never appeared.

There is an easy answer to this puzzle from the perspective of Chern-Simons theory. The partition function of complex Chern-Simons theory can be written as a sum of contributions of different flat connections \cite{gukov-2003, Wit-anal}
\be \CZ(\text{CS on $M_3$}) = \sum_{\text{flat $\CA$}} \frac{1}{\text{Vol}(\text{Stab}(\CA))} \CZ_\CA\,. \label{CS-sum} \ee
Each contribution, however, is weighted by volume of subgroup of the gauge group $G_\C$ that preserves $\CA$, \emph{i.e.} the volume of the stabilizer. For irreducible $\CA$ the volume is trivial and for reducible $\CA$ the volume is infinite. Therefore, if $M_3$ is such that it admits both reducible and irreducible flat connections (for example, if $M_3$ is a hyperbolic knot complement), then the non-perturbative partition function will be dominated by the irreducible connections, with reducible connections contributing exactly zero.

A better answer would be that the current definitions of partition functions in complex Chern-Simons theory are missing a grading --- namely, the $U(1)_t$ that played a major role in knot homology and that is the star of this paper. We expect that properly incorporating this grading will regularize any infinities in \eqref{CS-sum}, and allow all flat connections to contribute on equal, finite, footing. As we will see in examples throughout the rest of the paper, such a grading seems essential for promoting complex Chern-Simons theory to a full TQFT, allowing generic cutting and gluing operations.

%%%%%%%%%%%%%%%%%%%%%%%%%%%%%%%%%%%%%%%%%%%%%%%%%%%%%%%%%%%%%%%%%%%%%%%%%%%%%%%%%%%%%%%%%%%%%%%%%%%%%%%%%%%%%%%%%%%%%%%%%%%%%%%%%%%%%%

\acknowledgments{We would like to thank S.~Nawata, S.~Razamat, B.~Willett, and E.~Witten for useful discussions.
Many ideas in this paper were developed at the 2013 Simons Summer Workshop in Mathematics and Physics, whose support and hospitality we gratefully acknowledge.
The work of H.J.C. and S.G. is supported in part by DOE Grant DE-FG02-92ER40701. The work of T.D. is supported in part by DOE grant DE-SC0009988. This work has also been supported by the ERC Starting Grant no. 335739 \emph{``Quantum fields and knot homologies''}, funded by the European Research Council under the European Union's Seventh Framework Programme.
}

\newpage

\bibliographystyle{JHEP_TD}
\bibliography{toolbox}

\end{document}